\documentclass[longauth]{aa} 

\usepackage{color}
\newcommand{\gtsim}{\protect\raisebox{-0.5ex}{$\:\stackrel{\textstyle >}
        {\sim}\:$}}
\newcommand{\ltsim}{\protect\raisebox{-0.5ex}{$\:\stackrel{\textstyle <}
        {\sim}\:$}}
\newcommand{\teff}{$T_{\rm eff}$}
\newcommand{\logg}{$\log g$}
\newcommand{\csi}{$\xi$}
\newcommand{\feh}{[Fe/H]} 

\newcommand{\vsini}{$v \sin i$}

\usepackage{graphicx}
\usepackage{txfonts}
\usepackage{aalongtable,lscape}   
\usepackage{natbib,twoopt}
\bibpunct{(}{)}{;}{a}{}{,} 
\makeatother

\begin{document}

\title{The GAPS Programme with HARPS-N at TNG}
\subtitle{XXXV. Fundamental properties of transiting exoplanet host stars\thanks{Based on observations made with the Italian {\it Telescopio Nazionale Galileo} (TNG), operated on the 
island of La Palma by the INAF - {\it Fundaci\'on Galileo Galilei} at the {\it Roque de los Muchachos} Observatory of the {\it Instituto de Astrof\'isica de Canarias} (IAC) in the 
framework of the large programme Global Architecture of Planetary Systems (GAPS; P.I. A. Sozzetti).}}
   
\author{
K. Biazzo\inst{\ref{oaroma}} \and 
V. D'Orazi\inst{\ref{oapadova}} \and 
S. Desidera\inst{\ref{oapadova}} \and 
D. Turrini\inst{\ref{oatorino}} \and 
S. Benatti\inst{\ref{oapalermo}} \and 
R. Gratton\inst{\ref{oapadova}} \and 
L. Magrini\inst{\ref{oaarcetri}} \and 
A. Sozzetti\inst{\ref{oatorino}} \and 
M. Baratella\inst{\ref{aip}} \and 
A. S. Bonomo\inst{\ref{oatorino}} \and 
F.~Borsa\inst{\ref{oabrera}} \and 
R. Claudi\inst{\ref{oapadova}} \and 
E. Covino\inst{\ref{oacapodimonte}} \and 
M. Damasso\inst{\ref{oatorino}} \and 
M. P. Di Mauro\inst{\ref{iaps}} \and
A.~F.~Lanza\inst{\ref{oacatania}} \and 
A.~Maggio\inst{\ref{oapalermo}} \and 
L. Malavolta\inst{\ref{unipadova}, \ref{oapadova}} \and 
J.~Maldonado\inst{\ref{oapalermo}} \and 
F. Marzari\inst{\ref{unipadova}} \and 
G. Micela\inst{\ref{oapalermo}} \and 
E.~Poretti\inst{\ref{tng}, \ref{oabrera}} \and 
F. Vitello\inst{\ref{irabologna}} \and 
L. Affer\inst{\ref{oapalermo}} \and 
A.~Bignamini\inst{\ref{oatrieste}} \and 
I. Carleo\inst{\ref{wesleyan}} \and 
R.~Cosentino\inst{\ref{tng}} \and 
A.~F.~M.~Fiorenzano\inst{\ref{tng}} \and 
P. Giacobbe\inst{\ref{oatorino}} \and 
A.~Harutyunyan\inst{\ref{tng}} \and 
G. Leto\inst{\ref{oacatania}} \and 
L.~Mancini\inst{\ref{unitor}} \and 
E.~Molinari\inst{\ref{oacagliari}} \and 
M.~Molinaro\inst{\ref{oatrieste}} \and 
D.~Nardiello\inst{\ref{oapadova}} \and 
V.~Nascimbeni\inst{\ref{oapadova}} \and 
I. Pagano\inst{\ref{oacatania}} \and 
M.~Pedani\inst{\ref{tng}} \and 
G.~Piotto\inst{\ref{unipadova}} \and 
M.~Rainer\inst{\ref{oabrera}} \and    
G. Scandariato\inst{\ref{oacatania}}
}
\offprints{K. Biazzo}
\mail{katia.biazzo@inaf.it}

\institute{INAF - Osservatorio Astronomico di Roma, Via Frascati 33, I-00040 Monte Porzio Catone (RM), Italy, \label{oaroma} \email{katia.biazzo@inaf.it}
\and INAF - Osservatorio Astronomico di Padova, Vicolo dell'Osservatorio 5, I-35122 Padova, Italy \label{oapadova}
\and INAF - Osservatorio Astrofisico di Torino, Via Osservatorio 20, I-10025 Pino Torinese, Italy \label{oatorino}
\and INAF - Osservatorio Astronomico di Palermo, Piazza del Parlamento 1, I-90134 Palermo, Italy \label{oapalermo}
\and INAF - Osservatorio Astrofisico di Arcetri, Largo E. Fermi 5, I-50125 Firenze, Italy \label{oaarcetri}
\and Leibniz-Institut f\"ur Astrophysik Potsdam, An der Sternwarte 16, D-14482 Potsdam, Germany \label{aip}
\and INAF - Osservatorio Astronomico di Brera, Via E. Bianchi 46, I-23807 Merate (LC), Italy \label{oabrera}
\and INAF - Osservatorio Astronomico di Capodimonte, Salita Moiariello 16, I-80131 Napoli, Italy \label{oacapodimonte}
\and INAF - Istituto di Astrofisica e Planetologia Spaziali, Via del Fosso del Cavaliere 100, I-00133 Roma, Italy \label{iaps}
\and INAF - Osservatorio Astrofisico di Catania, Via S. Sofia 78, I-95123 Catania, Italy  \label{oacatania}
\and Dipartimento di Fisica e Astronomia, Universit$\grave{\rm a}$ di Padova, Vicolo dell'Osservatorio 2, I-35122 Padova, Italy  \label{unipadova}
\and Fundaci{\'o}n Galileo Galilei - INAF, Rambla Jos{\'e} Ana Fernandez P{\'e}rez 7, E-38712 Bre$\tilde{\rm n}$a Baja (TF), Spain  \label{tng}
\and INAF - Istituto di Radioastronomia, Via Gobetti 101, I-40127 Bologna, Italy \label{irabologna}
\and INAF - Osservatorio Astronomico di Trieste, Via Tiepolo 11, I-34143 Trieste, Italy \label{oatrieste}
\and Astronomy Department, 96 Foss Hill Dr, Van Vleck Observatory, Wesleyan University, Middletown, CT-06459, US  \label{wesleyan}
\and Dipartimento di Fisica, Universit\`a di Roma Tor Vergata, Via della Ricerca Scientifica 1, I-00133 Roma, Italy \label{unitor}
\and INAF - Osservatorio di Cagliari, Via della Scienza 5, I-09047 Selargius (CA), Italy  \label{oacagliari}}

\date{Received 03/03/2022/ accepted 23/05/2022}

\abstract
{Exoplanetary properties strongly depend on stellar properties: "to know the planet" with accuracy and precision it is necessary "to know the star" as accurately and precisely as possible.}
{Our immediate aim is to characterize in a homogeneous and accurate way a sample of 27 transiting planet-hosting stars observed within the Global Architecture of Planetary System program. 
For the wide visual binary XO-2, we considered both components (N: hosting a transiting planet; S: without a known transiting planet). Our final goal is to widely analyze the sample by 
deriving several stellar properties, abundances of many elements, kinematic parameters, and discuss them in a context of planetary formation.}
{
We determined stellar parameters (effective temperature, surface gravity, rotational velocity) and abundances of 26 elements (Li, C, N, O, Na, Mg, Al, Si, S, Ca, Sc, Ti, V, Cr, Fe, Mn, Co, Ni, Cu, Zn, Y, 
Zr, Ba, La, Nd, Eu). Our study is based on high-resolution HARPS-N at TNG and FEROS at ESO spectra and uniform techniques. Depending on stellar parameters and chemical elements, we used 
line equivalent widths or spectral synthesis methods. We derived kinematic properties taking advantage of Gaia data and estimated for the first time in exoplanet host stars ages using 
elemental ratios as chemical clocks.
}
{ 
Effective temperature of our stars is of $\sim$4400-6700\,K, while iron abundance [Fe/H] is within $-$0.3 and 0.4\,dex. Lithium is present in seven stars. [X/H] and [X/Fe] 
abundances versus [Fe/H] are consistent with the Galactic Chemical Evolution. The dependence of [X/Fe] with the condensation temperature is critically analyzed with 
respect to stellar and kinematic properties. All targets with measured C and O abundances show C/O<0.8, compatible with Si present in rock-forming minerals. Mean C/O 
and [C/O] are slightly lower than the Sun. Most of targets show 1.0<Mg/Si<1.5, compatible with Mg distributed between olivine and pyroxene, and mean Mg/Si lower than the Sun. 
HAT-P-26, the target hosting the lowest-mass planet, shows the highest Mg/Si ratio. From our chemo-dinamical analysis we find agreement between ages and position within the 
Galactic disk. Finally, we note a tendency for higher density planets to be around metal-rich stars and hints of higher stellar abundances of some volatiles (e.g., O) for 
lower mass planets. We cannot exclude that part of our results could be also related to the location of the stars within the Galactic disk.
}
{We try to trace the planetary migration scenario from the composition of the planets related to the chemical composition of the hosting stars. This kind of study 
will be useful for upcoming space missions data to get more insights into the formation/migration mechanisms.}

	   
\titlerunning{Fundamental properties of transiting exoplanet host stars}
\authorrunning{K. Biazzo et al.}
\maketitle

\section{Introduction}
\label{sec:intro}
Many of the known extrasolar planets until around ten years ago have been unveiled by the use of the Doppler radial velocity (RV). Alone, the RVs only yield partial information on orbital elements of the planets and their minimum masses, and no insight is obtained about the planetary physical properties, like their true masses, radii, and mean densities. Such additional information is available in the case of planetary transits. In recent years, major planet-search programs using the photometric transit technique started to deliver interesting results, giving a new breath to the study of exoplanets. Now, the known transiting extrasolar planets are growing in number, giving us information about the physical properties of orbiting planets. Complementary follow-up observations of the transits have further permitted us access to the atmospheres of these worlds, giving important clues about the physics of these atmospheres (see, e.g., \citealt{Charbonneauetal2002, Madhusudhan2019}, and references therein). Therefore, transiting exoplanets provide us with unique laboratories to test theories of exoplanet formation and evolution with relatively high precision.

When deriving properties for a statistically significant sample of exoplanets, a precise and homogeneous determination of the stellar parameters is crucial for an accurate characterization of the parent stars and, in turn, of the planet properties. The connection between star and planet is perpetually interweaved, so one cannot be studied without accounting for the other. For instance, when the planet is transiting, from the bulk planetary density it is possible to have hints of the internal planetary structure and the gas/ice/rock ratios. However, the required accurate estimates of planetary mass and radius/density necessarily rely on precise determination of mass and radius of the hosting star. The derivation of stellar mass (and radius) is strongly connected to the effective temperature (\teff), surface gravity (\logg), and iron abundance (\feh) of the star, besides to be dependent on the evolutionary tracks considered. Therefore, planetary properties are critically dependent on the properties of the hosting stars (see, e.g., \citealt{Torresetal2012, Santosetal2013, Sousaetal2015, Maldonadoetal2018, Maldonadoetal2019}). Moreover, several studies have pointed out the existence of correlations between characteristics of the host stars and properties/frequencies of their planetary systems, in particular for giant transiting planets, for which the present work is focused on. As a result, the evidences supporting the correlation between stellar metallicity and occurrence rate of giant planets (e.g., \citealt{Santosetal2004, Valentietal2005}) and the weakening of this correlation towards lower regimes of planetary mass (e.g., \citealt{Sousaetal2008, Ghezzietal2010}) or for wide-orbit planets (\citealt{Swastiketal2021}), the connection between planet radius and stellar metallicity (\citealt{Buchhaveetal2014}), the correlation between stellar metallicity and planetary heavy-metal content (e.g., \citealt{Guillotetal2006}), the trend between orbit eccentricity and star metallicity (\citealt{DawsonMurray2013}), the role of the abundances of $\alpha$-elements (\citealt{Robinsonetal2006, Gonzalez2009, Adibekyanetal2012c, Adibekyanetal2015}) or of the Li depletion observed for massive planet-hosting stars (e.g., \citealt{delgadomenaetal2015}) are becoming more and more numerous thanks to the constant increase of new planet discoveries. 

Again, information on the hosting star chemical composition is important to separate the signatures left on the planet during its formation and migration from those due to the star. In fact, the various physical processes participating in forming giant planets (e.g., planet-disk interaction, planet-planet scattering, in-situ formation, multiple generation of embryos) are thought to result in differences in atmospheric composition depending on the enrichment by chemical elements present at the formation site or accreted during migration (\citealt{Voelkeletal2022}, and references therein). In particular, some studies suggest that planetary carbon-to-oxygen ratio and metallicity with respect to the corresponding stellar values could provide constraints on the original formation region of the planet with respect to the H$_2$O, CO$_2$, and CO snow lines, and on the time when planet migrated to its present orbit (e.g., \citealt{Obergetal2011}). For instance, enhanced C/O ratio of the planet compared to its host star was found to be produced when the planet formed far beyond the water snowline, predominantly by gas accretion, and then underwent a subsequent disk-free (high-eccentricity) inward migration (\citealt{Madhusudhanetal2014}). Therefore, C/O ratio in planet-host stars can provide key information about the protoplanetary disk regions in which the planet was formed, as abundances of the volatiles in the disk gas and solids are heavily affected by the disk radial temperature profile and, therefore, by the distance from the host star while planets accrete. On the contrary, other elemental ratios, like magnesium-to-silicon (Mg/Si) which governs the distribution of silicates in the protoplanetary disk, do not depend so strongly on the distance to the stars as the C/O ratio does (\citealt{Thiabaudetal2015b}). However, recent studies highlighted how the use of multiple elemental ratios involving elements with a high contrast in volatility (like S/N, Si/N, S/O, N/O) can provide more detailed and robust constraint on the formation and migration history of giant planets than possible with C/O alone (\citealt{Turrinietal2021a, Turrinietal2021b}). The same studies argued how planetary elemental ratios normalized to stellar abundances can provide more unequivocal indications and allow for a more straightforward comparison between different planets orbiting different host stars (\citealt{Turrinietal2021a, Turrinietal2021b, KoleckiWang2022}). 

With this in mind, the requirement for homogeneity and precision for the stellar parameters and elemental abundances becomes even more crucial: fundamental parameters of large samples of planet-hosting stars are often found in the literature as the result of analysis performed by different methodologies, resulting in an inhomogeneous census of stars with planets. All these arguments highlight how for achieving a comprehensive characterization of exoplanets, the homogeneous and precise determination of the fundamental properties and elemental abundances of the hosting stars is pivotal. We therefore analyzed a sample of transiting planet host stars with the HARPS-N spectrograph at the {\it Telescopio Nazionale Galileo} (TNG) within the Global Architecture of Planetary System (GAPS; \citealt{Covinoetal2013}) project. Our immediate aim is twofold: $i.$ we apply an accurate procedure to derive stellar parameters, global properties, abundances of multiple elements, and kinematic properties of transiting planet host stars in a homogeneous and as precise as possible way using high-quality data; $ii.$ we study possible relationships between astrophysical, kinematic, chemical parameters of exoplanet host stars and properties of their transiting planets, thus providing necessary information for future studies of their exoplanets with new facilities. We are aware that our procedure is based on non automatic tools and therefore it is time consuming, but we think that such a kind of approach can be used as a benchmark analysis for interpreting the composition, the origin and evolution of planets with current/future theoretical models and statistical studies. For instance, over the last decades, thanks to the successful photometric space missions ({\it CoRoT}, {\it Kepler/K2} and {\it TESS}), a remarkable synergy has emerged between ground-based spectroscopy and asteroseismic techniques for the determination of accurate fundamental parameters of exoplanet host stars (e.g., \citealt{DiMauroetal2011, Chaplinetal2013}). Similarly, the efforts done for large sample of data applying automatic procedures to derive stellar parameters and iron abundances, like, e.g., those presented by \cite{Sousaetal2021}, are very useful at the same level for the statistical approach.
 
The outline of this paper is as follows. We first present in Sect.\,\ref{sec:obs} the spectroscopic dataset. In Sect.\,\ref{sec:data_analysis}, we describe the measurements of stellar parameters and elemental abundances for 26 species. We then present our results and discuss the behavior of the elemental abundances of the stars with respect to their kinematic or global properties and with respect to planetary properties in Sect.\,\ref{sec:result_discussion}. In Sect.\,\ref{sec:conclusions} we draw our conclusions. 

\section{Stellar sample, observations, and data reduction}
\label{sec:obs}
The stellar sample was selected within two GAPS sub-programs aimed at searching for additional companions in known systems and at determining the Rossiter-McLaughlin effect in transiting systems (see, e.g., \citealt{bonomoetal2017}). Within this sample, we selected targets with spectral types from F5 to K7 (see Table\,\ref{tab:basic_info}) and with rotational velocity ($v \sin i$) known from the NASA Exoplanet Archive smaller than $10$\,km/s. This 
was done to avoid strong problems due to line blending or the presence of molecular lines, which must be dealt with different procedures than those adopted in this work.
In the end, we analyzed a total of 28 targets with $9.3 \ltsim V \ltsim 13.4$ mag, of which 13 of the HATNet Exoplanet Survey (HAT-P; \citealt{bakosetal2004}) in the north, 5 
of the Wide Angle Search for Planets (WASP; \citealt{kaneetal2003}), two in common between HATNet and WASP (i.e. WASP11 or HAT-P-10A, HAT-P-30 or WASP51), one of the 
Kilodegree Extremely Little Telescope survey (KELT; \citealt{pepperetal2007}), two of the XO Project (\citealt{mcculloughetal2005}), two of the Qatar Exoplanet 
Survey (\citealt{alsubaietal2013}), and one of the Trans-Atlantic Exoplanet Survey (TrES; \citealt{odonovan2007}). Twenty-seven stars host massive planets with 
masses ($M_{\rm p}$) from $\sim 0.2$ to $\sim 7.3\,M_{\rm Jup}$, and one star (HAT-P-26) hosts a Neptune-size mass planet, with $\sim 0.06\,M_{\rm Jup}$. All stars but one (namely, XO-2S, observed as a wide companion of the known planet host, \citealt{desideraetal2014}) host transiting planets. Table\,\ref{tab:basic_info} lists basic information on the final sample taken from the literature, together with some characteristics of the planets. Interestingly, the transits of some of our targets (e.g., HAT-P-12, HAT-P-26, WASP-43) will be observed within the Cycle\,1 of the Guaranteed Time Observations of the {\it James Webb Space Telescope}\footnote{see https://www.stsci.edu/jwst/science-execution/approved-programs/cycle-1-gto} (JWST).

Observations were performed between the end of 2012 and 2016 with the high resolution HARPS-N at TNG spectrograph ($R \sim 115\,000$, $\lambda \sim 3\,900-6\,900$\,\AA; 
\citealt{cosentinoetal2012}). Solar spectra were also obtained through observations of the asteroid Vesta. Spectra reduction was obtained using the HARPS-N 
instrument Data Reduction Software pipeline (see \citealt{bonomoetal2017} which present the radial velocity curves of the sample). Standard steps for data reduction and appropriate cross-correlation masks were applied to each target. Final high signal-to-noise ratios ($SNR$) merged spectra for each target star and for the Sun were obtained by co-adding, after proper shift to the rest frame by the corresponding RV. All individual spectra of the given star and of Vesta, respectively, reached a $SNR$ (per pixel at $\lambda \sim$6000\,\AA) between 100 and 300 for the targets (except for Qatar-2, the faintest one in the sample, with mean $SNR$ around 50) and $SNR\sim300$ for the solar spectrum.

Due to the importance to derive accurate oxygen abundance for as many targets as possible (see, e.g., discussions in Sects.\,\ref{sec:CNOS}, \ref{sec:mgsi_co_fe}, \ref{sec:abundances_plmass}, \ref{sec:traceformation}), and since very useful oxygen 
lines like those of the \ion{O}{i} triplet at $\sim$777\,nm are not present within the HARPS-N spectral range, we decided to search for processed FEROS at ESO spectra ($R \sim 48\,000$, $\lambda \sim 3\,600-9\,000$\,\AA; \citealt{kauferetal1999}) available 
in the ESO archive. In the end, we found FEROS spectra for the Sun and seven stars (namely, HAT-P-30, WASP-54, HAT-P-17, HAT-P-26, HAT-P-20, Qatar\,2, WASP-43). Typical $SNR$ of these spectra was greater than $\sim$50 at $\lambda \sim$6000\,\AA. 
These spectra allowed us also to derive nitrogen abundance from high-excitation permitted lines (see Sect.\,\ref{sec:abun_anal}). For homogeneity reasons, we verified that, for the targets observed with FEROS and using the same method applied for HARPS-N
spectra based on spectral synthesis of oxygen $\lambda$6300.3\,\AA\,line and CN molecule (see Sects.\,\ref{sec:abun_anal}), we derived O and N abundances very close (within the uncertainties) to those obtained with HARPS-N spectra. On the other hand, 
we verified that for the two targets for which we could measure oxygen and nitrogen both using FEROS and HARPS-N diagnostics we obtained very similar results, within the errors.

We point out that for 16 targets we performed within the GAPS project a preliminary analysis of some astrophysical parameters (mainly only effective temperature, surface gravity, and iron abundance), but it was non homogeneous and based on previous version of tools, model atmospheres, and line lists (see, \citealt{Covinoetal2013, desideraetal2014, Espositoetal2014, damassoetal2015a, biazzoetal2015, Sozzettietal2015, Mancinietal2015, damassoetal2015b, Espositoetal2017, Mancinietal2018}). We remark here that the analysis performed in this work is absolutely new and innovative, homogeneous, and aimed at a as global as possible characterization of our sample of stars. 

\section{Data analysis}
\label{sec:data_analysis}
\subsection{Stellar parameters and iron abundance}
\label{sec:stellar_parameters}
We derived stellar parameters and iron abundance using the code MOOG (\citealt{sneden1973}; version 2017) with the driver {\it abfind}, that assumes local thermodynamic equilibrium (LTE) and where the radiative and Stark broadenings are treated in a standard way. For collisional broadening, we used the \cite{barklemetal2000} prescriptions for damping values. We used plane-parallel model 
atmospheres linearly interpolated from the ATLAS9 grids of \cite{castellikurucz2003}, with solar-scaled chemical composition and new opacities ({\sc odfnew}). 

Effective temperature \teff, surface gravity \logg, microturbulence velocity \csi, and iron abundance \feh\, were measured through a method based on equivalent widths (EWs) of 
iron lines. We adopted the list of iron lines by \cite{biazzoetal2012} and corrected the atomic parameters of the \ion{Fe}{ii} line at $\lambda$6516.08\,\AA\,with the NIST most recent values 
(\citealt{kramidaetal2020}) because they led to most reliable solar iron abundance measurements. The iron line list was built to minimize potential correlations between the atmospheric 
parameters, by including lines of different strengths at a given excitation potential and by having a wide distribution of excitation potentials. 
In addition, we keep in our line list only iron lines that could be reliably measured at our spectral resolution. In the end, a total of 82\,\ion{Fe}{i}+11\,\ion{Fe}{ii} lines were considered. The EWs 
of the target stars were measured by means of a direct integration or Gaussian fitting procedure using the IRAF\footnote{IRAF is distributed by the National Optical Astronomy 
Observatories, which are operated by the Association of Universities for Research in Astronomy, Inc., under the cooperative agreement with the National Science Foundation. 
NOAO stopped supporting IRAF, see \tt{https://iraf-community.github.io/}} {\sc splot} task. We discarded strong lines ($EW>150$\,m\AA) and those lines with fitting errors larger 
than $3\sigma$. Each line equivalent width was controlled and measured several times and particular attention was paid to 
the continuum tracing. In fact, the continuum placement of each stellar line was defined looking at the continuum position of the same line in the solar spectrum and the same 
criteria (both for the continuum definition and the intervals selected for the integration) were adopted. \teff\,and \csi\,were determined by imposing the condition that the \ion{Fe}{i} abundance 
does not depend on the excitation potential and equivalent width of the lines, while \logg\,was obtained from the \ion{Fe}{i}/\ion{Fe}{ii} ionization equilibrium. We used an iterative 
procedure, by changing the parameters at steps of 5\,K in \teff, 0.01\,km/s in \csi, and 0.01\,dex in \logg, and requiring that the slope of the \ion{Fe}{i} abundance with respect to 
the excitation potential (for \teff) or EW (for \csi) was close to zero and that the difference between the mean iron abundance obtained from the \ion{Fe}{i} and \ion{Fe}{ii} lines was 
lower than 0.01\,dex (for \logg). We therefore derived stellar parameters with internal accuracy (at 3$\sigma$) ranging from 15 to 90\,K in \teff, from 0.09 to 0.19\,dex in \logg, 
from 0.02 to 0.41\,km/s in \csi, and from 0.07 to 0.15\,dex in [Fe/H] (see Table\,\ref{tab:final_stellar_param_abund}). As a sanity check, final iron abundances of each target was plotted 
as a function of stellar \teff, \logg, and \csi, and \vsini\,(for the measurement of the rotational velocity see Sect.\,\ref{sec:abun_anal}) to look for possible presence of spurious trends due, e.g., to line blending. 

Our analysis was performed differentially with respect to the solar spectrum, which was used as a reference to minimize errors due to uncertainties in measurements of EWs, 
continuum definition, atomic parameters, and model atmospheres. We therefore analyzed the co-added spectrum of Vesta, using our line list, imposing the solar effective temperature 
$T_{\rm eff \odot}=5770$\,K and surface gravity $\log g_\odot=4.44$\,dex, and leaving the microturbulence free to vary. Final optimization was obtained for $\xi_\odot=0.99$\,km/s, 
leading to $\log n{\rm (\ion{Fe}{i})}_\odot =7.49 \pm 0.05$ and $\log n{\rm (\ion{Fe}{ii})}_\odot =7.49 \pm 0.05$.

\subsection{Abundance of other elements and rotational velocity}
\label{sec:abun_anal}

For elements other than iron, we applied spectroscopic techniques based on line EWs and spectral synthesis depending of the element. We also measured the rotational velocity through spectral 
synthesis.

\subsubsection{Analysis based on equivalent widths}

Once stellar parameters and iron abundance were measured, we computed abundances of other elements ([X/H]\footnote{Throughout the paper, the abundance of the X 
element is given as [X/H]=$\log \frac{\epsilon{\rm (X)}}{\epsilon{\rm (H)}} + 12$, where $\log \epsilon(X)$ is the absolute abundance.}) using the MOOG code (\citealt{sneden1973}; version 2017) 
and the drivers {\it abfind} and {\it blends} for the treatment of the lines without and with hyperfine structure (HFS). In particular, in addition to Fe we computed the abundance of 25 elements: Li, C, N, O, Na, Mg, Al, Si, S, Ca, Sc, Ti, V, Cr, Mn, Co, Ni, Cu, Zn, Y, Zr, Ba, La, Nd, and Eu. 
As done for Fe, also for Ti and Cr we measured the abundances of two ionization states, while for the other elements only the abundance of one species (first or second ionization state) was measured (see Table\,\ref{tab:final_stellar_abund}).

As done for the iron lines, the EWs of each spectral line were measured on the one-dimensional spectra interactively using the {\it splot} task in IRAF. The location of the local 
continuum was carefully selected tracing as much as possible the same position for each spectral line of all stars and the asteroid Vesta. This was done with the aim to minimize 
the error in the selection of the continuum. We also excluded features affected by telluric absorption. 

Following the prescriptions by \cite{biazzoetal2015}, we started from the line list of \ion{Na}{i}, \ion{Mg}{i}, \ion{Al}{i}, \ion{Si}{i}, \ion{Ca}{i}, \ion{Ti}{i}, \ion{Ti}{ii}, \ion{Cr}{i}, \ion{Cr}{ii}, \ion{Ni}{i}, and 
\ion{Zn}{i} by \cite{biazzoetal2012} complemented with additional lines and atomic parameters for \ion{Na}{i}, \ion{Al}{i}, \ion{Si}{i}, \ion{Ti}{i}, \ion{Ti}{ii}, \ion{Cr}{i}, \ion{Ni}{i}, and 
\ion{Zn}{i} taken from \cite{schuleretal2011b} and \cite{Sozzettietal2006}. For the \ion{Mg}{i} line at $\lambda$4730\,\AA\,and the \ion{Al}{i} line at $\lambda$6698.67\,\AA\,we considered the NIST 
(\citealt{kramidaetal2020}) and the \cite{melendezetal2014} atomic parameters. 
In the cases of \ion{C}{i}, \ion{S}{i}, \ion{Sc}{ii}, \ion{V}{i}, \ion{Mn}{i}, \ion{Co}{i}, and \ion{Cu}{i}, we considered the line lists by \cite{kurucz1993}, \cite{schuleretal2011b}, \cite{kramidaetal2020}, 
\cite{johnsonetal2006}, and \cite{scottetal2015}, where the hyperfine structure by \cite{johnsonetal2006} and \cite{kurucz1993} was adopted for Sc, V, Mn, Cu, and Co. 
Solar isotopic ratios by \cite{andersgrevesse1989} were considered for Cu (i.e. 69.17\% for $^{63}$Cu and 30.83\% for $^{65}$Cu). For the $s$-process elements 
\ion{Y}{ii}, \ion{Zr}{ii} (first peak) and \ion{Ba}{ii}, \ion{La}{ii} (second peak), the mixed $s/r$-process element \ion{Nd}{ii}, and the $r$-process element \ion{Eu}{ii} we considered the line lists by 
\cite{johnsonetal2006}, \cite{ljungetal2006}, \cite{prochaskaetal2000}, \cite{lawleretal2001b}, and \cite{denhartogetal2003}, where the HFS by \cite{gallagheretal2010} and 
\cite{lawleretal2001a} was adopted for Ba and La, respectively. Solar isotopic ratios by \cite{andersgrevesse1989} were considered for Ba (i.e. 2.417\% for $^{134}$Ba, 
7.854\% for $^{136}$Ba, 71.70\% for $^{138}$Ba, 6.592\% for $^{135}$Ba, and 11.23\% for $^{137}$Ba) and Eu (i.e. 47.8\% for $^{151}$Eu and 52.2\% for $^{153}$Eu). 

For the targets observed also with FEROS, we measured oxygen and nitrogen abundances through line EWs. Oxygen abundance was estimated using the \ion{O}{i} triplet of permitted lines 
at $\lambda$7771.94, $\lambda$7774.17, $\lambda$7775.39\,\AA\,with atomic parameters by NIST (\citealt{kramidaetal2020}), and considering the non-LTE (NLTE) corrections by \cite{amarsietal2015}. 
To our knowledge, no NLTE correction are present for [Fe/H]>0 and for \teff<5000 K, and the effect is important for \teff\,differences also of 50\,K (while it is negligible for [Fe/H] differences within 0.4\,dex). 
This is why we considered the \ion{O}{i} abundances measured for the Sun and the stars WASP-54, HAT-P-17, HAT-P-26, and the slightly metal-rich star HAT-P-30, while we excluded from the analysis the 
other (cooler) targets observed with FEROS (i.e. HAT-P-20, Qatar-2, and WASP-43). The importance of these effects has been also demonstrated for hot exoplanetary atmospheres (\citealt{Borsaetal2022}).

Thanks to FEROS spectra, we also measured nitrogen abundance using three high-excitation permitted lines (e.g., $\lambda$7442.3, $\lambda$7468.3, $\lambda$8216.3\,\AA), considering the atomic parameters by \cite{caffauetal2009}. Stars with \teff \ltsim 5200\,K are too cool to have detectable \ion{N}{i} lines, while within the warmer targets observed with FEROS only the spectra of Vesta and HAT-P-30 had sufficient $SNR$ to measure nitrogen abundance. No correction for NLTE effects was applied for N abundance of our sample because at our knowledge it is only available for the Sun (see \citealt{caffauetal2009}).

Besides \ion{O}{i}, due to the relatively wide range in stellar parameters (mainly \teff\, and \feh) of our targets, we also applied NLTE corrections to each line abundance of \ion{C}{i}, \ion{Na}{i}, \ion{Mn}{i}, and \ion{Co}{i}, following the prescriptions given by \cite{amarsietal2019}, \cite{lindetal2011}, \cite{bergemanngehren2008}, and \cite{bergemannetal2010}, respectively. For the other elements, no NLTE departure was considered because corrections were negligible or not reported in the literature for the lines used in this work and for the \teff\, and \feh\,ranges of our targets. As for the iron lines, final elemental abundance of each target was plotted as a function of stellar \teff, \logg, \csi, and \vsini\,to look for possible trends due to, e.g., line blendings. When trends were found for a specific element (in particular at lower and higher \teff), each line of that element was plotted as a function of \teff, \logg, \csi, and \vsini\,to recognize the presence of effects due, mainly, to line blending or 
bad quality of the spectrum for that specific line, that was afterwards removed. In the end, no trend was found for the final abundances and stellar parameters, with the exception of \ion{Cr}{ii} and \ion{C}{i} for which a residual trend with effective temperature remained. These two elements will be discussed in the next paragraph.

\subsubsection{Analysis based on spectral synthesis}

The projected rotational velocity (\vsini) and the lithium abundance ($\log n$(Li)) were measured by applying the spectral synthesis technique. We used the {\it synth} driver within the MOOG code 
(\citealt{sneden1973}; version 2017) and considered synthetic spectra obtained from \cite{castellikurucz2003} grids of model atmospheres at the stellar parameters (\teff, \logg, \csi, \feh) derived in 
Sect.\,\ref{sec:stellar_parameters}. We applied the same method also to derive the carbon, nitrogen, and oxygen abundances, which are elements analyzed through line EWs, too. 
In particular, we convolved the synthetic spectra with a Gaussian profile corresponding to the resolution of HARPS-N of $R\sim115\,000$, 
taking into account the optical limb-darkening coefficients by \cite{Claret2019} at the stellar \teff, \logg, \csi, and \feh. Moreover, empirical relationships obtained using asteroseismic rotational velocities by \cite{doyleetal2014} 
from {\it Kepler} data were used for deriving the macroturbulence velocity ($V_{\rm macro}$) for $T_{\rm eff}$>5700 K, while empirical relationships by \cite{Breweretal2016} were considered for 
$T_{\rm eff}$<5700 K\footnote{This was done because these relationships are not valid for the whole parameter space of our targets, but we verified that in the $T_{\rm eff}$ range in common, 
the two calibrations are in very good agreement (mean difference in $V_{\rm macro}$ of $\sim 0.5$\,km/s).}.

For the \vsini, we synthesized spectral lines in two regions around 6200 and 6700\,\AA, as done in \cite{barbatoetal2019}, until the minimum of residuals between stellar and synthetic spectra were 
reached. Final values are listed in Table\,\ref{tab:final_stellar_param_abund}, where the errors take into account uncertainties in stellar parameters ($T_{\rm eff}$, $\log g$, $\xi$, [Fe/H]) and spectral continuum definition (for the uncertainties in the continuum position, see next paragraph).

The forbidden [\ion{O}{i}] line at 6300.3\,\AA\,was considered for deriving O abundance through spectral synthesis and using the aforementioned code and 
model atmospheres. Atomic parameters for the O line and the nearby (blended) \ion{Ni}{i} line were taken from \cite{caffauetal2008} and \cite{johanssonetal2003}, respectively (see also \citealt{bertrandelisetal2015}, 
and references therein). Nickel abundances was fixed to the values we derived above. No NLTE correction was made because the $\lambda$6300.3\,\AA\,line is not affected by deviation from LTE (\citealt{caffauetal2008}).

\begin{figure} 
\begin{center}
\includegraphics[width=8.5cm]{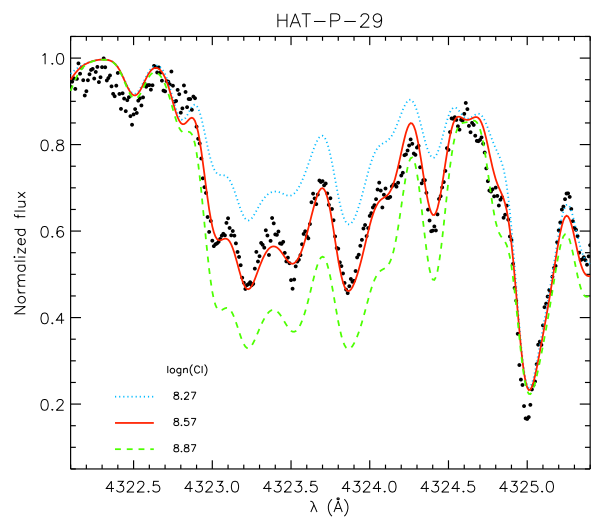}
\caption{Comparison between observed (black dots) and synthetic spectra for the target HAT-P-29 in the region around the CH band at $\sim$4320\,\AA. The shown synthetic spectra were computed for $\log n$(C)=8.57 
(red solid line; best fit), 8.87 (green dashed line), 8.27\,dex (blue dotted line), respectively.}
\label{fig:CH_HATP29} 
\end{center}
\end{figure}

Then, adopting the estimated O abundance, C was measured from $^{12}$CH and $^{13}$CH molecular features around 4320\,\AA\,(see the example in Fig.\,\ref{fig:CH_HATP29}). Molecular parameters were taken 
from \cite{masseronetal2014}. Atomic lines were re-adjusted on the solar spectrum. C abundance from these CH bands were measured for all targets but two cool stars (namely, Qatar-2 and WASP-43), 
mainly because of the very critical continuum placement.

Finally, N abundance was measured from CN features at $\sim$4215\,\AA, adopting the measured C and O abundances derived through spectral synthesis. The fitting procedure was differential, i.e. adopting the same spectral range and features in each star, and it was repeated until the minimum of residuals was obtained. The line lists for the $^{12}$C$^{14}$N, $^{13}$C$^{14}$N, and $^{12}$C$^{15}$N isotopologues were taken from \citet{Snedenetal2014}.
Given the sensitivity to \teff, N abundance from CN band was measured for stars in the $4600-5300$\,K \teff\,range.

\begin{figure} 
\begin{center}
\includegraphics[width=5.5cm,angle=90]{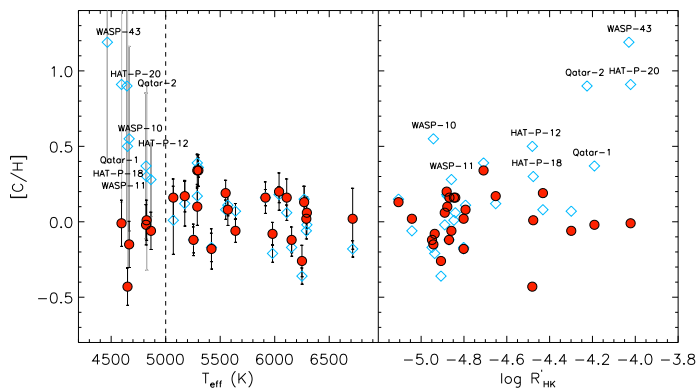}
\caption{{\it Left panel}. Carbon abundances as a function of \teff, as derived from \ion{C}{i} lines (open diamonds) and the CH band (filled circles). Dashed line marks the position at $T_{\rm eff}=5000$\,K. Stars with $T_{\rm eff} < 5000$\,K are highlighted. {\it Right panel}. C abundances as a function of the activity index 
$\log R'_{\rm HK}$. Symbols as in the left panel. }
\label{fig:Cabun_vs_Teff_Rhk} 
\end{center}
\end{figure}

Since CNO abundances were derived with both EWs and spectral synthesis, as final abundance we considered the weighted average coming from these two methods. 
In particular, Fig.\,\ref{fig:Cabun_vs_Teff_Rhk} shows the C abundances derived from \ion{C}{i} 
lines (open diamonds) and those derived from the CH band (filled circles). We find a good agreement between [C/H] measured from these two diagnostics for \teff>5000\,K, with the largest differences for cooler stars, for which 
the atomic carbon lines provide abundances larger than twice the typical error in [C/H]. Similar results were found by, e.g., \cite{baratellaetal2020} and \cite{delgadomenaetal2021}. The latter authors mentioned possible dependence 
of the trend with the metallicity (slight increase of [C/Fe] towards lower metallicity for [C/H] derived through atomic lines, and flatter trend for abundances obtained from molecular lines). We do not find similar trends, most 
probably due to our small stellar sample. We find instead an evident trend with the Mittag chromospheric activity index $\log R'_{\rm HK}$ derived by Claudi et al. (in prep). Similar results were obtained by \cite{baratellaetal2020}, 
who also found a positive correlation between atomic [C/H] and $\log R'_{\rm HK}$, justified as possible unknown blends in the optical lines becoming more important in active stars than in quiet stars. On the contrary, 
we do not find any trend between $\log R'_{\rm HK}$ and other abundances or $\xi$, thus further validating our method to derive elemental abundances (see also discussion in \citealt{baratellaetal2020}). In the end, we decided to consider a weighted 
average between C abundances derived with both atomic and CH band for \teff>5000\,K, while for \teff<5000\,K we considered only the values from the CH band as C abundances\footnote{We also find some  discrepancy in Cr abundances, for which we obtained mean differences of 0.15\,dex between \ion{Cr}{i} and \ion{Cr}{ii} for stars cooler than 5000\,K. Similar findings were observed in \cite{baratellaetal2020}, for which blendings at low \teff\, and chromospheric activity effects were invoked as possible reasons of the observed over-ionization. No similar trends were observed between \ion{Ti}{i} and \ion{Ti}{ii}. Whatever the reason is, we decided from now on to use only \ion{Cr}{i} and \ion{Ti}{i} as abundances of chromium and titanium, respectively.}. 

Lithium abundance ($\log n$(Li)) was derived through spectral synthesis of the absorption line at $\lambda$6707.8\,\AA, which was present in 7 stars (namely, KELT-6, HAT-P-14, WASP-38, HAT-P-30, 
WASP-13, HAT-P-4, HAT-P-3; see Sect.\,\ref{sec:lithium}). In particular, we used the lithium line list by \cite{Reddyetal2002} at the vicinity ($\pm$0.5\,\AA) of the Li\,6707.8\,\AA\,line, implemented with the VALD 
database (\citealt{kupkaetal2000}) for wavelengths farther from the line center. The line list by \cite{Reddyetal2002} considers the isotopes $^6$Li and $^7$Li of the  $\lambda$6707.8\,\AA\,line allowing
us to estimate the lithium isotopic $^6$Li/$^7$Li ratio. Lithium abundance as derived through the MOOG code was then corrected for the departure from LTE considering the non-LTE calculation 
of \cite{lindetal2009}. Double check of the final results was done computing the lithium abundances also using the Li equivalent widths and converting them in abundances through the 
curves-of-growth by \cite{lindetal2009}. We found consistent values with both methods within the uncertainties. In Table\,\ref{tab:final_stellar_param_abund} we list only the results obtained through spectral synthesis. Errors in $\log n$(Li) were 
derived by adding quadratically the uncertainties due to the abundance measurements and those due to stellar parameters. 

The results for all solar elemental abundances are given in Table \ref{tab:solar_abundances} together with those given by \cite{Asplundetal2021} and obtained using a 3D radiative-hydrodynamic model of the solar 
surface convection and atmosphere, and correction for departures from LTE conditions when necessary. This table shows good agreement between our solar abundance values and the literature results, with a mean difference 
between values derived in this work and in the literature of $0.00\pm0.04$.

\setlength{\tabcolsep}{4pt}
\begin{table*}
\tiny
\caption{Final stellar parameters (effective temperature, surface gravity, microturbulence velocity, rotational velocity), iron and lithium abundances as derived in this work (for \ion{Fe}{i} and \ion{Fe}{ii} we list in brackets the number of lines used to derive the abundances), together with macroturbulence velocity $V_{\rm macro}$ and mean radial velocity $<V_{\rm rad}>$.}
\label{tab:final_stellar_param_abund}
\begin{center}
\begin{tabular}{lcccrrcccr}
\hline\hline
Name & $T_{\rm eff}$  & $\log g$ & $\xi$ & [\ion{Fe}{i}/H] & [\ion{Fe}{ii}/H] & $V_{\rm macro}^\ast$ & $v\sin i$ & $\log n{\rm (Li)}$ & $<V_{\rm rad}>^{\ast \ast}$\\
 & (K)    & 	(dex)  & (km/s) & (dex) & (dex) &  (km/s) & (km/s) & (dex) & (km/s) \\
\hline
HAT-P-3  & 5175$\pm$75 & 4.52$\pm$0.09 & 0.63$\pm$0.04 &    0.25$\pm$0.09(70) &    0.25$\pm$0.12(10) & 2.0 & 1.4$\pm$0.5 &    $<0.5$	 & $-$23.378$\pm$0.018\\
HAT-P-4  & 5915$\pm$15 & 4.15$\pm$0.12 & 1.23$\pm$0.02 &    0.29$\pm$0.08(77) &    0.29$\pm$0.08(10) & 4.1 & 5.6$\pm$0.5 & 2.84$\pm$0.03 &  $-$1.366$\pm$0.012\\
HAT-P-12 & 4650$\pm$70 & 4.50$\pm$0.09 & 0.49$\pm$0.15 & $-$0.25$\pm$0.09(67) & $-$0.25$\pm$0.12(7)  & 1.6 & 0.5$\pm$0.5 &     ...	 & $-$40.458$\pm$0.002\\
HAT-P-14 & 6715$\pm$20 & 4.12$\pm$0.15 & 1.50$\pm$0.06 &    0.05$\pm$0.08(71) &    0.05$\pm$0.10(11) & 7.8 & 8.1$\pm$0.3 & 2.02$\pm$0.15 & $-$20.341$\pm$0.039\\
HAT-P-15 & 5570$\pm$90 & 4.39$\pm$0.15 & 0.59$\pm$0.22 &    0.26$\pm$0.09(73) &    0.26$\pm$0.09(10) & 2.8 & 2.0$\pm$0.3 &     ...	 &    31.755$\pm$0.003\\
HAT-P-17 & 5255$\pm$45 & 4.50$\pm$0.15 & 0.60$\pm$0.19 &    0.03$\pm$0.08(73) &    0.03$\pm$0.09(11) & 2.1 & 1.3$\pm$0.3 &     ...	 &    20.200$\pm$0.007\\
HAT-P-18 & 4825$\pm$75 & 4.43$\pm$0.13 & 0.67$\pm$0.30 &    0.09$\pm$0.09(69) &    0.09$\pm$0.09(10) & 1.7 & 1.3$\pm$0.4 &     ...	 & $-$11.105$\pm$0.009\\
HAT-P-20 & 4595$\pm$60 & 4.49$\pm$0.09 & 0.40$\pm$0.40 &    0.17$\pm$0.10(64) &    0.17$\pm$0.14(8)  & 1.5 & 2.6$\pm$0.3 &     ...	 & $-$18.192$\pm$0.215\\
HAT-P-21 & 5640$\pm$25 & 4.24$\pm$0.11 & 0.98$\pm$0.04 &    0.04$\pm$0.08(77) &    0.04$\pm$0.08(11) & 3.0 & 3.9$\pm$0.5 &     ...	 & $-$52.978$\pm$0.079\\
HAT-P-22 & 5290$\pm$50 & 4.35$\pm$0.15 & 0.58$\pm$0.26 &    0.26$\pm$0.08(70) &    0.26$\pm$0.09(11) & 2.2 & 1.8$\pm$0.5 &     ...	 &    12.669$\pm$0.037\\
HAT-P-26 & 5070$\pm$80 & 4.49$\pm$0.09 & 0.43$\pm$0.32 &    0.02$\pm$0.09(74) &    0.02$\pm$0.11(11) & 2.0 & 0.8$\pm$0.4 &     ...	 &    13.844$\pm$0.001\\
HAT-P-29 & 6110$\pm$70 & 4.35$\pm$0.13 & 1.08$\pm$0.03 &    0.24$\pm$0.08(76) &    0.24$\pm$0.09(11) & 4.5 & 4.5$\pm$0.7 &     ...	 & $-$21.629$\pm$0.009\\
HAT-P-30 & 6290$\pm$60 & 4.30$\pm$0.14 & 1.05$\pm$0.04 &    0.10$\pm$0.08(75) &    0.10$\pm$0.07(10) & 5.2 & 3.0$\pm$0.5 & 3.04$\pm$0.05 &    44.727$\pm$0.018\\
HAT-P-36 & 5550$\pm$80 & 4.33$\pm$0.16 & 0.67$\pm$0.22 &    0.27$\pm$0.09(74) &    0.27$\pm$0.09(10) & 2.7 & 3.6$\pm$0.3 &     ...	 & $-$16.243$\pm$0.019\\
KELT-6   & 6250$\pm$20 & 3.93$\pm$0.13 & 1.38$\pm$0.05 & $-$0.28$\pm$0.07(76) & $-$0.28$\pm$0.06(11) & 5.8 & 4.5$\pm$0.5 & 1.19$\pm$0.06 &     1.165$\pm$0.007\\
Qatar-1  & 4820$\pm$85 & 4.45$\pm$0.12 & 0.69$\pm$0.35 &    0.16$\pm$0.09(68) &    0.16$\pm$0.09(10) & 1.7 & 1.9$\pm$0.7 &     ...	 & $-$38.034$\pm$0.056\\
Qatar-2  & 4645$\pm$60 & 4.52$\pm$0.10 & 0.45$\pm$0.40 &    0.12$\pm$0.10(59) &    0.12$\pm$0.11(7)  & 1.6 & 2.1$\pm$0.4 &     ...	 & $-$23.971$\pm$0.037\\
TRES-4   & 6270$\pm$45 & 4.09$\pm$0.17 & 1.56$\pm$0.03 &    0.26$\pm$0.09(75) &    0.26$\pm$0.10(11) & 5.5 & 9.5$\pm$0.8 &     ...	 & $-$16.100$\pm$0.009\\
WASP-10  & 4665$\pm$60 & 4.40$\pm$0.13 & 0.53$\pm$0.38 &    0.13$\pm$0.09(66) &    0.13$\pm$0.09(8)  & 1.6 & 3.3$\pm$0.5 &     ...	 &  $-$4.228$\pm$0.019\\
WASP-11  & 4865$\pm$25 & 4.43$\pm$0.12 & 0.67$\pm$0.11 &    0.07$\pm$0.09(72) &    0.07$\pm$0.11(10) & 1.7 & 1.2$\pm$0.3 &     ...	 &     4.911$\pm$0.012\\
WASP-13  & 5980$\pm$50 & 4.09$\pm$0.12 & 1.11$\pm$0.04 &    0.10$\pm$0.08(78) &    0.10$\pm$0.08(10) & 4.5 & 4.0$\pm$0.7 & 2.05$\pm$0.07 &     9.855$\pm$0.010\\
WASP-38  & 6295$\pm$20 & 4.27$\pm$0.18 & 1.31$\pm$0.05 &    0.09$\pm$0.09(79) &    0.09$\pm$0.10(11) & 5.3 & 8.3$\pm$0.8 & 1.97$\pm$0.04 & $-$9.7317$\pm$0.047\\
WASP-39  & 5420$\pm$20 & 4.30$\pm$0.11 & 0.73$\pm$0.03 &    0.00$\pm$0.07(73) &    0.00$\pm$0.09(11) & 2.4 & 1.8$\pm$0.4 &     ...	 & $-$58.443$\pm$0.003\\
WASP-43  & 4465$\pm$90 & 4.45$\pm$0.19 & 0.68$\pm$0.41 & $-$0.04$\pm$0.15(49) & $-$0.04$\pm$0.14(4)  & 1.5 & 2.6$\pm$0.4 &     ...	 &  $-$3.661$\pm$0.041\\
WASP-54  & 6155$\pm$25 & 4.02$\pm$0.11 & 1.20$\pm$0.03 & $-$0.09$\pm$0.08(78) & $-$0.09$\pm$0.09(10) & 5.2 & 3.6$\pm$0.9 &     ...	 &  $-$3.107$\pm$0.014\\
WASP-60  & 6040$\pm$30 & 4.21$\pm$0.11 & 1.14$\pm$0.02 &    0.23$\pm$0.07(77) &    0.23$\pm$0.09(11) & 4.4 & 3.3$\pm$0.6 &     ...	 & $-$26.526$\pm$0.002\\
XO-2N    & 5290$\pm$70 & 4.46$\pm$0.10 & 0.59$\pm$0.15 &    0.37$\pm$0.10(74) &    0.37$\pm$0.10(10) & 5.2 & 1.8$\pm$0.4 &     ...	 &    46.920$\pm$0.009\\
XO-2S    & 5300$\pm$80 & 4.41$\pm$0.13 & 0.60$\pm$0.20 &    0.32$\pm$0.10(72) &    0.32$\pm$0.11(10) & 5.2 & 1.7$\pm$0.4 &     ...	 &    46.574$\pm$0.005\\
\hline
\end{tabular}
\end{center}
\footnotesize{Note: $^\ast$ The macroturbulence velocity $V_{\rm macro}$ was computed from empirical relationships taken from the literature and useful to derive parameters and abundances through the spectral synthesis method 
(see Sect.\,\ref{sec:abun_anal}). 
$^{\ast \ast}$ The mean radial velocity $<V_{\rm rad}>$ was computed by us to derive kinematic properties (see Sect.\,\ref{sec:spacemotion}). 
As final errors in radial velocities we considered both those in the mean values of the HARPS-N measurements listed here and a typical zero-point level of 0.2\,km/s.
}
\end{table*}

\begin{table}[ht]
\caption{Comparison between our measured solar abundances and standard values from \cite{Asplundetal2021}. $Z$ represents the atomic number and $T_{\rm cond}$ the 50\% condensation temperature values derived by \cite{Lodders2003}.} 
\label{tab:solar_abundances}
\begin{center}
\begin{tabular}{llrcc}
\hline
\hline
Species & $Z$ & $T_{\rm cond}$ & $\log n_{\rm This\,work}$ & $\log n_{\rm Literature}$ \\ 
\hline
\ion{Li}{i}   & 2 & 1142 & $1.04 \pm 0.09$  & $0.96 \pm 0.06$  \\
\ion{C}{i}(EW) & 6 & 40 & $8.43 \pm 0.05$  & \\
\ion{C}{i}(synt) & & & $8.41 \pm 0.03$  & $8.46 \pm 0.04$   \\
\ion{N}{i}(EW) & 7 & 123 & $7.95 \pm 0.03$  &   \\
\ion{N}{i}(synt) & & &  ...   & $7.83 \pm 0.07$  \\
\ion{O}{i}(EW)   & 8 & 180 &  $8.66 \pm 0.03$  &    \\
\ion{O}{i}(synt) & & & $8.69 \pm 0.10$  & $8.69 \pm 0.04$  \\
\ion{Na}{i}   & 11 & 958 & $6.21 \pm 0.01$  &    $6.22 \pm 0.03$  \\
\ion{Mg}{i}   & 12 & 1336 & $7.59 \pm 0.06$  & $7.55 \pm 0.03$  \\
\ion{Al}{i}   & 13 & 1653 & $6.42 \pm 0.06$  & $6.43 \pm 0.03$\\
\ion{Si}{i}   & 14 & 1310 & $7.52 \pm 0.05$  & $7.51 \pm 0.03$  \\
\ion{S}{i}    & 16 & 664 & $7.13 \pm 0.03$  & $7.12 \pm 0.03$ \\
\ion{Ca}{i}   & 20 & 1517 & $6.30 \pm 0.06$  & $6.30 \pm 0.03$  \\
\ion{Sc}{ii}  & 21 & 1659 & $3.10 \pm 0.06$  & $3.14 \pm 0.04$  \\
\ion{Ti}{i}   & 22 & 1582 & $4.95 \pm 0.04$  & $4.97 \pm 0.05$  \\
\ion{Ti}{ii}  & & & $4.95 \pm 0.05$  &		   	 \\
\ion{V}{i}    & 23 & 1429 & $3.86 \pm 0.04$  & $3.90 \pm 0.08$  \\
\ion{Cr}{i}   & 24 & 1296 & $5.62 \pm 0.04$  & $5.62 \pm 0.04$  \\
\ion{Cr}{ii}  &   & & $5.62 \pm 0.05$  &		   	\\
\ion{Mn}{i}   & 25 & 1158 & $5.42 \pm 0.04$  & $5.42 \pm 0.06$   \\
\ion{Fe}{i}   & 26 & 1334 & $7.49 \pm 0.05$  & $7.46 \pm 0.04$ \\
\ion{Fe}{ii}  & & & $7.49 \pm 0.05$  &		   	\\
\ion{Co}{i}   & 27 & 1352 & $4.96 \pm 0.06$  & $4.94 \pm 0.05$  \\
\ion{Ni}{i}   & 28 & 1353 & $6.24 \pm 0.05$  & $6.20 \pm 0.04$  \\
\ion{Cu}{i}   & 29 & 1037 & $4.22 \pm 0.09$  & $4.18 \pm 0.05$ \\
\ion{Zn}{i}   & 30 & 726 & $4.56 \pm 0.05$  & $4.56 \pm 0.05$  \\
\ion{Y}{ii}   & 39 & 1659 & $2.18 \pm 0.06$  & $2.21 \pm 0.05$ \\
\ion{Zr}{ii}  & 40 & 1741 & $2.57 \pm 0.02$  & $2.59 \pm 0.04$  \\
\ion{Ba}{ii}  & 56 & 1455 & $2.19 \pm 0.05$  & $2.27 \pm 0.05$  \\
\ion{La}{ii}  & 57 & 1578 & $1.06 \pm 0.01$  & $1.11 \pm 0.04$  \\
\ion{Nd}{ii}  & 60 & 1602 & $1.41 \pm 0.01$  & $1.42 \pm 0.04$  \\
\ion{Eu}{ii}  & 63 & 1356 & $0.50 \pm 0.03$  & $0.52 \pm 0.04$  \\
\hline
\end{tabular}
\end{center}
\end{table}

\subsubsection{Uncertainties in elemental abundances}

Derived abundances are mainly affected by uncertainties in atomic parameters, stellar parameters, and measured EWs (or continuum position when spectral synthesis was applied). 

Uncertainties in atomic parameters, such as the transition probability ($\log gf$), should cancel out, since our analysis is carried out differentially with respect to the Sun. 
Errors due to uncertainties in stellar parameters ($T_{\rm eff}$, $\xi$, $\log g$, [Fe/H]) were estimated first by assessing errors in stellar parameters 
themselves and then by varying each parameter separately, while keeping the other two unchanged. As shown in Table\,\ref{tab:final_stellar_param_abund} and explained in Sect.\,\ref{sec:stellar_parameters}, uncertainties 
in stellar parameters are in the range 15-90\,K for $T_{\rm eff}$ (with a standard deviation $\sigma$ of 25\,K), 0.09-0.19\,dex for $\log g$ ($\sigma=0.03$\,dex), 0.02-0.41\,km/s for $\xi$ ($\sigma=0.14$\,km/s), and 0.07-0.15\,dex 
for [Fe/H] ($\sigma=0.02$\,dex). Due to the small values of the standard deviations across the wide range of stellar effective temperature for the errors in $T_{\rm eff}$, $\log g$, and [Fe/H], and due to the smaller influence of errors in 
$\xi$ for almost all elements, when compared to the other sources of uncertainties, we decided to consider typical uncertainties in stellar parameters of $60$\,K, $0.10$\,dex, $0.15$\,km/s, and $\sim 0.08$\,dex in $T_{\rm eff}$, $\log g$, 
$\xi$, and [Fe/H], respectively. We thus assumed these values as uncertainties in stellar parameters. Errors in abundances [X/H] due to uncertainties in stellar parameters are summarized in Table~\ref{tab:errors} for two different stars in 
our sample from which we derived all elemental abundances: one of the coolest one (HAT-P-12 or WASP-10) and one of the warmest stars (HAT-P-30). Total errors in $\log n$(Li) are instead reported in Table\,\ref{tab:final_stellar_param_abund}.

As for the errors due to uncertainties in EWs, they are well represented by the standard deviation around the mean abundance determined from all the lines. These errors are listed in Tables~\ref{tab:final_stellar_param_abund} and 
\ref{tab:final_stellar_abund}, where uncertainties in [X/H] were obtained by quadratically adding the error for the target and the error for the Sun. When only one line was measured, the error in [X/H] is the standard deviation of three independent 
EW measurements obtained taking different position of the continuum. The number of lines employed for the abundance analysis is listed in Tables~\ref{tab:final_stellar_param_abund} and \ref{tab:final_stellar_abund} in brackets.

As for the uncertainties due to the definition of continuum position when spectral synthesis was applied, random errors affecting our best-fitting procedure were evaluated by changing the continuum position until the standard deviation (observed minus synthetic spectra) was two times larger than best fit value, where residuals of our best-fit solutions are typically smaller than 0.02. These error budgets are listed in Tables~\ref{tab:final_stellar_abund}.

\setlength{\tabcolsep}{6pt}
\begin{table*}  
\caption{Internal errors in abundance determination due to uncertainties in stellar parameters for one of the coolest 
star (namely, HAT-P-12 for all elements but C and N; WASP-10 for C and N) and for one of the warmest star (namely, HAT-P-30) in our samples.}
\label{tab:errors}
\small
\begin{center}
\begin{tabular}{lcccc}
\hline
\hline
HAT-P-12 & $T_{\rm eff}=4650$ K & $\log g=4.50$\,dex & $\xi=0.49$ km/s & [Fe/H]=$-0.25$\,dex\\
\hline
 & $\Delta T_{\rm eff}=-/+60$ K & $\Delta \log g=-/+0.10$\,dex & $\Delta \xi=-/+0.15$ km/s & $\Delta$[Fe/H]$=-/+0.08$\,dex\\
\hline
$[$\ion{Fe}{i}/H$]$  &   0.00/0.00  & 0.01/$-$0.01 & 0.01/$-$0.02 &   .../...  \\
$[$\ion{Fe}{ii}/H$]$ & 0.08/$-$0.07 & $-$0.03/0.09 & 0.01/$-$0.01 &   .../...  \\
$[$\ion{C}{i}/H$]$   & 0.09/$-$0.09 &    0.01/$-$0.01 & 0.01/$-$0.01 &  $-$0.01/0.01  \\
$[$\ion{N}{i}/H$]$   & 0.05/$-$0.05 &    0.00/0.00   &  0.00/0.00 &  0.00/0.00     \\
$[$\ion{O}{i}/H$]$   & 0.08/$-$0.08 &  0.05/$-$0.05  &  0.00/0.00 &  0.00/0.00     \\
$[$\ion{Na}{i}/H$]$  & $-$0.03/0.05 &    0.03/0.00 & 0.01/$-$0.01 &  $-$0.01/0.01 \\
$[$\ion{Mg}{i}/H$]$  &   0.01/0.00  &    0.01/0.00 &   0.00/0.00  &  $-$0.02/0.02 \\
$[$\ion{Al}{i}/H$]$  & $-$0.02/0.04 &    0.02/0.00 & 0.01/$-$0.01 &  $-$0.01/0.01 \\
$[$\ion{Si}{i}/H$]$  & 0.04/$-$0.04 & $-$0.01/0.00 &   0.00/0.00  &  $-$0.02/0.02 \\
$[$\ion{S}{i}/H$]$   & 0.04/$-$0.06 & $-$0.02/0.00 &   0.00/0.00  &    0.00/0.00  \\
$[$\ion{Ca}{i}/H$]$  & $-$0.03/0.07 &    0.04/0.00 & 0.01/$-$0.01 &  $-$0.02/0.02 \\
$[$\ion{Sc}{ii}/H$]$ & 0.01/$-$0.06 & $-$0.03/0.03 &   0.00/0.00  &  $-$0.03/0.03 \\
$[$\ion{Ti}{i}/H$]$  & $-$0.06/0.08 &    0.03/0.00 & 0.02/$-$0.03 &  $-$0.02/0.02 \\
$[$\ion{Ti}{ii}/H$]$ & $-$0.01/0.01 & $-$0.02/0.00 & 0.02/$-$0.03 &  $-$0.04/0.03  \\ 
$[$\ion{V}{i}/H$]$   & $-$0.08/0.09 & 0.01/$-$0.01 & 0.02/$-$0.03 &  $-$0.01/0.01 \\
$[$\ion{Cr}{i}/H$]$  & $-$0.04/0.05 &    0.01/0.00 & 0.02/$-$0.02 &  $-$0.01/0.02 \\
$[$\ion{Cr}{ii}/H$]$ & 0.03/$-$0.05 & $-$0.03/0.00 & 0.01/$-$0.01 &  $-$0.03/0.02  \\
$[$\ion{Mn}{i}/H$]$  & $-$0.03/0.03 & 0.02/$-$0.02 & 0.02/$-$0.02 &  $-$0.04/0.03 \\
$[$\ion{Co}{i}/H$]$  & $-$0.01/0.01 & $-$0.02/0.03 & 0.00/$-$0.01 &  $-$0.03/0.03 \\
$[$\ion{Ni}{i}/H$]$  & 0.00/$-$0.01 & $-$0.01/0.00 & 0.00/$-$0.02 &  $-$0.03/0.02 \\    
$[$\ion{Cu}{i}/H$]$  &   0.00/0.00  & $-$0.01/0.03 & 0.01/$-$0.01 &  $-$0.03/0.03 \\
$[$\ion{Zn}{i}/H$]$  & 0.03/$-$0.04 &    0.00/0.00 & 0.01/$-$0.02 &  $-$0.03/0.03 \\
$[$\ion{Y}{i}/H$]$   & $-$0.03/0.00 & $-$0.02/0.00 & 0.03/$-$0.03 &  $-$0.03/0.03  \\
$[$\ion{Zr}{ii}/H$]$ & $-$0.04/0.00 & $-$0.04/0.00 & 0.02/$-$0.02 &  $-$0.03/0.03  \\
$[$\ion{Ba}{ii}/H$]$ & $-$0.02/0.02 &    0.00/0.00 & 0.03/$-$0.04 &  $-$0.05/0.05  \\
$[$\ion{La}{ii}/H$]$ & $-$0.02/0.02 & $-$0.04/0.05 &   0.00/0.00  &  $-$0.03/0.03  \\
$[$\ion{Nd}{ii}/H$]$ & $-$0.05/0.02 & $-$0.03/0.00 & 0.01/$-$0.01 &  $-$0.03/0.03  \\
$[$\ion{Eu}{ii}/H$]$ &   0.00/0.00  & $-$0.03/0.05 &   0.00/0.00  &  $-$0.03/0.03  \\
\hline	\\	
HAT-P-30 & $T_{\rm eff}=6290$ K & $\log g=4.30$\,dex & $\xi=1.05$ km/s & [Fe/H]=0.10\,dex  \\
\hline
 & $\Delta T_{\rm eff}=-/+60$ K & $\Delta \log g=-/+0.10$\,dex & $\Delta \xi=-/+0.15$ km/s & $\Delta$[Fe/H]$=-/+0.08$\,dex \\

\hline
$[$\ion{Fe}{i}/H$]$  & $-$0.04/0.04 &  0.01/$-$0.01 & 0.02/$-$0.03  &    .../... \\
$[$\ion{Fe}{ii}/H$]$ &    0.00/0.00 &  $-$0.04/0.03 & 0.05/$-$0.02  &    .../... \\
$[$\ion{C}{i}/H$]$   & 0.04/$-$0.04 &  $-$0.03/0.03 & 0.02/$-$0.02  &   0.00/0.00 \\
$[$\ion{N}{i}/H$]$   & 0.04/$-$0.04 &  $-$0.03/0.03 &   0.00/0.00   & $-$0.01/0.01\\
$[$\ion{O}{i}/H$]$   & 0.04/$-$0.04 &. 0.01/$-$0.01 & 0.02/$-$0.02  &   0.00/0.00 \\
$[$\ion{Na}{i}/H$]$  & $-$0.03/0.03 &  0.01/$-$0.01 & 0.01/$-$0.01  &   0.00/0.00 \\
$[$\ion{Mg}{i}/H$]$  & $-$0.03/0.02 &    0.00/0.00  & 0.01/$-$0.01  &   0.00/0.00 \\
$[$\ion{Al}{i}/H$]$  & $-$0.02/0.02 &    0.00/0.00  & 0.00/0.00     &   0.00/0.00 \\
$[$\ion{Si}{i}/H$]$  & $-$0.02/0.02 &    0.00/0.00  & 0.01/$-$0.01  &   0.00/0.00 \\
$[$\ion{S}{i}/H$]$   & 0.01/$-$0.01 & $-$0.03/0.03  & 0.00/0.00     &   0.00/0.00 \\
$[$\ion{Ca}{i}/H$]$  & $-$0.04/0.04 &  0.02/$-$0.02 & 0.03/$-$0.03  &   0.00/0.00 \\
$[$\ion{Sc}{ii}/H$]$ &    0.00/0.01 & $-$0.04/0.04  & 0.01/$-$0.01  & $-$0.02/$-$0.05 \\
$[$\ion{Ti}{i}/H$]$  & $-$0.05/0.05 &    0.00/0.00  & 0.02/$-$0.02  &   0.00/0.00 \\
$[$\ion{Ti}{ii}/H$]$ &    0.00/0.01 & $-$0.04/0.04  & 0.04/$-$0.04  & $-$0.01/0.01 \\
$[$\ion{V}{i}/H$]$   & $-$0.05/0.05 &    0.00/0.00  & 0.00/0.00     &   0.00/0.00 \\
$[$\ion{Cr}{i}/H$]$  & $-$0.04/0.04 &    0.00/0.00  & 0.02/$-$0.02  &   0.00/0.00 \\
$[$\ion{Cr}{ii}/H$]$ & 0.01/$-$0.01 & $-$0.04/0.03  & 0.03/$-$0.03  & $-$0.01/0.01 \\
$[$\ion{Mn}{i}/H$]$  & $-$0.05/0.05 &    0.00/0.00  & 0.02/$-$0.02  &   0.00/0.00 \\
$[$\ion{Co}{i}/H$]$  & $-$0.05/0.05 &    0.00/0.00  & 0.00/0.00     &   0.00/0.00 \\
$[$\ion{Ni}{i}/H$]$  & $-$0.04/0.04 &    0.00/0.00  & 0.01/$-$0.02  &   0.00/0.00 \\	
$[$\ion{Cu}{i}/H$]$  & $-$0.05/0.05 &    0.00/0.00  & 0.01/$-$0.01  &   0.00/0.00 \\
$[$\ion{Zn}{i}/H$]$  & $-$0.02/0.02 &    0.00/0.01  & 0.06/$-$0.06  &   0.00/0.00 \\
$[$\ion{Y}{i}/H$]$   & $-$0.01/0.01 & $-$0.04/0.04  & 0.07/$-$0.06  & $-$0.02/0.02 \\
$[$\ion{Zr}{ii}/H$]$ & $-$0.01/0.01 & $-$0.04/0.04  & 0.04/$-$0.03  & $-$0.02/0.02 \\
$[$\ion{Ba}{ii}/H$]$ & $-$0.03/0.03 & $-$0.02/0.02  & 0.08/$-$0.08  & $-$0.02/0.02 \\
$[$\ion{La}{ii}/H$]$ & $-$0.02/0.02 & $-$0.04/0.04  & 0.00/0.00     & $-$0.02/0.02 \\
$[$\ion{Nd}{ii}/H$]$ & $-$0.02/0.02 & $-$0.04/0.04  & 0.01/$-$0.01  & $-$0.02/0.02 \\
$[$\ion{Eu}{ii}/H$]$ & $-$0.01/0.01 & $-$0.04/0.04  & 0.00/0.00     & $-$0.02/0.02 \\
\hline	\\	
\end{tabular}
\end{center}
Notes. Numbers refer to the differences between the abundances obtained with ($-$ and $+$) and without the uncertainties in stellar parameters.
\end{table*}

\subsection{Final parameters and comparison with previous works}
\label{sec:comparison}

\begin{figure} 
\begin{center}
\includegraphics[width=8.5cm,angle=90]{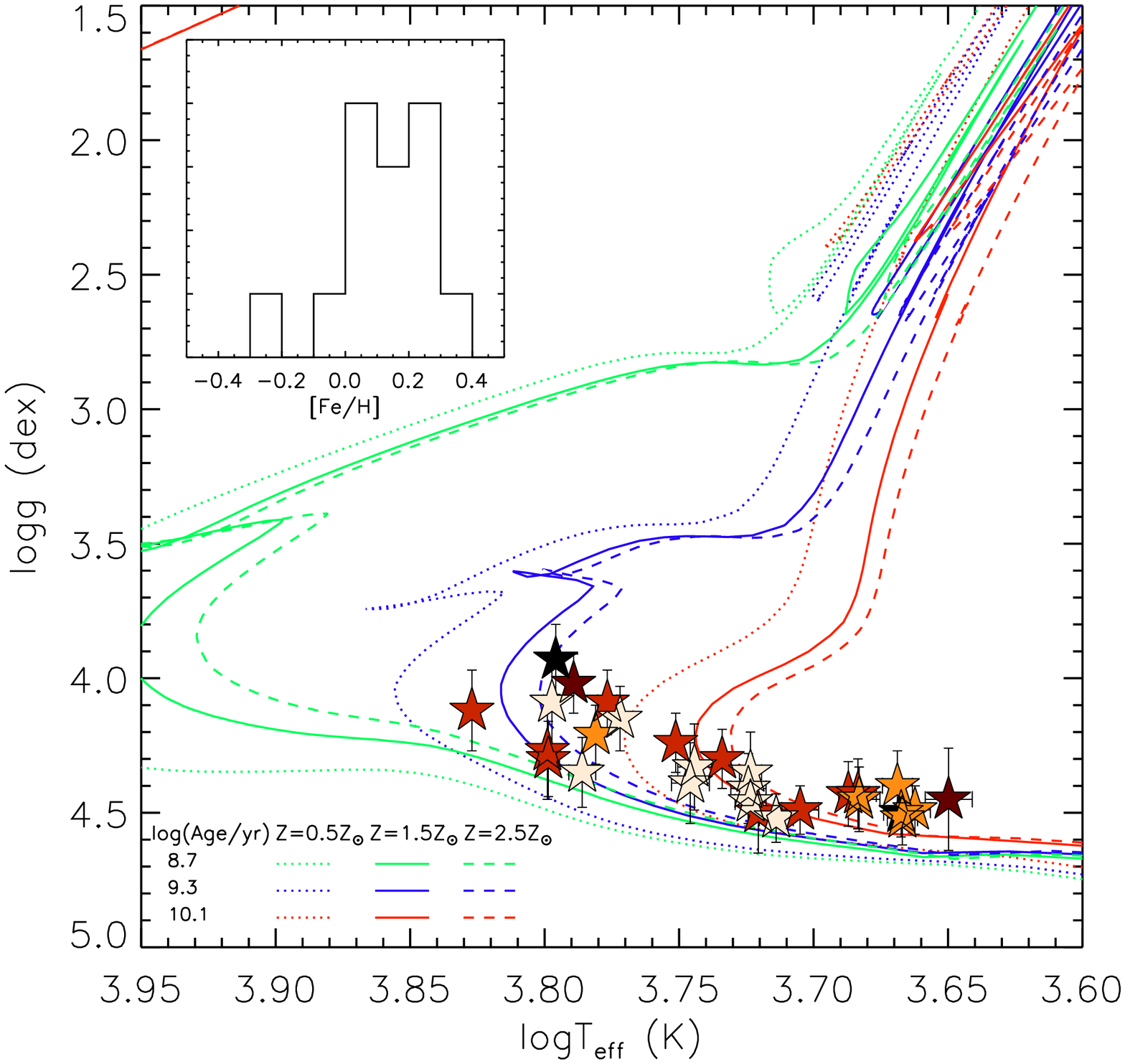}
\caption{\teff-\logg\,diagram for our stellar sample. Green, blue, and red lines are the PARSEC isochrones at $\sim$500\,Myr, $\sim$2\,Gyr, and $\sim$12.5\,Gyr. Dotted, solid, and dashed lines represent three different values of metallicities, as labelled in the lower part of the main panel. The inset represents the histogram distribution in [Fe/H] of our targets. Star symbols are filled in darker colors going from the most metal-rich toward the most metal poor stars, with the following four bins: [Fe/H]$\ge +0.24$\,dex, $-0.12 <$ [Fe/H] $\le +0.24$\,dex, $-0.15 <$ [Fe/H] $\le -0.02$\,dex, and [Fe/H]$\le -0.15$\,dex.}
\label{fig:logg_teff} 
\end{center}
\end{figure}

Fig.\,\ref{fig:logg_teff} shows the position of our targets in the Kiel  \teff-\logg\,diagram along with stellar model tracks for three different metallicities (i.e. $Z=0.5Z_\odot$, $1.5Z_\odot$, $2.5Z_\odot$) and ages (i.e. $\log (Age/{\rm yr})=8.7, 9.3, 10.1$) spanning the values of our stellar sample (see Sect.\,\ref{sec:stellar_parameters} for our iron abundance measurement and  Sect.\,\ref{sec:ages} for our age estimation using chemical clocks). The stars are color-coded with respect to [Fe/H], from $\sim -0.3$\,dex up to $\sim +0.4$\,dex, as shown in the histogram. Our homogeneous procedure led to stellar parameters agreeing with theoretical tracks within the uncertainties. Some evidence for our \logg\,to be slightly greater than that derived from the tracks at a mean level of $\sim$0.1\,dex seems to be present for stars cooler than $\sim 4900$\,K, but this is compatible with the internal error in $\log g$ and probably due to the smaller number of \ion{Fe}{ii} lines for lower \teff. Similar findings were obtained by \cite{Brucalassietal2021} and \cite{Magrinietal2022} for targets of the {\it Ariel} mission.

The comparison of our stellar parameters with those derived in the literature when at least five targets were in common are plotted in Fig.\,\ref{fig:param_comp} and shown in Table\,\ref{tab:param_abund_compar}. In particular, we consider the comparisons with the values of the exoplanet discovery papers derived through different methods, the stellar parameters obtained by \cite{Torresetal2012} averaging three methods based on spectral synthesis, line equivalent widths, and cross-correlation against a library of spectra, the values by \cite{Breweretal2016} derived using spectral synthesis, and those listed in the SWEET-Cat (Stars With ExoplanETs Catalogue; \citealt{Santosetal2013, Sousaetal2021}). SWEET-Cat, for the first time described in \cite{Santosetal2013}, is a continuously updated catalogue of stellar parameters for planet-hosting stars derived, whenever possible, using the same methodology. A very recent version of the catalogue, whose parameters were re-derived using better quality spectra and following the same homogeneous procedure, is available (see \citealt{Sousaetal2021}). In the new version of the catalogue, recent Gaia\,EDR3 parallaxes were also considered to derive accurate surface gravity of the host stars. 

As a result from the comparison of our \teff\,, \logg\,, and \feh\,, and \vsini\,with those given by the literature and listed in Table\,\ref{tab:param_abund_compar}, the standard deviation of the difference is of $60-90$\,K, $0.07-0.23$\,dex, $0.04-0.11$\,dex, and $\sim 0.5-0.7$\,km/s respectively. Concerning the microturbulence velocity, small differences are present among the comparisons with the two versions of SWEET-Cat (see Table\,\ref{tab:param_abund_compar}), while within the discovery papers, most values of this parameter were fixed, therefore we cannot make a meaningful comparison. However, we can consider the relationship by \cite{Adibekyanetal2012a}, which is based on the dependence of \csi\,on \teff\, and \logg\, for stars with $4500-6500$\,K, $3.0-5.0$ in \logg, and $-1.4$<[Fe/H]<0.5\,dex. We find a very good agreement between the values computed through the mentioned relationship and those derived through our MOOG analysis, with mean difference of $0.10\pm0.18$\,km/s.

Surface gravities were also derived by \cite{Mortieretal2013} and \cite{Sousaetal2021} using alternative methods to those based on spectroscopy. In particular, considering the surface gravities derived by \cite{Mortieretal2013} for 11 targets in common with us using a method based on photometric light curves and the previous version of the SWEET-Cat, the difference with respect to our values is of $0.04\pm0.11$\,dex (see the inset in the second panel of Fig.\,\ref{fig:param_comp}), while considering the \logg\,derived by \cite{Sousaetal2021} through Gaia parallaxes, the standard deviations strongly improves with respect to the spectroscopic values listed by the authors, becoming $0.09\pm0.09$\,dex. Therefore, our analysis based on a careful method leads to \logg\,results that are very close to the values derived in accurate ways (through transit light curves or Gaia parallaxes), even if we are aware that the kind of analysis performed in this work is strongly time-consuming and cannot be easily applied to surveys of hundreds-thousand targets. Similar method to derive trigonometric \logg\, was already applied by \cite{Brucalassietal2021} for $\sim 150$ targets within the {\it Ariel} reference sample (see also \citealt{Tinettietal2021, Danielskietal2021, Magrinietal2022}). Due to the high quality of the Gaia parallax, the authors suggest to adopt trigonometric \logg\,as a viable possibility for big stellar samples for which some spectroscopic methods based on automatic tools tend to under/over-estimate the surface gravity at low and high temperatures. 

Concerning the iron abundances, from Fig.\,\ref{fig:param_comp} we recognize a presence of most discarding values when the comparison is made with respect to the discovery papers. In particular, the targets discarding by more than 1$\sigma$ are those from which the microturbulence velocity was not derived or was fixed (i.e. XO-2S, TRES-4, Qatar-2, HAT-P-20) or for part of the WASP sample. For this sub-sample of WASP targets (namely, WASP-38, WASP-60, WASP-54, WASP-39), $T_{\rm eff}$ and $\log g$ were derived with a method based on different strong lines; by chance, this method seems to led lower $T_{\rm eff}$, and therefore lower [Fe/H] than our values. We do not go deeper in details because we are aware that these analyses date back to more than ten years ago. Here, we only remark that most recent analysis points toward findings more similar to our results.

\cite{Mortieretal2013} and \cite{Breweretal2016} measured, respectively, also abundances of 12 and 14 elements other than iron. For the 11 targets in common with \cite{Mortieretal2013}, we find a mean difference in [X/H] of $-0.01\pm0.07$\,dex for Al, Ca, Co, Cr, \ion{Cr}{ii}, Mg, Mn, Na, Ni, \ion{Sc}{ii}, Si, Ti, \ion{Ti}{ii}, and V, while for the five targets in common with \cite{Breweretal2016} we find a mean difference in elemental abundances of C, N, O, Na, Mg, Al, Si, Ca, Ti, V, Cr, Mn, Ni, and Y of $0.02\pm0.04$\,dex. Furthermore, \cite{Burkeetal2007} derived Na, Si, Ti, Ni abundances for the XO-2 binary components, \cite{Teskeetal2014} measured C, O, Ni for XO-2N, XO-2S, and TRES-4, while \cite{Teskeetal2019} obtained C, O, Mg, Si, Ni abundances for HAT-P-15 and HAT-P-17. Comparing our results with those achieved by these authors, we find mean differences of $0.03\pm0.09$, $-0.04\pm0.08$, and $0.00\pm0.07$\,dex with respect to \cite{Burkeetal2007}, \cite{Teskeetal2014}, and \cite{Teskeetal2019}, respectively. Therefore, a general agreement between our abundance values and those from the literature is obtained for the targets in common.

\begin{table*}[ht]
\caption{Mean difference and standard deviation between the values of the atmospheric parameters derived in the literature and those obtained in this work ($n$ is the number of stars in common).} 
\label{tab:param_abund_compar}
\begin{center}
\begin{tabular}{rrrrrcl}
\hline
\hline
$\Delta T_{\rm eff}$ & $\Delta \log g$ & $\Delta$[Fe/H] & $\Delta \xi$ & $\Delta v\sin i$ & $n$ & Reference \\ 
(K) & (dex) & (dex) & (km/s) & (km/s) &   &   \\ 
\hline
$-25\pm82$ & $0.06\pm0.07$ & $-0.04\pm0.11$ & $0.03\pm0.33^\ast$ & $-0.1\pm0.7$ & 28 & Discovery papers$^\diamond$\\
 $20\pm60$ & $0.06\pm0.06$ &  $0.00\pm0.08$ &      ...      & $-0.1\pm0.5$ & 15 & \cite{Torresetal2012}\\
 $35\pm90$ & $0.04\pm0.21$ &  $0.00\pm0.07$ & $0.21\pm0.20$ & ... & 28 & \cite{Santosetal2013}\\
$-57\pm85$ &$-0.07\pm0.09$ &  $0.04\pm0.08$ &      ...      &$-0.3\pm0.6$ & 5 & \cite{Breweretal2016}\\
 $51\pm85$ &$-0.04\pm0.23$ &  $0.01\pm0.04$ & $0.23\pm0.22$ & ... & 28 & \cite{Sousaetal2021}\\
\hline
\end{tabular}
\end{center}
Notes: $^\ast$ $\xi$ in the discovery papers was often fixed. $^\diamond$ For the discovery papers, see the NASA Exoplanet Archive.
\end{table*}

\section{Results and discussion}
\label{sec:result_discussion}

\subsection{[X/H] versus [Fe/H]}
\begin{figure*} 
\begin{center}
\includegraphics[width=14cm,angle=90]{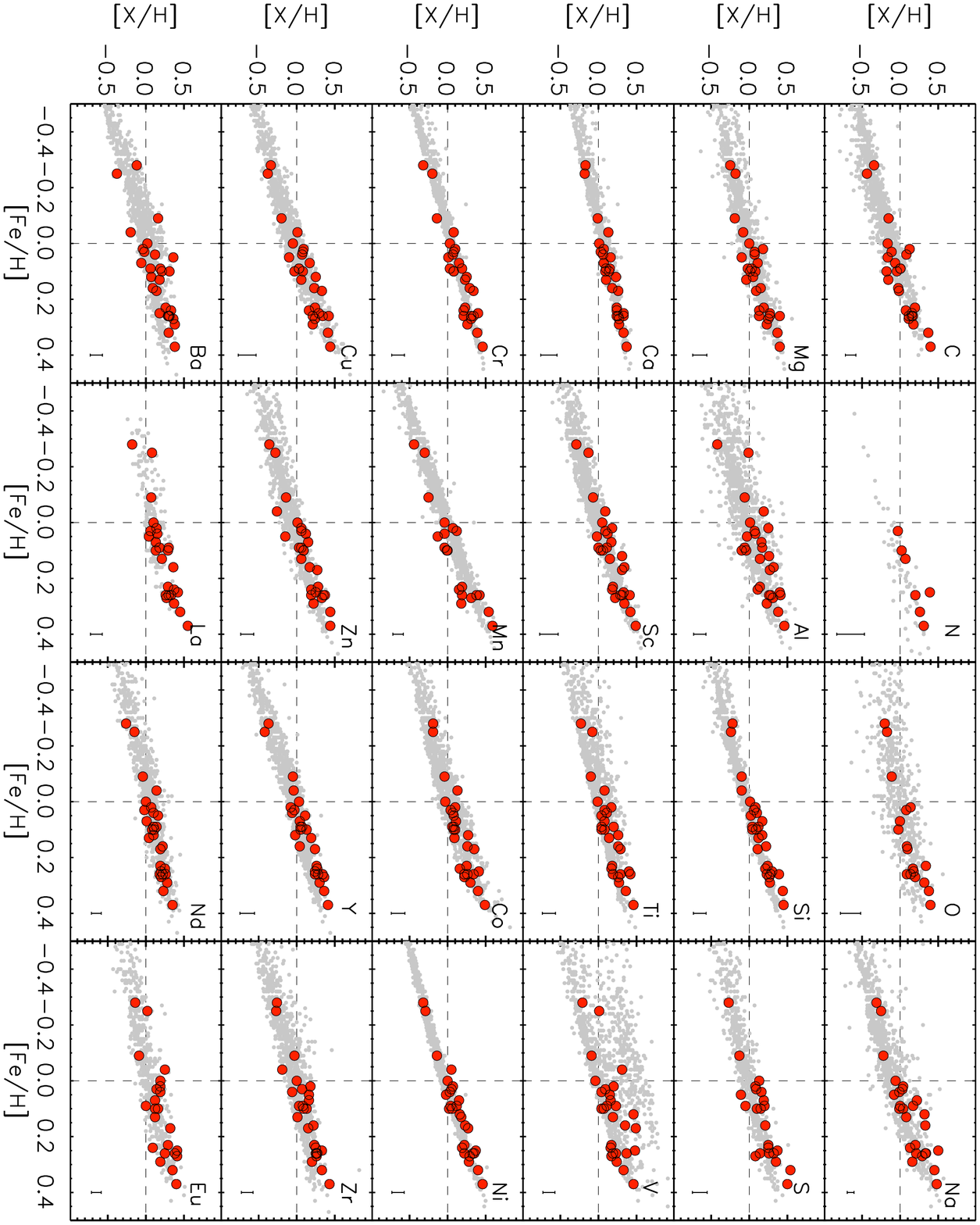}
\caption{[X/H] versus [Fe/H] for our sample. C by \cite{suarezandresetal2017} for $\sim 4400-6800$\,K warm targets, N by \cite{suarezandresetal2016} for stars with \teff\,$\sim 4500-6500$\,K, O by \cite{bertrandelisetal2015} 
for \teff\,$\sim 5200-6800$\,K, Na, Mg, Al, Si, Ca, Sc, Ti, V, Cr, Mn, Co, and Ni by \cite{Adibekyanetal2012c} for targets with \teff\,around 4500-6800\,K, S by \cite{costasilvaetal2020} for 4900-6800\,K, Cu, Zn, Y, Zr, Ba, Nd, and Eu by \cite{delgadomenaetal2018} for stars with 4500-6800\,K, and La by \cite{misheninaetal2016} for 4800-6200\,K warm stars are overplotted with grey points. The La values of the latter authors were shifted by 0.2\,dex. Mean error in [X/H] are shown on the bottom-right of each panel, 
while solar values are marked with dashed lines.}
\label{fig:abundances_FeH} 
\end{center}
\end{figure*}

The [X/H] (and [X/Fe]) versus [Fe/H] relations are generally used to study the Galactic chemical evolution since iron is a good chronological indicator of nucleosynthesis. In Fig.\,\ref{fig:abundances_FeH} we show the [X/H] values for all derived elemental abundances versus [Fe/H] (while in Fig.\,\ref{fig:abundancesFe_FeH} we show [X/Fe] vs [Fe/H]). We also overplot in grey the abundances of FGK dwarf stars observed within the HARPS at ESO GTO planet search program or with ELODIE at OHP. 
Studying FGK dwarfs is very useful because they contain information about the history of the evolution of chemical abundances in the Galaxy. Low-mass stars have indeed long lifetime and their atmospheres have preserved much of their original chemical 
composition (e.g., \citealt{Adibekyanetal2012c}). 

We note that for some iron-peak elements, like [Cr/H] and [Ni/H], which are synthesized by SNIa explosions, and some $\alpha$-elements (like [Si/H] and [Ca/H]), which are mostly produced in the aftermath of type II SNe explosions with small contributions 
from type Ia SNe, we obtain low dispersion for the whole range of [Fe/H], as for the dwarfs in the Galactic disk. Other elements, like sodium and aluminum, mostly produced through Ne and Mg burning in massive stars, show greater dispersion, as observed 
in the literature. Moreover, abundance ratios like [Al/Fe], [Sc/Fe], [V/Fe], [Co/Fe], and [Ni/Fe] show a rise for [Fe/H]$\gtsim$0.2, as observed in thin disk stars (see \citealt{Adibekyanetal2012c}). Here, we cannot conclude if the observed dispersions are 
intrinsic or due, at least in part, to statistical or methodological reasons (e.g., number of lines, $SNR$, etc). We only highlight how for all elements the general pattern of [X/H] and [X/Fe] with the iron abundance is similar to those of the sample of dwarf 
stars in the Galactic disk, without clear evidences of peculiar behavior for our planet-host stars when compared with field stars. Therefore, the [X/H]-[Fe/H] trends of our sample seem to reflect the Galactic chemical evolution in the solar neighborhood. 

\subsection{Elemental abundance versus condensation temperature} 
\label{sec:condensation}
\begin{figure*} 
\begin{center}
\includegraphics[width=17cm]{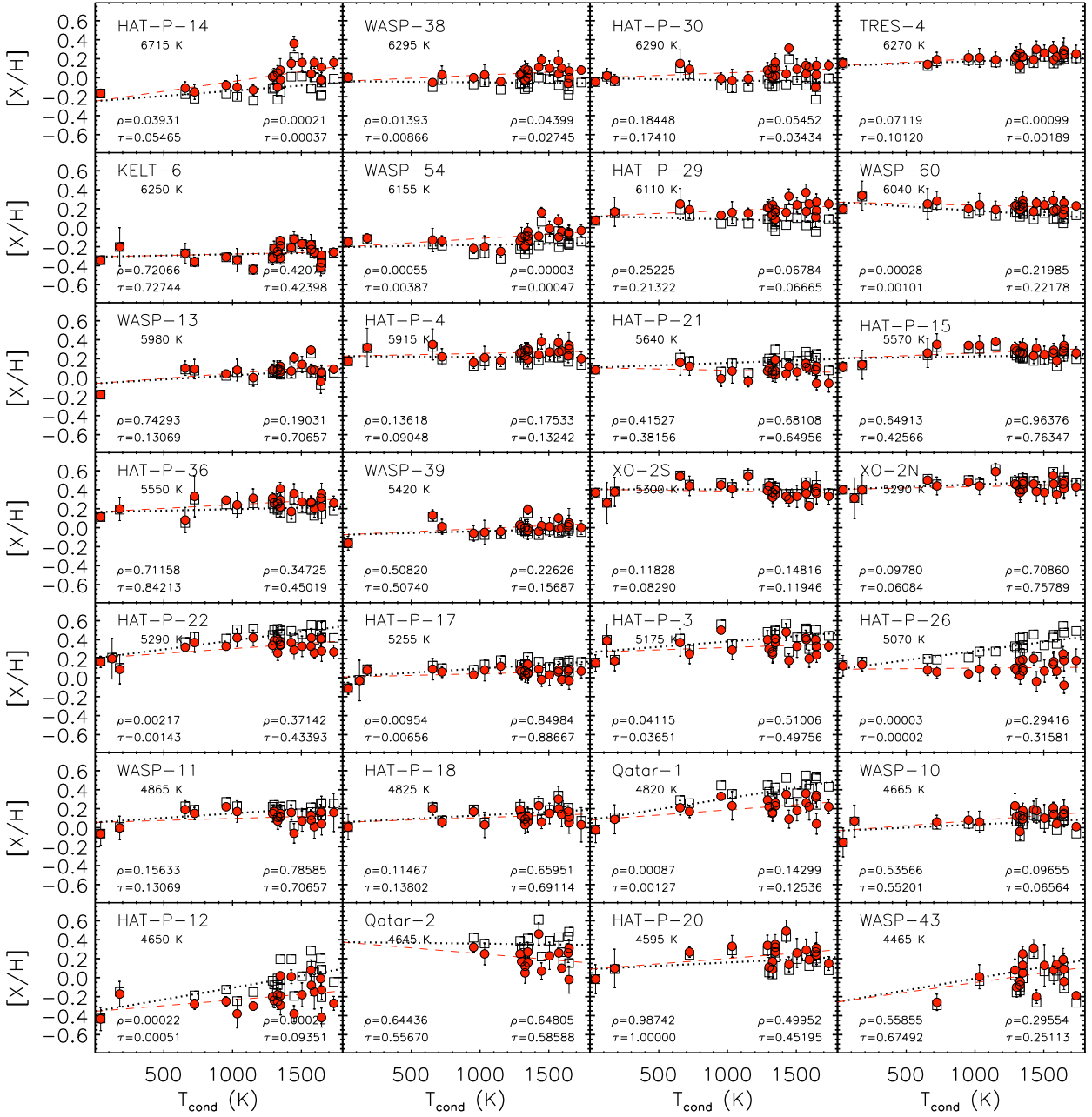}
\caption{[X/H] versus $T_{\rm cond}$ for all targets from the warmest HAT-P-14 to the coolest WASP-43. Open squares (together with black dot line) and filled dots (together with red dashed line) correspond to the abundance values with and without the GCE correction. $\rho$ and $\tau$ in all panels are the Spearman and Kendall significance after (left corner) and before (right corner) the GCE removal.}
\label{fig:abund_Tcond} 
\end{center}
\end{figure*}

In Fig.\,\ref{fig:abund_Tcond} we plot for each star the elemental abundance [X/H] as a function of the condensation temperature $T_{\rm cond}$, from the warmest target to the coolest one. It was indeed reported that stars hosting high-mass planets are expected to be more enriched in refractory elements, i.e. elements with high condensation temperature, and deficient in low $T_{\rm cond}$ volatile elements (e.g., \citealt{Smithetal2001}). This should happen because any accretion event, occurring very close to the star, i.e. in high-temperature environments, would add refractory elements, which condense at high $T_{\rm cond}$ with respect to volatiles (see, e.g., \citealt{Sozzettietal2006}, and references therein). Similarly, \cite{melendezetal2009} concluded that solar twins without close-in giant planets chemically resemble the Sun, with depletion of refractory elements relative to the volatiles, suggesting that the presence of such planets might prevent the formation of Earth-sized planet. Revealing these trends requires developing a very accurate differential analyses, which is precise for binary systems (e.g., \citealt{grattonetal2001, ramirezetal2011, liuetal2014, teskeetal2015, biazzoetal2015, tuccimaiaetal2019}) and members of stellar clusters (e.g., \citealt{Yongetal2013}) or through the use of set of comparison stars (\citealt{Liuetal2020, Tautvaisieneetal2022}), because many observational uncertainties can be considered common-mode effects. If no comparing targets is available, an unambiguous explanation for these trends is difficult to reach because they could also reflect the wide diversity of exoplanetary systems, as well as a variety of scenarios which could occur within the circumstellar disk (\citealt{Spinaetal2016a}), or could be associated with the correlation of elemental abundances with the age and birthplace in the Galaxy (e.g., \citealt{Adibekyanetal2014}). \cite{gonzalezhernandezetal2013} have found that after removing the Galactic chemical evolution effects from a sample of main-sequence objects, stars with and without planets show similar mean abundance patterns. We applied a similar approach, adopting \cite{Spinaetal2016b} for the correction of Galactic Chemical Evolution (GCE). In Fig.\,\ref{fig:abund_Tcond} we show the [X/H] abundances as a function of $T_{\rm cond}$ before (filled dots and dashed lines) and after (open squares and dotted lines) the removal of the GCE effects. The Spearman ($\rho$) and Kendall ($\tau$) statistical significance computed with IDL\footnote{IDL (Interactive Data Language) is a registered trademark of Exelis Visual Information Solutions.} for the linear regression analysis before and after the GCE removal are displayed respectively in the bottom-right and bottom-left corners of the plots. Excluding Qatar-2 and WASP-43 for which we could not measure abundance of volatile elements, within the other 26 targets, those showing a trend after the GCE removal are ten, with both $\rho$ and $\tau$ lower than 0.05 and wide range of planetary masses (from 0.06\,$M_{\rm J}$ up to 2.6\,$M_{\rm J}$) and effective temperature (from 4650 up to 6715\,K). In particular, HAT-P-12 and HAT-P-26, respectively hosting planets with $M_{\rm p} \sim 0.2\,M_{\rm J}$ and 0.06\,$M_{\rm J}$, show the most pronounced significant trend at a level of $\rho \sim \tau \sim 2-3\times10^{-5}$ up to $2-5\times10^{-4}$ and among the highest values of positive slopes, while WASP-60 (with a planet of 0.5\,$M_{\rm J}$) is the only within this sub-sample of 10 targets showing a negative trend, i.e. a decreasing refractory-to-volatile abundance ratios. From one side, we note that HAT-P-26 and HAT-P-12 are also targets with kinematic properties consistent with transition thin-thick disk (see Sect.\,\ref{sec:spacemotion}) and with high values of [$\alpha$/Fe] and Mg/Si (see Sect.\,\ref{sec:mgsi_co_fe}), which, together with the possible [X/H]-$T_{\rm cond}$ trend, could indicate some pattern in the formation of the nuclei of their planets. On the other hand, we note that the binary components XO-2N and XO-2S, here analyzed as single stars, do not show reliable correlation between elemental abundances and $T_{\rm cond}$, while precise and accurate differential analysis demonstrated that the difference in elemental abundance between these two binary components shows a trend with the condensations temperature indicating possible ingestion of material by XO-2N or depletion in XO-2S (see \citealt{teskeetal2015, biazzoetal2015}). Finally, excluding Qatar-2 and WASP-43 for which we could not measure elemental abundance of volatiles, we find higher values of $T_{\rm cond}$-[X/H] slopes for cooler and older stars (see Sect.\,\ref{sec:ages}) with higher $\log g$, regardless of planetary mass and not evident without the GCE removal. We also find tentative evidence that stars with smaller Galactocentric distance and greater Galactic eccentricity (see Sect.\,\ref{sec:spacemotion}) have steeper slopes. Similar results were also found by other authors (e.g., \citealt{Adibekyanetal2014, Tautvaisieneetal2022}). In other words, stellar parameters and Galactic position are determinant to establish the stellar chemical pattern of the stars and only a strictly differential analysis (like those performed for binary stars) can remove spurious trends and help to draw definitive conclusions, which is not the case of this work.

\subsection{Chemical and kinematic properties}
\label{sec:spacemotion}

\begin{figure*} 
\begin{center}
\includegraphics[width=17cm]{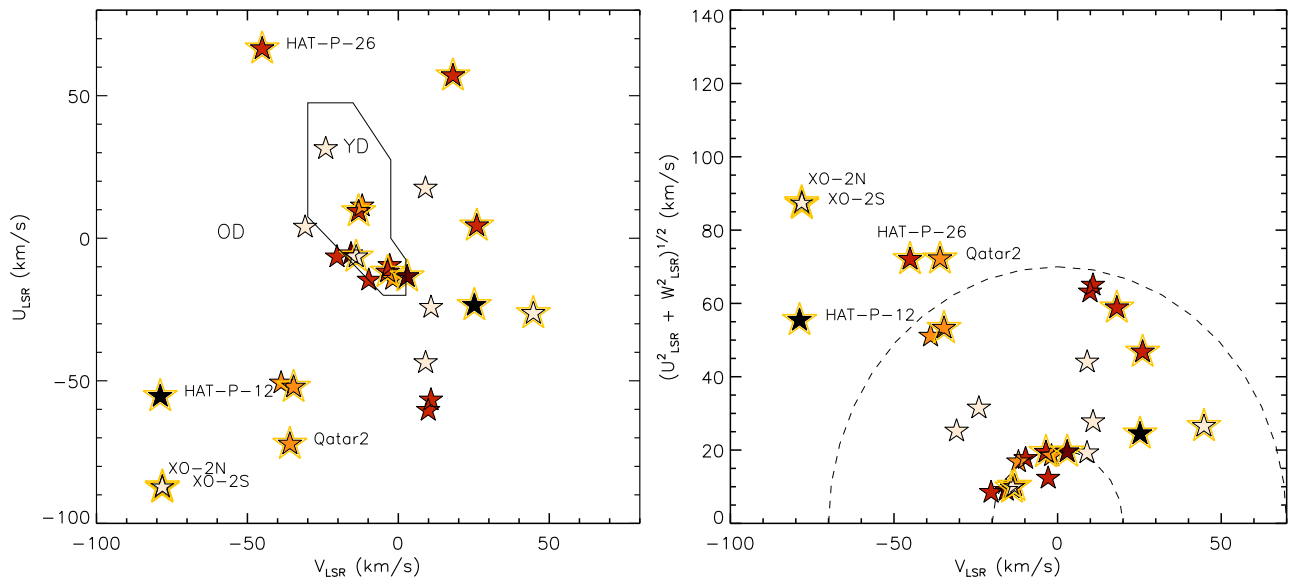}
\caption{{\it Left panel:} space velocities of our exoplanet-hosting stars in the ($U,V$) plane. Star symbols are filled in darker colors following the same [Fe/H] bins as Fig.\,\ref{fig:logg_teff}. Empty yellow stars surround those targets with stellar 
age greater than the median value of the sample, i.e. $\sim 5.5$\,Gyr. The solid line represents the boundary separating young-disk (YD) and old-disk (OD) stars according to \cite{eggen1996}. {\it Right panel:} Toomre diagram for our planet-hosting 
stars. Dashed lines indicate constant peculiar total velocities $v_{\rm tot}=(U_{\rm LSR}^2+V_{\rm LSR}^2+W_{\rm LSR}^2)^{1/2}$= 20 and 70 km/s. Stars with $v_{\rm tot} > 70$\,km/s are labeled. Errors in both panels are within the symbol dimension.}
\label{fig:space_motion} 
\end{center}
\end{figure*}

To study possible chemo-kinematic peculiarities of planet-hosting stars, we computed the stellar Galactic space velocities. The space velocity components $UVW$ were derived with respect to the 
local standard of rest (LSR), correcting for the solar motion derived by \cite{coskunogluetal2011}, i.e. ($U_\odot$, $V_\odot$, $W_\odot$)=($-8.50$, 13.38, 6.49) km/s. Parallaxes ($\pi$) and propers motions ($\mu_{\alpha}, \mu_{\delta}$) 
were taken from {\it Gaia}\,EDR3 (\citealt{gaiacollaborationetal2016, gaiacollaborationetal2021}), mean radial velocities ($<V_{\rm rad}>$) were obtained from the HARPS-N spectra, and ICRS coordinates at epoch=2000 were taken from the 
SIMBAD Astronomical Database. We considered the general outline of \cite{johnsonsoderblom1987} in a left-hand coordinate system, i.e. with $U$ pointing towards the Galactic anti-center, $V$ towards the local direction of rotation in the 
plane of the Galaxy, and $W$ towards the North Galactic Pole. The uncertainties were obtained considering the prescriptions by \cite{gagneetal2014}, thus we used the full covariance matrix taking into account the error contributions of 
$V_{\rm rad}$, $\pi$, $\mu_{\alpha}$, and $\mu_{\delta}$. Combining the measurement errors in parallaxes, proper motions, and radial velocities, the resulting average uncertainties in the $U$, $V$, $W$ velocities are of about 0.15\,km/s. 
The values derived for each target are listed in Table\,\ref{tab:space_motion} and the Boettlinger Diagram in the ($U$, $V$) plane is shown in the left panel of Fig.\,\ref{fig:space_motion}, where the boundary separating of the young-disk (YD) and the old-disk 
(OD) stars according to \cite{eggen1996} is displayed with a solid line. The YD locus contains associations younger than $\sim 1$\,Gyr (see \citealt{gagneetal2018}). Nine targets, namely HAT-P-14, TRES-4, WASP-54, WASP-13, HAT-P-3, WASP-11, HAT-P-18, WASP-10, and HAT-P-20 seem to belong to the young disk, and other three (WASP-43, HAT-P-4, WASP-38) are very close to the YD boundary. On the right panel of the same figure we show the Toomre diagram, which is a representation of the combined 
vertical and radial kinetic energies versus the rotational energy. Low-velocity stars, within a total velocity $v_{\rm tot}=(U_{\rm LSR}^2+V_{\rm LSR}^2+W_{\rm LSR}^2)^{1/2}$= 50\,km/s are, to a first approximation, mainly thin disk stars, while stars with $70 \ltsim v_{\rm tot} \ltsim 180$\,km/s are likely to be thick disk stars (see \citealt{bensbyetal2014}, and references therein). Eleven of our targets, namely HAT-P-14, WASP-38, HAT-P-29, WASP-54, WASP-13, HAT-P-3, WASP-11, HAT-P-18, WASP-10, HAT-P-20, and WASP-43 are very close to the Sun, with $v_{\rm tot} \ltsim 20$\,km/s, while five exoplanet-hosting stars (XO-2N, XO-2S, Qatar-2, HAT-P-26, HAT-P-12) have $v_{\rm tot} > 70$\,km/s, compatible with thick disk stars. 

We also calculated the thick-to-thin disk probability ratios. In particular, we considered the prescriptions of \cite{bensbyetal2014} for the Gaussian distributions of random velocities of the different stellar populations. To get the probability $D$ and $TD$ for the thin and thick disk that a given star belongs to a specific population, we considered the asymmetric drift, the velocity dispersion, and the fractions of each population listed in their Table\,A.1. By then dividing the thick disk probability with the thin disk probability, we get the probability for the thick disk-to-thin disk ($TD/D$) membership. \cite{bensbyetal2014} require that for a star to be a candidate thick disk star its probability 
must be at least twice that of being a thin disk star (i.e. $TD/D > 2$), and vice versa for a candidate thin disk star $TD/D < 0.5$. All our targets show $TD/D < 0.5$ with the exception of 5 stars: $i.$ two targets with probability ratios in between 
the thin and thick disks, i.e. HAT-P-21 and HAT-P-26, with $TD/D \sim 0.5$; $ii.$ three targets with probabilities consistent with the thick disk, i.e. the star HAT-P-12 (with a value of $\sim 3.7$), and the XO-2 binary system 
(with $TD/D \sim 13$). All targets within or nearby the YD boundary show $TD/D$ close to zero. The same occurs for targets with $v_{\rm tot} \ltsim 20$\,km/s.

\begin{figure} 
\begin{center}
\includegraphics[width=8.5cm]{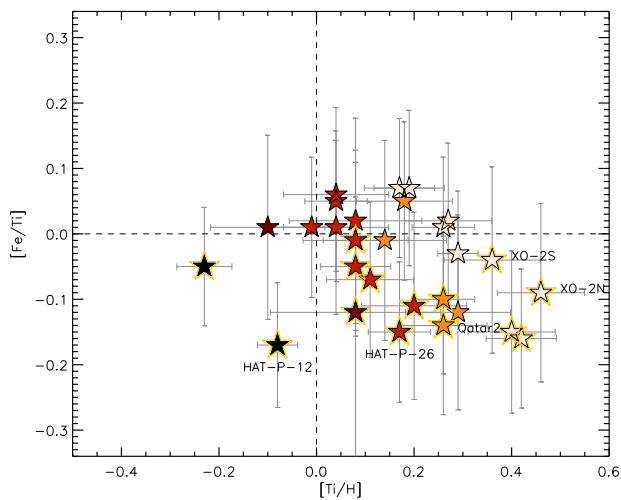}
\caption{[Fe/Ti] versus [Ti/H] abundance trends. Stars have been color-coded depending on their [Fe/H], as done in Fig.\,\ref{fig:logg_teff}. Empty yellow stars surround those targets with stellar age greater than the median value 
of the sample (i.e. $\sim 5.5$\,Gyr). Dashed lines represent the solar values. Stars with $v_{\rm tot} > 70$\,km/s are labeled. }
\label{fig:FeTi_Ti_Age} 
\end{center}
\end{figure}

Following \cite{bensbyetal2014}, with the aim to further investigate the chemo-kinematic properties of our sample, we show in Fig.\,\ref{fig:FeTi_Ti_Age} the [Fe/Ti]-[Ti/H] abundance trend. All targets with $TD/D > 0.5$ (XO-2N, XO-2S, HAT-P-21, HAT-P-26, HAT-P-12) are placed below [Fe/Ti]=0. Moreover, we have also coded the symbols with an empty yellow star when the chemical stellar age is greater than the median value of the entire sample (i.e. $\sim 5.5$\,Gyr; see Sect.\,\ref{sec:ages} for the age determination). It is evident how the [Fe/Ti] abundance signature has the same structure as the age, with stars older than $\sim 5.5$\,Gyr having [Fe/Ti]<0.0. Moreover, most of the targets with [Fe/Ti]<0.0 show different kinematic positions in the Toomre diagram when compared to the targets with [Fe/Ti]$\ge$0.0, with mean $v_{\rm tot}$ greater than about 20\,km/s, with few exceptions. 
Hence, there are kinematically cold stars that are older and $\alpha$-enhanced, i.e. with higher values of Ti, as well as kinematically hot stars that are younger and less $\alpha$-enhanced, in line with what found by \cite{bensbyetal2014} for dwarf thin/thick disk stars in the solar neighborhood. We also note that within the four stars hosting planets with masses smaller than $5\,M_{\rm Nep}$ (i.e. XO-2S, HAT-P-26, HAT-P-18, HAT-P-12) all but HAT-P-18 show $v_{\rm tot} > 20$ km/s. Moreover, their mean distance from the Sun is smaller (by more than 60\,pc) than that of stars hosting planets with higher masses. Similar 
results were found by \cite{Adibekyanetal2012b}, who noted that, as expected, low-mass planets are easier to find at smaller distances due to the higher apparent magnitudes of their hosts. We therefore performed a similar approach 
as that proposed by these authors. In particular, we calculated the maximal height a star can reach above the Galactic plane ($Z_{\rm max}$), the current Galactic eccentricity ($e_{\rm G}$), the peri-/apo- ($R_{\rm peri}$, $R_{\rm apo}$) center radii 
of an orbit, and the Galactocentric distance ($R_{\rm GC}$) with the \texttt{galpy}\footnote{\url{http://github.com/jobovy/galpy}} package, a python package for galactic-dynamics calculations. We assumed the built-in model \texttt{MWpotential2014} for the Milky Way's gravitational potential (see \citealt{Bovy2015}). 
We set the distance from the Galactic center to $R_{0} = 8.0$\,kpc (\citealt{Bovyetal2012}) and the height above the plane to $z_{0} = 0.025$\,kpc (\citealt{Juricetal2008}), and used the parallaxes and proper motions from {\it Gaia}\,EDR3 to 
transform the celestial coordinates in galactocentric radius and height above the Galactic plane. We also used our mean radial velocities from HARPS-N to obtain the orbital parameters and considered the chemical ages as derived in 
Sect.\,\ref{sec:ages}. Table\,\ref{tab:space_motion}, together with $UVW$ velocities and $TD/D$ probabilities, lists also the output results from the \texttt{galpy} package.

First, we note that all targets with $TD/D \gtsim 0.5$ show also $|\Delta (R_{\rm mean} - R_{\rm GC})| \gtsim 1$, where $R_{\rm mean}$ is the mean position of the stellar galactic orbits (calculated as the average of $R_{\rm peri}$ and $R_{\rm apo}$). 
This could indicate that stars with $R_{\rm mean}$ very different from $R_{\rm GC}$ could have experienced greater migration and therefore could have higher probability to belong to the thin-thick 
disk transition or thick disk (see \citealt{Magrinietal2022}; and references therein). Then, we find that on average stars hosting planets with masses lower than $5\,M_{\rm Nep}$ have higher Galactic eccentricity (by $\sim 0.2$) and greater age (by $\sim 5$\,Gyr), while the difference in $Z_{\rm max}$ of only $\sim 45$\,pc is within the uncertainties in the average (of around 150\,pc). These low-mass planet-hosting stars have also lower $V_{\rm LSR}$ and $W_{\rm LSR}$ space velocity components than stars hosting higher mass planets. Moreover, comparing their [Mg/Fe], [Si/Fe], and [Ti/Fe] ratios, we find that the stars with $M_{\rm p} < 5\,M_{\rm Nep}$ planets are more enhanced by Mg (with a difference of 0.09 dex), by Si (difference of 0.04\,dex), and Ti (difference of 0.09\,dex) when compared to stars with higher mass planets. These elements are all tracers of the rocky component of the cores of the Neptune-mass planets, thus possibly indicating that Neptunian/super-Earth host stars tend to belong to the "thicker" disk when compared with Jupiter mass hosting stars. Even if we are cautious about these findings because they could depend on several selection effects (e.g., stellar magnetic activity), our results seem to give support to the findings by \cite{Adibekyanetal2012b}, according to which stars hosting low-mass planets tend to belong to a "thicker" disk.

\begin{figure} 
\begin{center}
\includegraphics[width=7.5cm,angle=90]{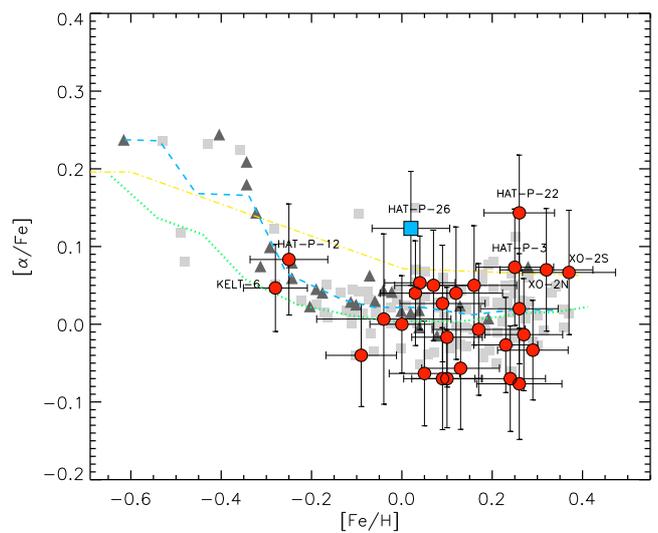}
\caption{[$\alpha$/Fe] versus [Fe/H] for our exoplanet-hosting sample. Dots refer to the massive-planet hosts, while the square is the position of the Neptunian host HAT-P-26 ($M_{\rm p} < 30\,M_\oplus$). The grey squares and triangles refer to the 
Jovian hosts and the stars hosting exclusively Neptunians and super-Earths, respectively, by \cite{Adibekyanetal2012b}. The dashed blue and dotted green lines represent, respectively, the mean distributions of the planet 
host and non-host samples by the same authors. The dot-dashed yellow line separates the stars with high- and low-$\alpha$ content as found by \cite{Adibekyanetal2012c}. The positions of the two most metal-poor and $\alpha$- enhanced targets within our sample are labeled.}
\label{fig:fe_alpha} 
\end{center}
\end{figure}

In Fig.\,\ref{fig:fe_alpha} we plot [$\alpha$/Fe]\footnote{The $\alpha$ index refers to the average abundance of Mg, Si, and Ti, i.e. [$\alpha$/Fe]=$\frac{1}{3}$([Mg/Fe]+[Si/Fe]+[Ti/Fe]). Calcium was not 
included because for [Fe/H]>0 the [Ca/Fe] trend for dwarf stars in the Galactic disk differs from that of Mg, Si, and Ti (see \citealt{Adibekyanetal2012c}).} against [Fe/H] for our sample, 
where the position of the star HAT-P-26 hosting a Neptunian/super-Earth planet (i.e. with masses $< 30\,M_\oplus$) is represented with a blue square. In the same plot, we show the results by \cite{Adibekyanetal2012b}, who analyzed 1111 nearby FGK dwarfs observed in the context of the HARPS GTO program, 135 of which hosting high-mass planets (grey squares), and Neptunians/super-Earths with masses $< 30\,M_\oplus$ (grey triangles). They have found that planet-hosting stars show a continuous increase 
in [$\alpha$/Fe] with decreasing [Fe/H] at metallicities from $-0.2$ to $-0.3$\,dex (starting from the thin disk, they rise toward the thick disk), while the thin and thick disk stars without planets are separated very 
well by their [$\alpha$/Fe] ratios (see dashed and dotted lines). In this metallicity regime we have two targets (namely, KELT-6 and HAT-P-12). KELT-6 shows $UV$ values compatible with the Galactic young disk, it is close 
the circle with total velocity of 20 km/s in the Toomre diagram, and the $TD/D$ ratio is very low ($\sim 0.02$), thus appearing to belong to the thin disk (see Fig.\,\ref{fig:space_motion}). HAT-P-12 shows a relatively high proper motion, it is compatible with thick disk stars in the Toomre diagram, and $TD/D \sim 3.7$, i.e. slightly higher than the thick disk threshold. Concluding, our chemo-kinematic analysis for these two targets seem to be consistent with KELT-6 clearly belonging to the thin disk and HAT-P-12 in between thin and thick disk. Moreover, our findings support the conclusion by \cite{Adibekyanetal2012b}, according to which planet-hosting metal-poor stars (like KELT-6) can have high [$\alpha$/Fe] even belonging to the thin disk. 

Furthermore, we note that the only target within our sample hosting a Neptunian/super-Earth planet, namely the solar-metallicity star HAT-P-26, shows the highest value of [$\alpha$/Fe], when compared to the sample by \cite{Adibekyanetal2012b} with similar 
planetary masses and iron abundance (triangles in Fig.\,\ref{fig:fe_alpha}). The relatively high abundance values of $\alpha$ elements for this target are similar to those of their Neptunian/super-Earth host stars with iron abundance around $-0.3$\,dex. 
Its position in Fig.\,\ref{fig:fe_alpha} is indeed consistent with the stars with high-$\alpha$ content as defined by \cite{Adibekyanetal2012c}. Again, this target shows high space velocities (see Fig.\,\ref{fig:space_motion}) 
and in the Toomre diagram, accordingly to \cite{bensbyetal2014}, it is in the region populated by thick disk stars. Finally, its $Z_{\rm max}$ is around 565 pc, i.e. one of the highest in the sample and the $TD/D$ ratio is $\sim 0.5$. 
Therefore, the chemo-kinematic analysis for this target seems to be consistent with a star close to the thin/thick disk transition, also supported by its relatively old age (see Sect.\,\ref{sec:ages}).

Other two targets with $\alpha$ content higher than the limit defined by \cite{Adibekyanetal2012c} are HAT-P-22 and HAT-P-3. The kinematic position in the $(U,V)$ plane and the Toomre diagram for the super-solar star HAT-P-3 are consistent with the thin disk, 
also supported by the $TD/D$ ratio close to zero. HAT-P-22 shows $v_{\rm tot} \sim 50$\,km/s and a $TD/D$ ratio of $\sim 0.08$. Again, both components of the XO-2 binary system are slightly above the aforementioned chemical limit and have also high Galactic space velocities. Moreover, their position in the Toomre diagram is within the locus of the thick disk stars, and the $TD/D$ ratio is $\sim 13$, i.e. higher than the threshold established by \cite{bensbyetal2014} for potential thick disk stars. 
From the analysis of the galactic orbits we find an eccentric orbit ($e_{\rm G} \sim 0.44$) with a maximum height above the Galactic plane of $\sim 104$\,pc. The relatively low $Z_{\rm max}$ led in the past to conclude that the binary system 
is confined to the Galactic thin disk (see \citealt{Burkeetal2007, damassoetal2015a}), but the other chemo-kinematic indicators suggest that the XO-2 binary system is at least in the thin/thick disk transition.

\subsection{Chemical and Isochronal Ages}
\label{sec:ages}
We computed stellar ages using elemental abundance ratios. It was indeed demonstrated that abundance ratios of pairs of elements produced over different timescales (e.g., [Y/Mg], [Y/Al]) can be used as valuable indicators of stellar age (\citealt{Nissen2015}). In fact, their [X/Fe] ratios show opposite behaviors with respect to stellar age (e.g., [Mg/Fe] and [Y/Fe] abundances respectively decrease and increase with stellar age). Therefore, their ratio, for instance [Y/Mg], shows a steep increasing trend with stellar age (see \citealt{casalietal2020}, and references therein). The latter authors derived relations in the form [A/B]=$c$+$x_1$ $\cdot$ [Fe/H]+$x_2$ $\cdot$ $Age$, with [A/B] generic abundance ratio used as a chemical clock (see their paper for a wide description of the method). Here, we considered the multivariate linear 
regression parameters $c$, $x_1$, and $x_2$ for all abundance ratios in common with \cite{casalietal2020}, i.e. [Y/Mg], [Y/Al], [Y/Ti], [Y/Ca], [Y/Si], [Y/Sc], [Y/V], [Y/Co], [Y/Zn], and [Zn/Fe] (see their Table\,6). We derived mean ages from all abundance 
ratios and rejected those values discrepant more than 1$\sigma$ from the average. 

Moreover, we also computed stellar ages (and masses) from isochrones fitting (and evolutionary tracks). We therefore considered the PARSEC\footnote{PAdova and TRieste Stellar Evolutionary Code} models by \cite{bressanetal2012} and the PARAM interface\footnote{http://stev.oapd.inaf.it/cgi-bin/param\_1.3} (version 1.3; \citealt{dasilvaetal2006}). 
This code considers as input some observational parameters (effective temperature, parallax, apparent $V$ magnitude, and iron abundance) to perform a Bayesian determination of the most 
likely stellar intrinsic properties, appropriately weighting all the isochrone sections that are compatible with the observational parameters. A flat distribution of ages with a range of 0.1-15\,Gyr was 
considered as priors for the analysis. We considered as effective temperature and iron abundance those values derived by us as described in Sect.\,\ref{sec:stellar_parameters}. The parallax was 
taken from {\it Gaia}\,EDR3, while the $V$ magnitude was computed from the  {\it Gaia}\,EDR3 $G$ magnitudes, $G_{\rm BP}$ and $G_{\rm RP}$ colors using the appropriate photometric relationships 
(\citealt{gaiacollaborationetal2021}) and the reddening maps by \cite{Capitanioetal2017}.

\begin{figure*} 
\begin{center}
\includegraphics[width=9cm, angle=90]{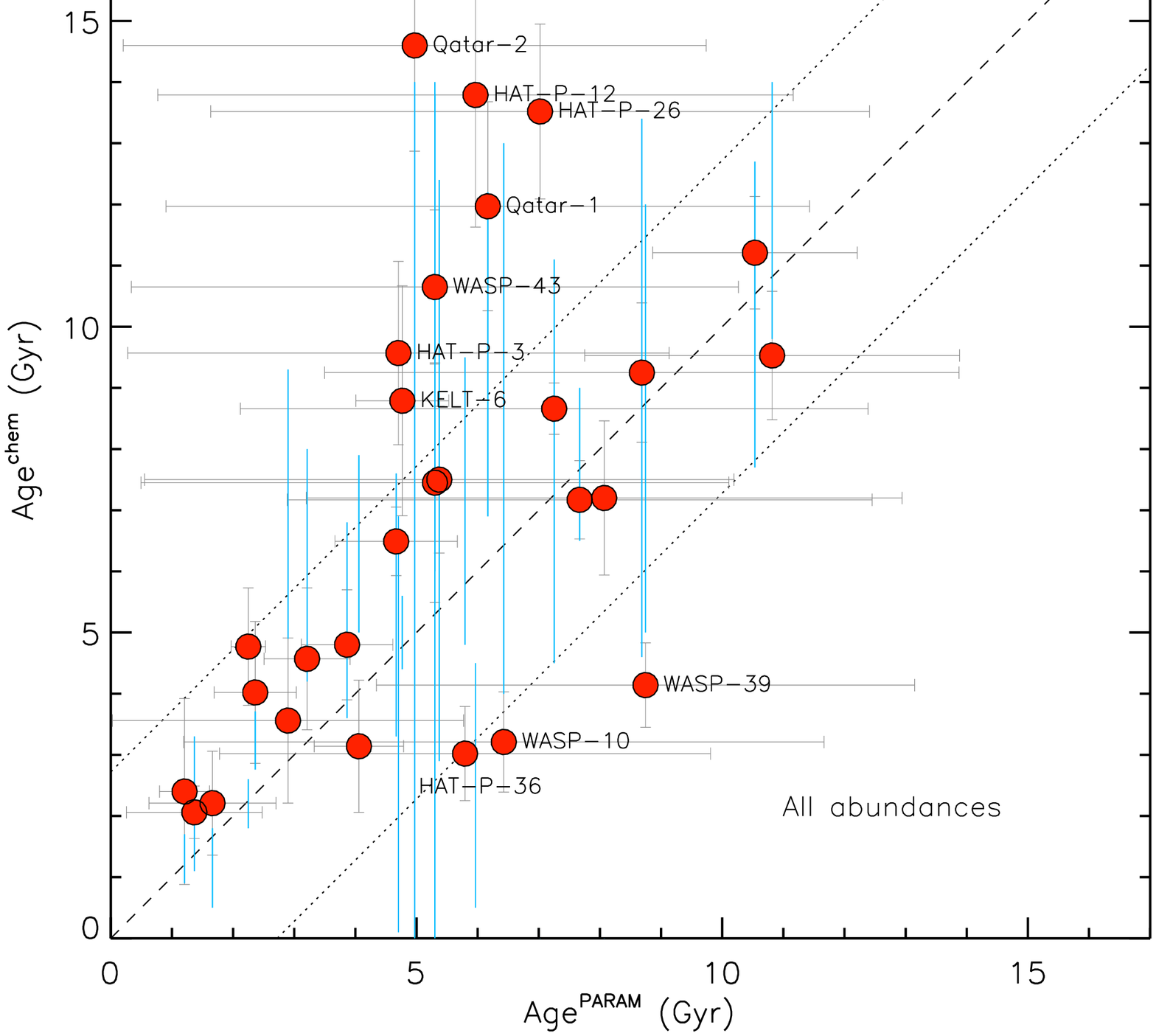}
\includegraphics[width=9cm, angle=90]{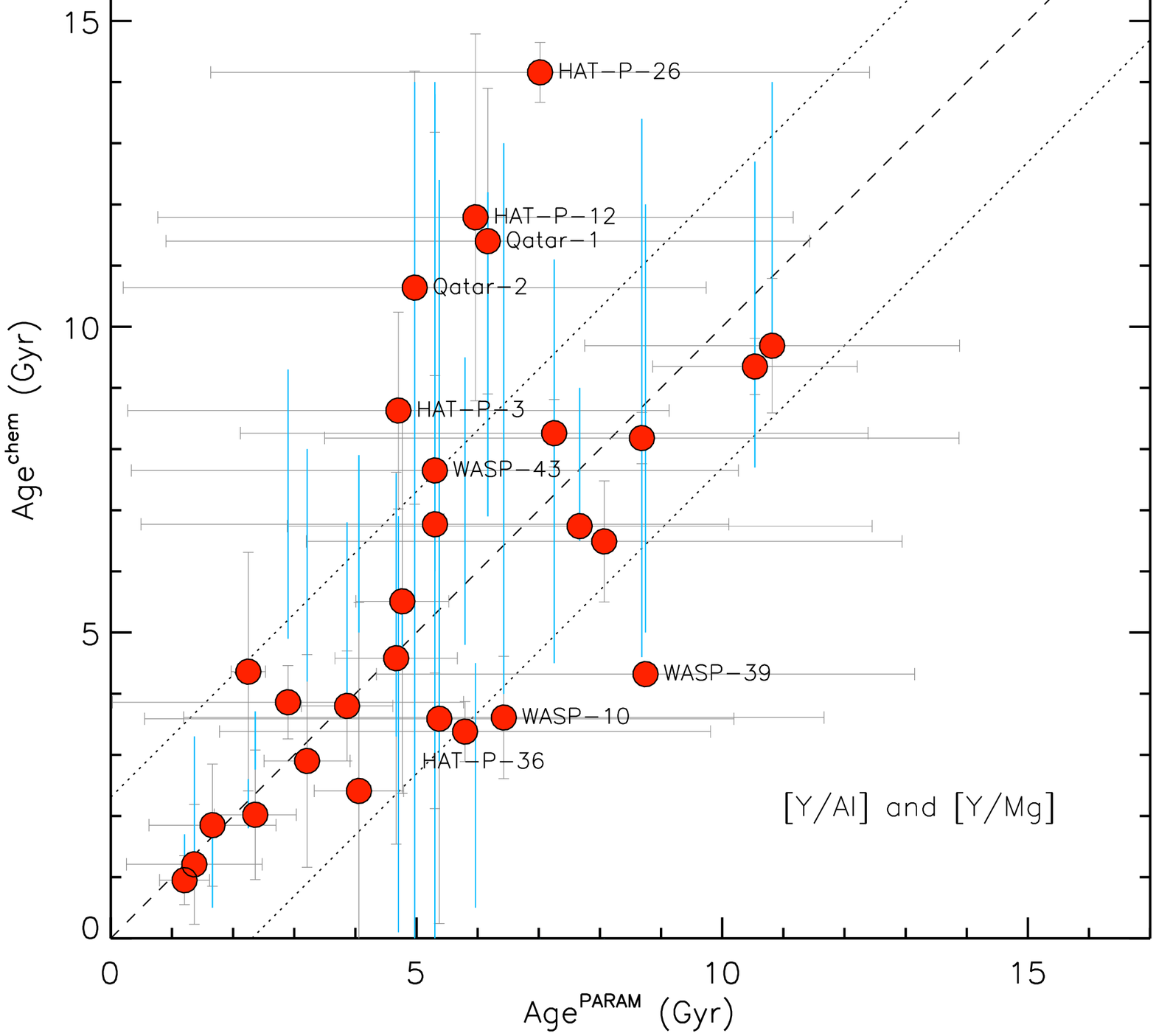}
\vspace{-0.5cm}
\caption{Comparison of the ages derived through abundance ratios and those inferred from the PARSEC models. We plot the results obtained considering all abundance ratios in common with \cite{casalietal2020} 
({\it left panel}) and the two best abundances ratios [Y/Al] and [Y/Mg] ({\it right panel}). Dashed line are the 1:1 relation, while the dotted lines represent the $\pm 1 \sigma$ levels from the average. Stars outside the $\pm 1 \sigma$ 
locus are marked. Light blue bars represent the range of values by \cite{bonomoetal2017} obtained through Yonsei-Yale evolutionary tracks.}
\label{fig:ages} 
\end{center}
\end{figure*}

In Fig.\,\ref{fig:ages} the comparison between the ages obtained with all aforementioned abundance ratios as a function of the isochronal ages obtained from PARSEC models is shown. In the 
same figure the range of ages as listed in \cite{bonomoetal2017} for each star and derived through stellar evolutionary tracks is shown, as comparison. We tried multiple combinations of abundance ratios to 
find the mean chemical ages by using the best abundance ratios discussed in \cite{casalietal2020}, and the better agreement with isochronal ages from PARSEC was found considering the two ratios [Y/Al] 
and [Y/Mg]. In Table\,\ref{tab:ages_masses}, we list the chemical ages obtained by using all abundances and the two aforementioned abundance ratios, together with the isochronal ages from the PARAM tool. In particular, we find a mean difference between ages derived from all the available abundance ratios and those obtained through PARSEC models of 2.7\,Gyr with a 
standard deviation of 2.4\,Gyr. Six targets (namely, Qatar-2, HAT-P-12, HAT-P-26, Qatar-1, HAT-P-3, WASP-43) fall outside the 1$\sigma$ limit, while three of them (KELT-6, WASP-10 and WASP-39) are close to it. The comparison with \cite{bonomoetal2017} gives an age difference of 2.9\,Gyr with a standard deviation of 2.7\,Gyr. If we consider the two best abundance ratios 
(i.e. [Y/Al] and [Y/Mg]; see, e.g., \cite{casalietal2020}, and references therein) the agreement is even better than that obtained using all abundance ratios, with a mean age difference of 2.0\,Gyr and a 
standard deviation of 2.0\,Gyr. In this case, the position of Qatar-2, HAT-P-12, WASP-43, and KELT-6 is closer to that of the isochronal age, with a general agreement visibly improved. The comparison 
with \cite{bonomoetal2017} in this case gives an age difference of 2.5\,Gyr with a standard deviation of 2.3\,Gyr. We note here that the most discrepant target (namely, HAT-P-26) is 
also one of the stars with high values of space velocities (see Sect.\,\ref{sec:spacemotion}), with $v_{\rm tot}>70$\,km/s, and with [$\alpha$/Fe] and $TD/D$ ratio compatible with the thin/thick disk transition. This justify its "chemically old" origin. Other stars (HAT-P-12, Qatar-1, Qatar-2) are in a region of the \logg-\teff\,diagram for which the age determination from both the isochrones and the chemical indicators is problematic mainly due to their relatively cool effective temperature.

\setlength{\tabcolsep}{4pt}
\begin{table}
\tiny
\caption{Chemical ages derived through all abundance ratios (column 2) and through [Y/Al] and [Y/Mg] ratios (column 3). Columns 4 and 5 list the stellar ages and masses obtained from the PARAM tool. }
\label{tab:ages_masses}
\begin{center}
\begin{tabular}{lrrrc}
\hline\hline
Name & $Age^{\rm chem}_{\rm all}$  & $Age^{\rm chem}_{\rm [Y/Al],[Y/Mg]}$ & $Age^{\rm PARAM}$ & $M_\star$\\
     & (Gyr)  & (Gyr) & (Gyr) & ($M_\odot$) \\
\hline
 HAT-P-3 &  9.6$\pm$1.5 &  8.6$\pm$1.6 &  4.7$\pm$4.4 & 0.88$\pm$0.03 \\ 
 HAT-P-4 &  4.8$\pm$0.9 &  3.8$\pm$0.9 &  3.8$\pm$0.7 & 1.28$\pm$0.05 \\ 
HAT-P-12 & 13.8$\pm$2.2 & 11.8$\pm$3.0 &  6.0$\pm$5.2 & 0.69$\pm$0.02 \\ 
HAT-P-14 &  2.4$\pm$1.5 &  1.0$\pm$0.4 &  1.2$\pm$0.4 & 1.41$\pm$0.03 \\ 
HAT-P-15 &  3.6$\pm$1.4 &  3.9$\pm$0.6 &  2.9$\pm$2.9 & 0.98$\pm$0.04 \\ 
HAT-P-17 &  8.7$\pm$0.4 &  8.3$\pm$0.6 &  7.3$\pm$5.1 & 0.87$\pm$0.04 \\ 
HAT-P-18 &  7.4$\pm$2.0 &  6.8$\pm$2.4 &  5.3$\pm$4.8 & 0.78$\pm$0.03 \\ 
HAT-P-20 &  7.5$\pm$1.2 &  3.6$\pm$3.4 &  5.4$\pm$4.9 & 0.74$\pm$0.02 \\ 
HAT-P-21 & 11.2$\pm$0.9 &  9.4$\pm$0.5 & 10.5$\pm$1.7 & 0.97$\pm$0.04 \\ 
HAT-P-22 &  9.5$\pm$1.1 &  9.7$\pm$1.1 & 10.8$\pm$3.1 & 0.92$\pm$0.03 \\ 
HAT-P-26 & 13.5$\pm$1.4 & 14.2$\pm$0.5 &  7.0$\pm$5.4 & 0.84$\pm$0.03 \\ 
HAT-P-29 &  2.1$\pm$0.4 &  1.2$\pm$1.0 &  1.4$\pm$1.1 & 1.20$\pm$0.03 \\ 
HAT-P-30 &  2.2$\pm$0.8 &  1.9$\pm$1.0 &  1.7$\pm$1.0 & 1.25$\pm$0.04 \\ 
HAT-P-36 &  3.0$\pm$0.8 &  3.3$\pm$0.5 &  5.8$\pm$4.0 & 1.00$\pm$0.05 \\ 
  KELT-6 &  8.8$\pm$1.9 &  5.5$\pm$3.1 &  4.8$\pm$0.8 & 1.14$\pm$0.05 \\ 
   Qatar-1 & 12.0$\pm$1.7 & 11.4$\pm$2.5 &  6.2$\pm$5.3 & 0.81$\pm$0.03 \\ 
 Qatar-2 & 14.6$\pm$1.7 & 10.6$\pm$3.5 &  5.0$\pm$4.8 & 0.74$\pm$0.02 \\ 
  TRES-4 &  4.8$\pm$1.0 &  4.4$\pm$2.0 &  2.2$\pm$0.3 & 1.46$\pm$0.02 \\ 
 WASP-10 &  3.2$\pm$0.8 &  3.6$\pm$1.0 &  6.4$\pm$5.2 & 0.77$\pm$0.02 \\ 
WASP-11 &  9.2$\pm$1.1 &  8.2$\pm$0.4 &  8.7$\pm$5.2 & 0.81$\pm$0.03 \\ 
 WASP-13 &  6.5$\pm$0.6 &  4.6$\pm$3.0 &  4.7$\pm$1.0 & 1.19$\pm$0.05 \\ 
 WASP-38 &  4.0$\pm$1.2 &  2.0$\pm$1.1 &  2.4$\pm$0.7 & 1.27$\pm$0.04 \\ 
 WASP-39 &  4.1$\pm$0.7 &  4.3$\pm$0.1 &  8.7$\pm$4.4 & 0.89$\pm$0.04 \\ 
 WASP-43 & 10.6$\pm$1.3 &  7.7$\pm$5.5 &  5.3$\pm$5.0 & 0.65$\pm$0.02 \\ 
 WASP-54 &  3.1$\pm$1.1 &  2.4$\pm$3.1 &  4.1$\pm$0.7 & 1.23$\pm$0.06 \\ 
 WASP-60 &  4.6$\pm$1.2 &  2.9$\pm$1.7 &  3.2$\pm$0.7 & 1.24$\pm$0.03 \\ 
  XO-2N &  7.2$\pm$0.6 &  6.7$\pm$0.1 &  7.7$\pm$4.8 & 0.94$\pm$0.04 \\ 
   XO-2S &  7.2$\pm$1.3 &  6.5$\pm$1.0 &  8.1$\pm$4.9 & 0.94$\pm$0.04 \\ 
 \hline
\end{tabular}
\end{center}
\end{table}

\subsection{Lithium abundance}
\label{sec:lithium}
\begin{figure*}
\begin{center}
\includegraphics[width=4.3cm]{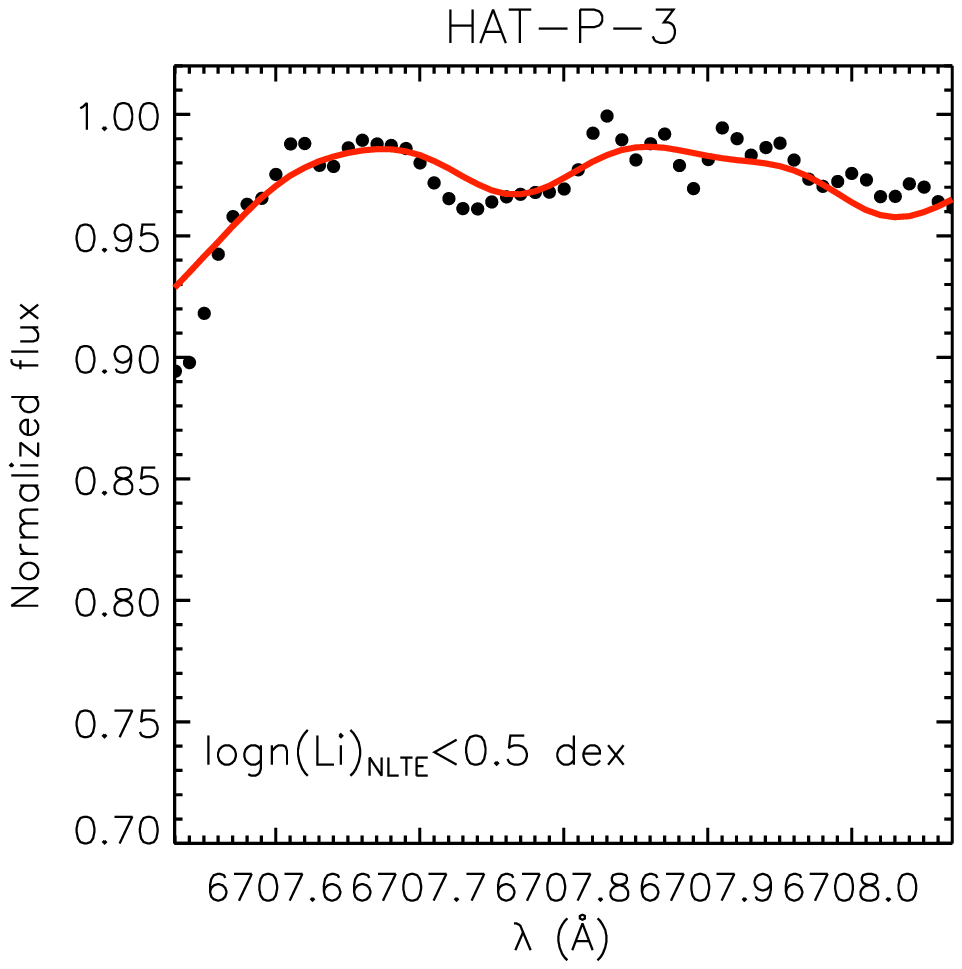}
\includegraphics[width=4.3cm]{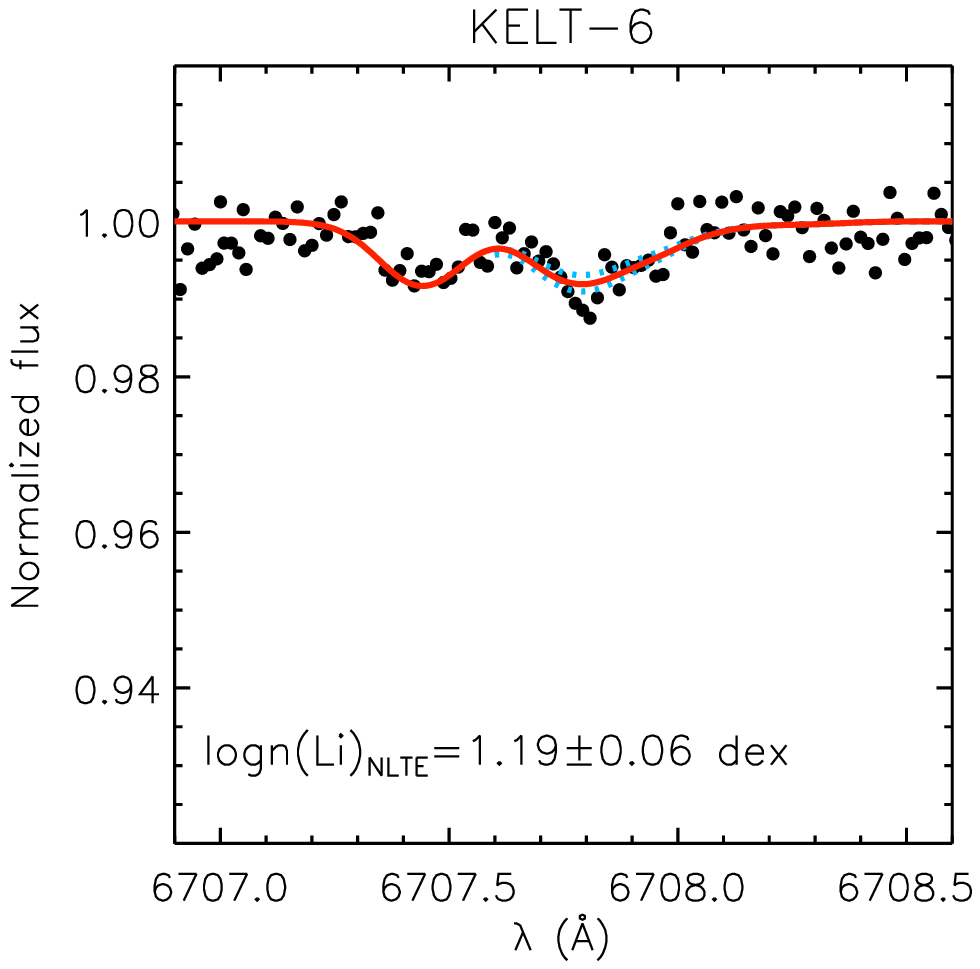}
\includegraphics[width=4.3cm]{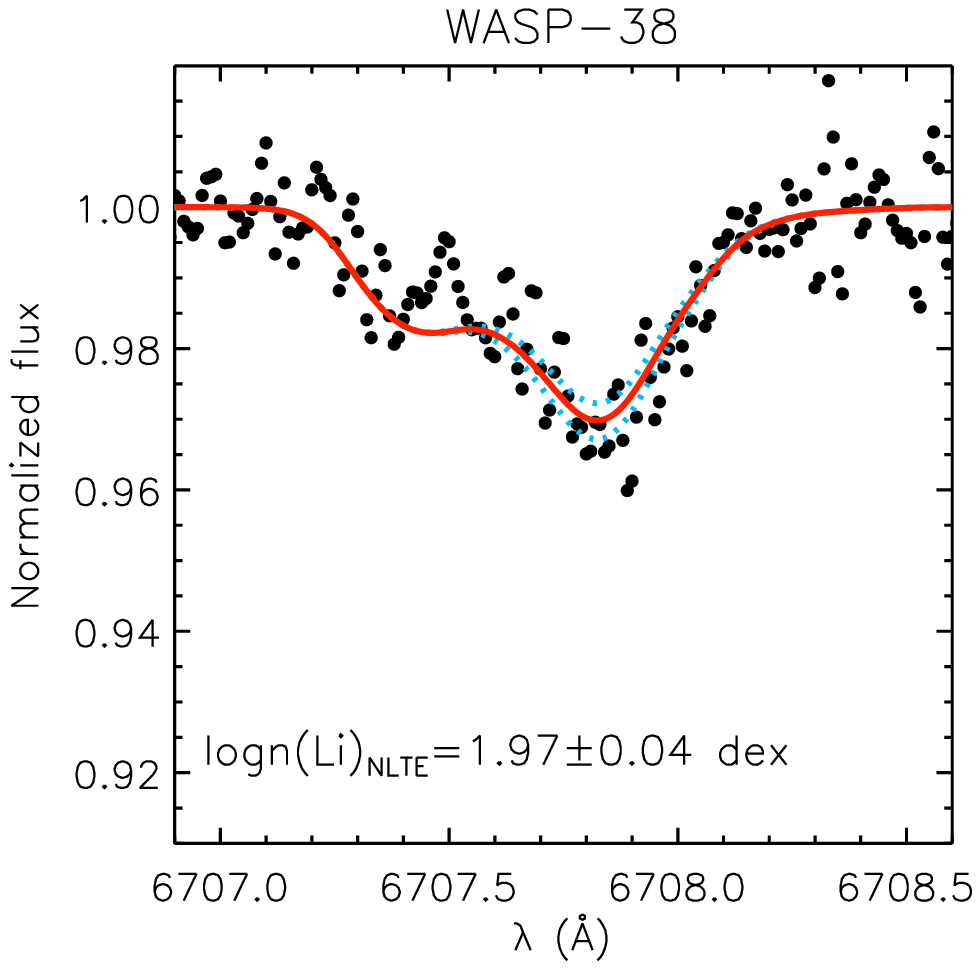}
\includegraphics[width=4.3cm]{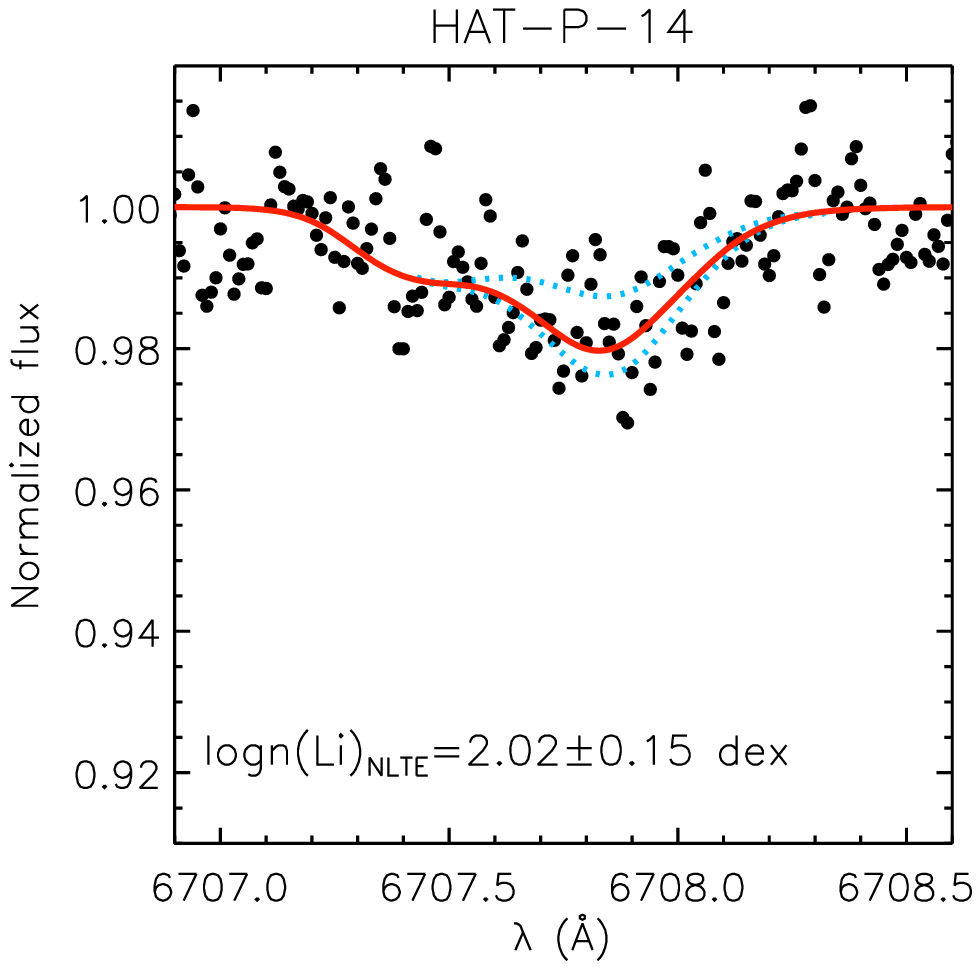}\\
\includegraphics[width=4.3cm]{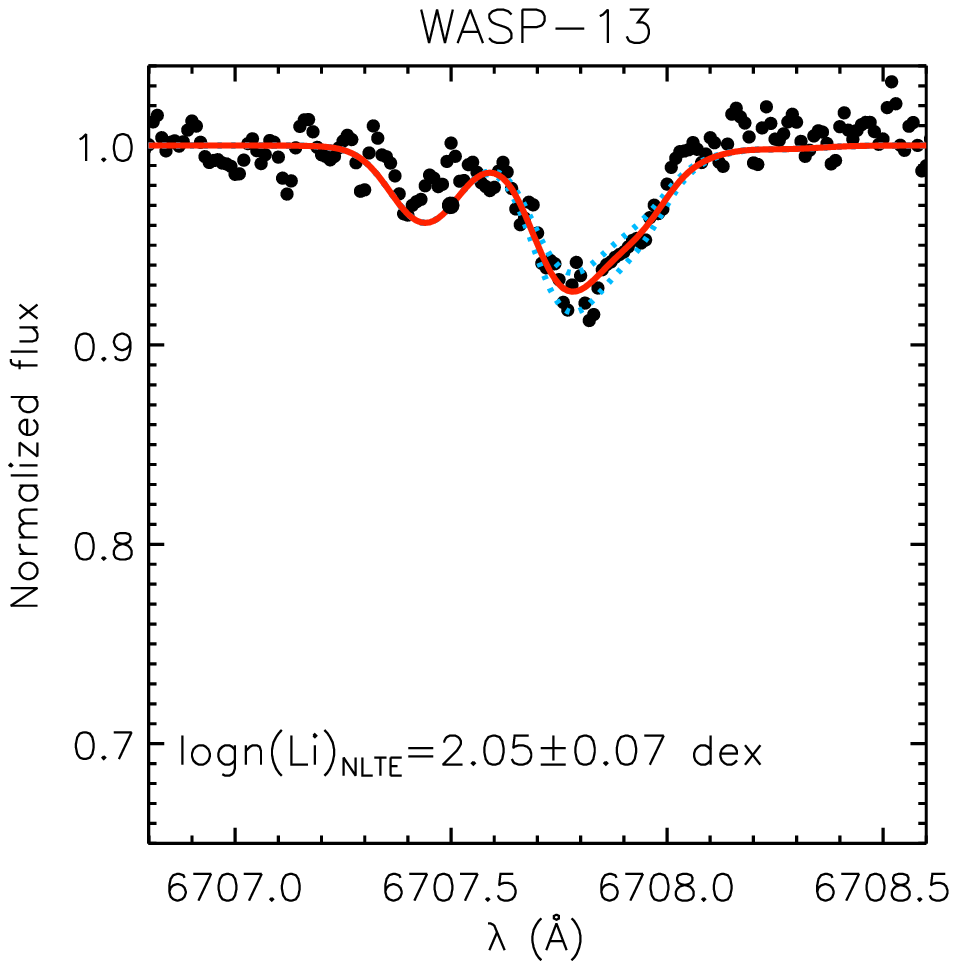}
\includegraphics[width=4.3cm]{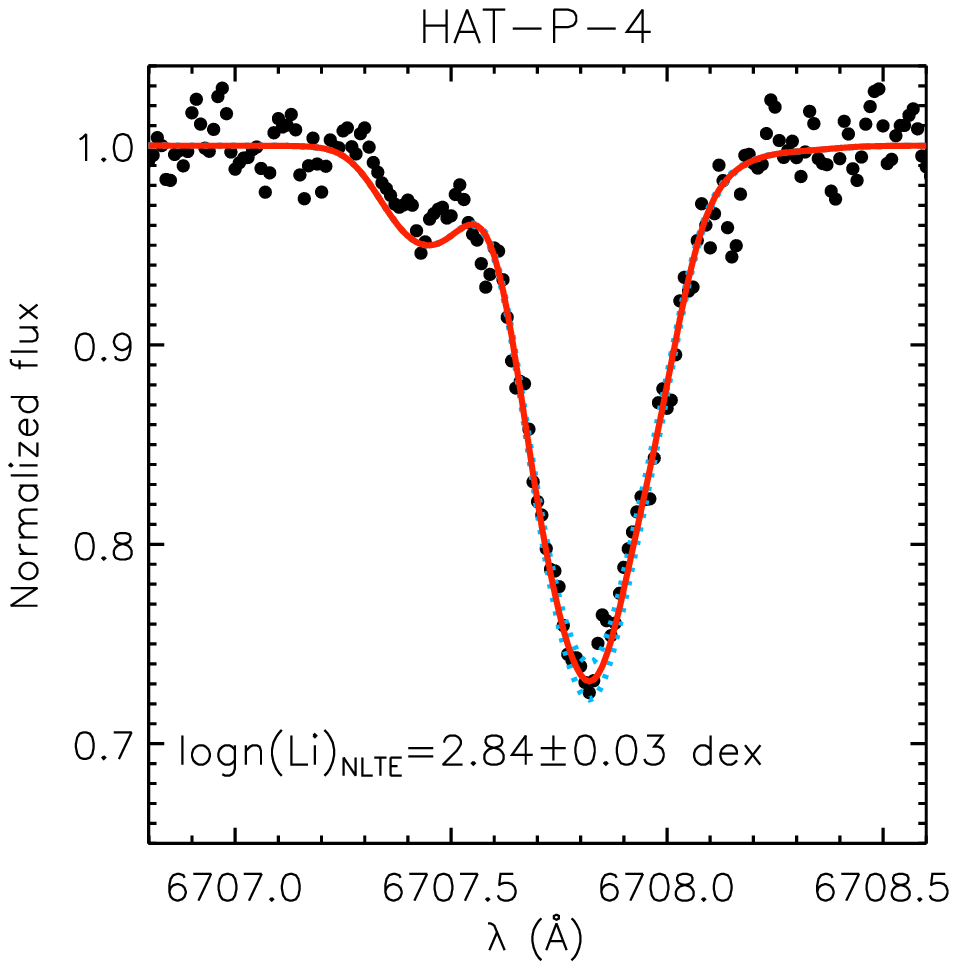}
\includegraphics[width=4.3cm]{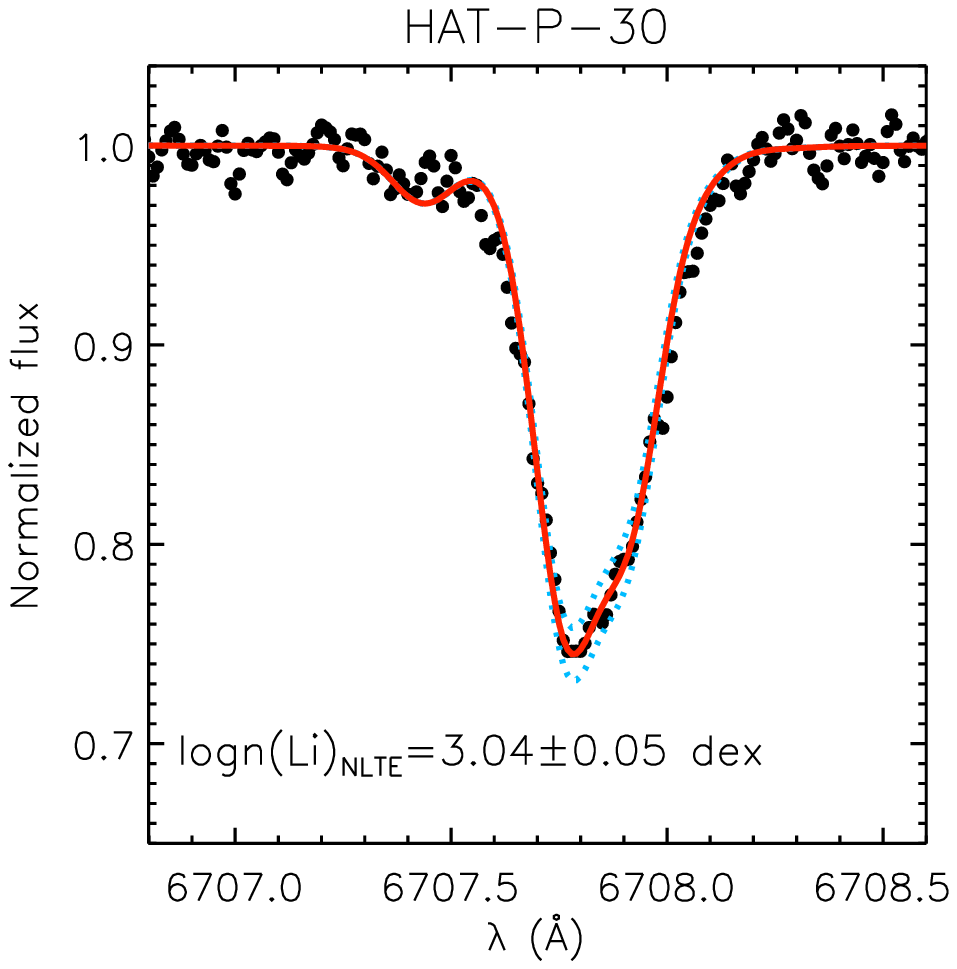}
\caption{Spectral synthesis around the Li region for the seven stars with detected lithium (with the exception of HAT-P-3 for which we could derive only an upper limit in Li abundance). The best-fit NLTE $\log n$(Li) and its margins of error superimposed on the observed spectrum 
are shown with continuum red and dotted light blue lines, respectively.}
\label{fig:lithium} 
\end{center}
\end{figure*}

\begin{figure*} 
\begin{center}
\includegraphics[width=9cm,angle=90]{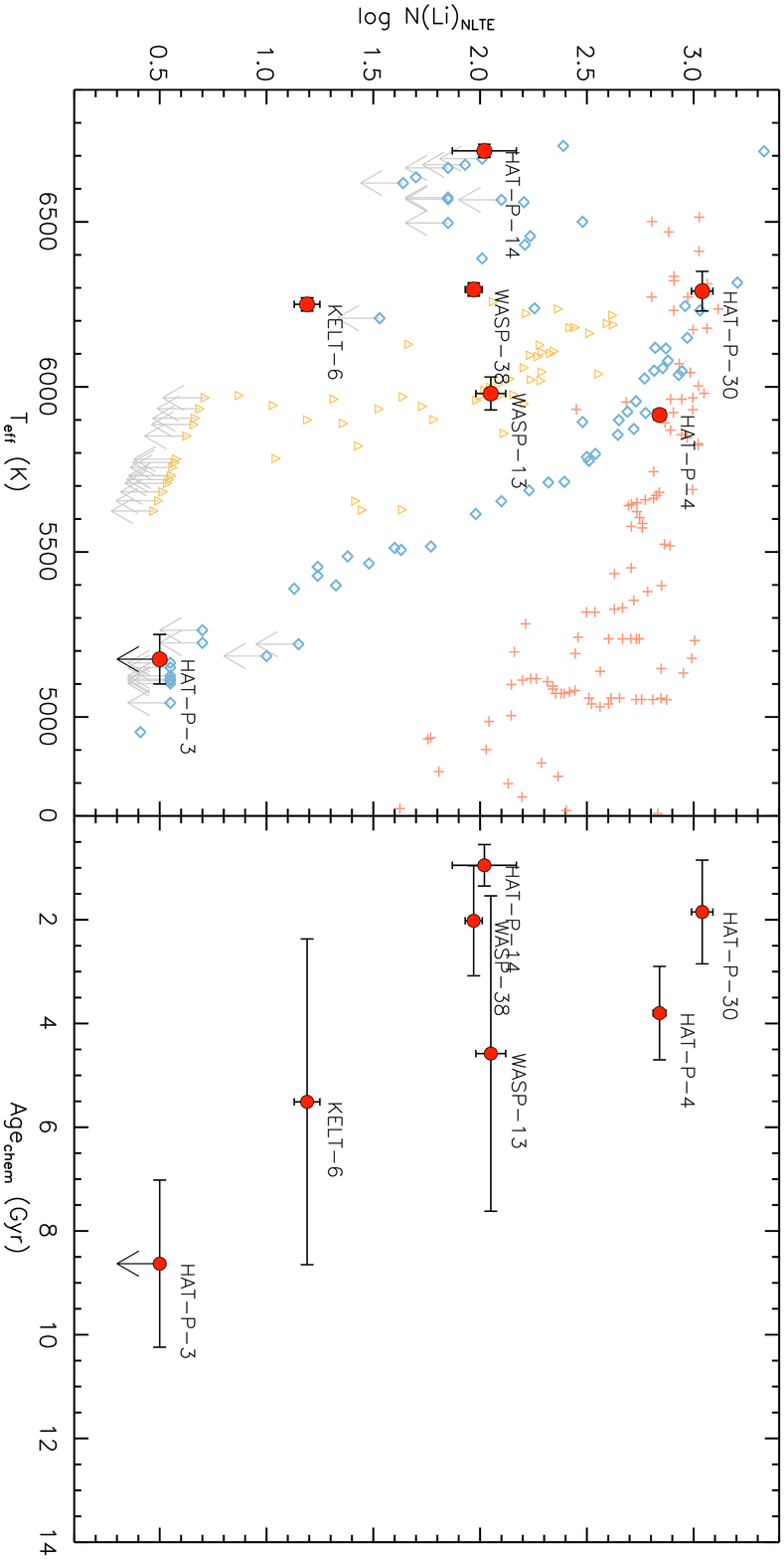}
\caption{{\it Left panel:} Li abundance versus effective temperature. The position of members in the Pleiades ($\sim 100$\,Myr; \citealt{SestitoRandich2005}), Hyades ($\sim 600$\,Myr; \citealt{Cummingsetal2017}), and M67 ($\sim 4.5$\,Gyr; \citealt{Pasquinietal2008}) clusters are also overplotted with red crosses, blue diamonds, and yellow triangles, respectively. Only the stars in our sample with presence of the Li line are plotted. {\it Right panel:} Li abundance versus chemical ages derived considering the [Y/Al] and [Y/Mg] ratios as chemical clocks.}
\label{fig:NLi_Teff_Age} 
\end{center}
\end{figure*}

We find that seven targets show lithium in their spectra, with HAT-P-3 the most uncertain case (see Fig.\,\ref{fig:lithium}). KELT-6 shows a value of $\sim 1.2$\,dex, WASP-38, HAT-P-14 and WASP-13 have lithium abundances around 2\,dex, 
while HAT-P-4 and HAT-P-30 have values $\log n{\rm (Li)} \sim 2.8-3.0$\,dex. Lithium in WASP-38, WASP-13, HAT-P-4, and HAT-P-30 was also detected by \cite{Mortieretal2013} and \cite{Enochetal2011}, who measured abundances very close to our values. Our targets with lithium are outside the region of the so called "lithium desert", located at $\sim 5900-6200$\,K and 
$\log n{\rm (Li)}<2$ dex (see Fig.\,\ref{fig:NLi_Teff_Age}), in agreement with \cite{ramirezetal2012} and \cite{Lopez-Valdiviaetal2015}. By chance, the number of stars with \teff\,$> 5900$ K with high lithium content higher than $\sim 2.0$\,dex (5) exceeds the number of stars with depleted lithium (1), in accordance with \cite{pavlenkoetal2018}, and justified by their thinner envelopes. On the other hand, all targets with lithium have $TD/D \le 0.05$, with the most metal-poor KELT-6 having the lower value of $\log n{\rm (Li)}$. This behavior could also reflect the pattern observed for Galactic thin stars, besides the depletion dependences related to their different stellar parameters (see \citealt{ramirezetal2012}).

We find as expected that the targets with lithium show on average higher \vsini\,($\sim$5.0 km/s) than stars without lithium detected ($\sim$2.6\,km/s), higher $T_{\rm eff}$ ($\sim$6090\,K against $\sim$5260\,K) and higher stellar mass 
($\sim$1.10\,$M_\odot$ against $\sim$0.97\,$M_\odot$), lower \logg\,(4.2 dex versus 4.4 dex), lower [Fe/H] (0.08 dex against 0.13\,dex), lower chemical age (3.9 Gyr versus 6.7 Gyr), and smaller planetary mass (1.1\,$M_{\rm Jup}$ vs 1.5\,$M_{\rm Jup}$). 
Moreover, within the sample with detected lithium, we recognize a tendency for the star with lower Li content (namely, KELT-6) to be also that with chemically derived older age. Similar dependence on lithium abundance on stellar parameters 
were also found by \cite{delgadomenaetal2014} for solar twins observed with HARPS and by \cite{pavlenkoetal2018} for CHEOPS dwarf stars. The dependence of Li abundance on \teff\,and on stellar mass is mainly due to the fact that 
high \teff\, (and therefore higher mass) stars have thinner envelopes, which naturally leads to higher Li, while cooler stars have deeper convective envelopes which allow for Li to get into hot enough regions for processing to occur 
(see, e.g., the pioneering work by \citealt{boesgaardetal1998}). The barely noticeable effect on \logg\,could be due to stronger lithium depletion in the atmospheres of older (and therefore of higher surface gravity) stars or enhanced 
mixing in stars with deeper convective envelopes; alternatively, those stars with lower \logg\,could be slightly evolved and thus their \teff\,at the main sequence could have been higher, therefore they didn't destroy so much Li due 
to their thinner convective envelopes (see \citealt{pavlenkoetal2018}). The slight lower mean [Fe/H] of stars with Li when compared to stars without detectable lithium could be caused by deeper convective envelopes expected 
for high opacities or Galactic chemical evolution (see \citealt{delgadomenaetal2014, delgadomenaetal2015}). Even if our sample is biased toward relatively slow rotating stars, we observe a difference in \vsini\,between targets with and 
without lithium compatible with depletion induced by rotation, as suggested in the rotational evolution models by \cite{Bouvier2008} and observed by \cite{pavlenkoetal2018} for CHEOPS dwarf stars with planets. 
We also find higher planetary mass for stars without lithium detected, which seems to indicate 
that destruction of Li is higher when the planet is more massive. Similar results were found by \cite{delgadomenaetal2014}, who suggested that this could be justified within a scenario where the disk is affecting the 
evolution of angular momentum, with a stronger effect for more massive disk, condition needed to form a giant planet (\citealt{Bouvier2008}). Moreover, when a giant planet forms in the disks, the accretion processes are expected 
to be more frequent and violent and produce Li destruction because of the temperature increase at the base of the convective envelope (\citealt{Baraffeetal2015}). 

In Fig.\,\ref{fig:NLi_Teff_Age} we plot the lithium abundance versus $T_{\rm eff}$ and stellar age derived through chemical indicators. Despite the relatively wide $T_{\rm eff}$ range ($\sim$5200$-$6700\,K), the plot shows an evident negative trend of Li abundance versus age, in which stars with higher lithium content show younger chemically-derived ages. For the chemically $\sim 8$\,Gyr old star HAT-P-3 we cannot make conclusions about possible age derived through the Li because of its upper limit in $\log n{\rm (Li)}$. KELT-6 is placed on the left of the warmest part of M67 members with similar \teff\, and its age derived through [Y/Al] and [Y/Mg] elemental abundance ratios is of $5.5$\,Gyr, consistent with the open cluster M67 (\citealt{Pasquinietal2008}). The position of WASP-13 within the $T_{\rm eff}$-$\log n{\rm (Li)}$ diagram is compatible with M67 stars, giving support to the chemically derived age of $\sim 4.6$\,Gyr. HAT-P-4 and HAT-P-30 are placed close to the Hyades group, while their chemical age is $\sim 2-4$\,Gyr, with HAT-P-4 older than HAT-P-30. Finally, HAT-P-14 and WASP-38 appear to be respectively in the warm and in the cool side of the Hyades Li dip, with a chemical age of $\sim 1-2$\,Gyr.

Finally, in our fit of the lithium line we tried to include the $^6$Li/$^7$Li ratio as a free parameter. This was done both because the inclusion of this isotopic ratio improves the fit of the $\lambda$6707.8\,\AA\,line and also because the determination of $^6$Li/$^7$Li would improve our knowledge of the stars in our sample. In fact, standard and non-standard stellar evolution models predict in solar-type stars a destruction of $^6$Li at the base of the convective envelope (\citealt{TalonCharbonnel2005}, and references therein), and hence the presence of $^6$Li in the atmosphere of a planet-hosting star has been justified as indication of an external pollution process, like planetary material accretion or super-flares around stars with hot Jupiters (e.g., \citealt{Cuntzetal2000, Israelianetal2001, Mottetal2017}). On the other hand, no $^6$Li/$^7$Li was detected in other stars with planets (\citealt{Reddyetal2002, Ghezzietal2009, Harutyunyanetal2018}). Unfortunately, our spectra are not of sufficiently high $SNR$ to determine $^6$Li/$^7$Li ratio with high enough precision. Usually, the $^6$Li/$^7$Li isotopic ratio is determined from analysis of spectra with $SNR>600$ (\citealt{Mottetal2017}). Anyway, trying to derive the $^6$Li/$^7$Li ratio from our MOOG analysis (see Sect.\ref{sec:abun_anal}), we tentatively find $^6$Li/$^7$Li of around $0.07$ for the solar spectrum, which is in between the values of $\sim$0.05 and $\sim$0.08 derived for the Sun by \cite{Asplundetal2021} and \cite{Lang1974}, respectively. For our sub-sample with lithium detected, we infer from our analysis a $^6$Li/$^7$Li lower than the solar value, with typical values ranging from $\sim 0.01-0.02$ for HAT-P-3, KELT-6, and WASP-38 up to $\sim 0.03$ for HAT-P-4, HAT-P-14, HAT-P-30, and WASP-13. Similar low values of $^6$Li/$^7$Li ratios are compatible with null results, in the sense that we do not find significant amount of $^6$Li in our sample of stars with detected Li. We tried to use other line lines (like that by \citealt{Melendezetal2012}) obtaining similar findings, therefore we think that higher $SNR$ is needed for more reliable $^6$Li/$^7$Li measurements, as suggested by \cite{Mottetal2017}.

\subsection{Stellar Carbon, Nitrogen, Oxygen, and Sulphur abundances}
\label{sec:CNOS}
\begin{figure}
\begin{center}
\includegraphics[width=12cm,angle=90]{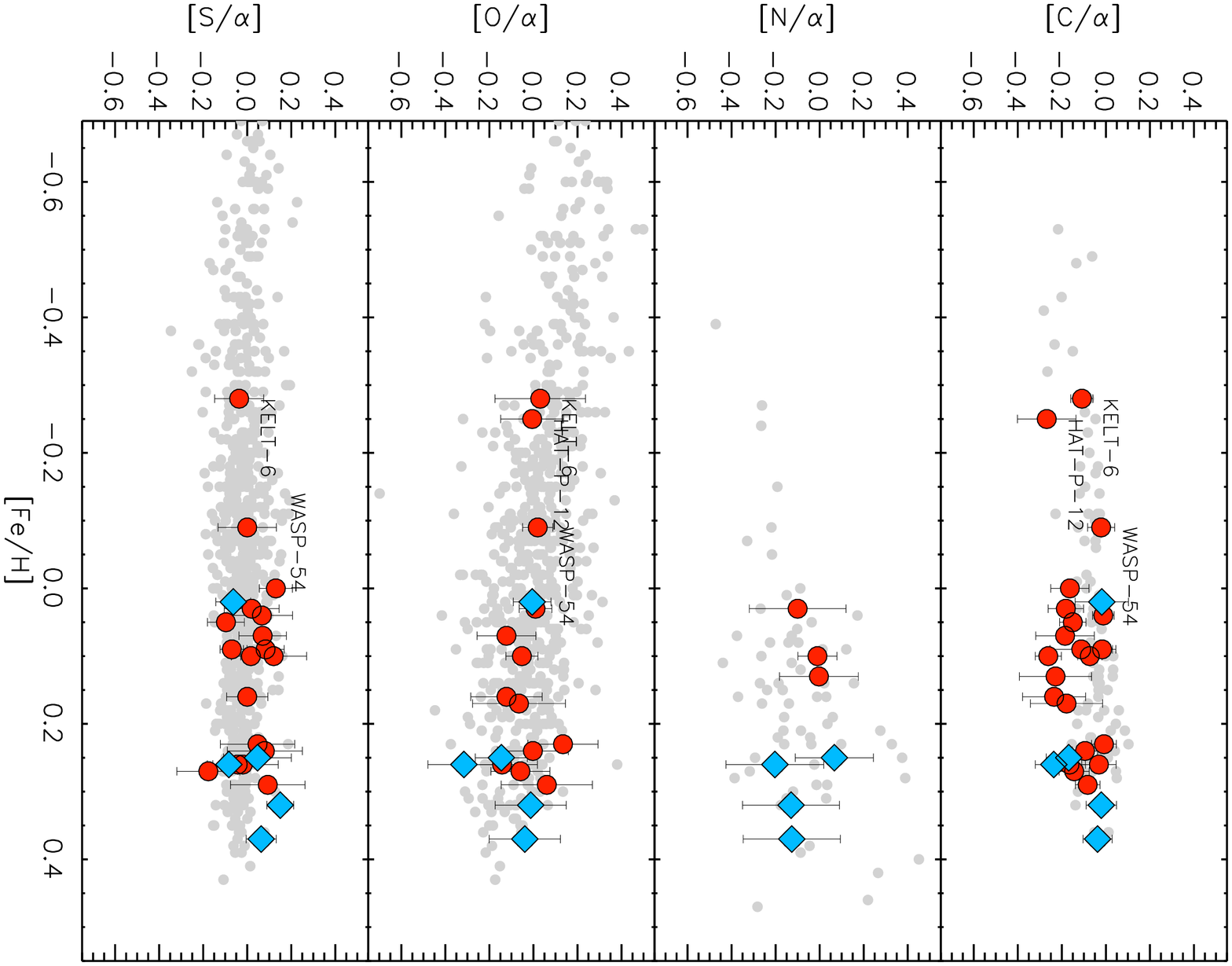}
\vspace{-1cm}
\caption{[C/$\alpha$], [N/$\alpha$], [O/$\alpha$], and [S/$\alpha$] ratios versus [Fe/H] for our targets. Blue diamonds mark stars with high content of [$\alpha$/Fe], as defined by the dot-dashed line in Fig.\,\ref{fig:fe_alpha}, namely HAT-P-26, HAT-P-22, XO-2N, XO-2S, and HAT-P-3. The overplotted grey points represent the values obtained by \cite{suarezandresetal2017}, \cite{suarezandresetal2016}, \cite{bertrandelisetal2015}, and \cite{costasilvaetal2020} for C, N, O, and S, respectively. Targets with [Fe/H]<0.0 are labeled.}
\label{fig:cnos_fe} 
\end{center}
\end{figure}

As shown in Fig.\,\ref{fig:abundances_FeH}, [C/H], [N/H], [O/H], and [S/H] abundances scale with iron, as we can expect because massive-planet host stars are statistically enhanced in Fe. In Fig.\,\ref{fig:cnos_fe} we plot the [C/$\alpha$], 
[N/$\alpha$], [O/$\alpha$], and [S/$\alpha$] ratios versus [Fe/H]. The figure shows that the position of our targets in the [X/$\alpha$]-[Fe/H] plots for these elements are similar to those found in the literature for nearby FGK stars by \cite{suarezandresetal2017} for C, \cite{suarezandresetal2016} for N, \cite{bertrandelisetal2015} for O, and \cite{costasilvaetal2020} for S. In particular, [C/$\alpha$] for our targets is almost constant for super-solar metallicity, as also observed by \cite{suarezandresetal2017} for solar-type stars. [O/$\alpha$] shows a possible decreasing trend (with a Spearman statistical significance $\rho \sim 0.2$) with increasing iron abundance, as also FG-type stars observed within the HARPS GTO (grey dots; \citealt{bertrandelisetal2015}). This kind of trend is expected, if O abundance scales with the iron abundance, as suggested above. Finally, hints of enhancements in sulphur are observed for the components of the XO-2 binary system (at [Fe/H]$\sim 0.3-0.4$\,dex).

As mentioned in the introduction, volatile elements like C, N, O, and S, can be used as proxy of the star-planet formation history (\citealt{Turrinietal2021a}). These authors have found that the joint use of the C/N, N/O, and C/O ratios both for the planets and their hosting stars breaks the degeneracy in the formation and migration tracks of giant planet, while the S/N ratio provides an additional independent probe into the metallicity of giant planets and their accretion of solids. We computed for our stellar sample the mean elemental ratios for two different stellar metallicity regimes around the iron abundance peak of our targets, which is [Fe/H]$\sim$0.12\,dex (see Table\,\ref{tab:CNOS_ratios}). As the comparison of these mean values highlights, the elemental ratios of interest for planetary studies can vary by also $\sim$10-30\% between stars with different metallicity, with, e.g., C/N going from values of around 2.0-2.8 for solar-metallicity or metal-poor targets up to 3.1-4.3 for super metal-rich targets. As such, the use of reference solar values would introduce significant biases in the interpretation of the planetary compositional data (see \citealt{Turrinietal2021a, Turrinietal2021b}). As an illustrative example, a planet with C/N value of $\sim$3.4 or C/O ratio of $\sim$0.6 orbiting a star with metallicity belonging to the $\le1.3\,Z_\odot$ group would be interpreted as possessing a solar value of this ratio when compared to the Sun, while its C/N value could actually be $\sim$1.2 times and $\sim$1.3 times superstellar in C/N and C/O, respectively. Similar conclusions were recently drawn also by \cite{Jorgeetal2022} for other elemental ratios (Fe/O, Si/O, Fe/S). The authors claim that planet formation scenarios should include the chemical abundance data of the host star and not impose solar abundance values to study the bulk exoplanetary properties.

\setlength{\tabcolsep}{3pt}
\begin{table}[ht]
\caption{Peak of the distributions of S/N, N/O, C/N, and C/O ratios for the targets divided into two different metallicity intervals around the mean [Fe/H] of our sample. 
Mean solar values are also shown.} 
\label{tab:CNOS_ratios}
\begin{center}
\tiny
\begin{tabular}{lccccc}
\hline
[Fe/H] & $Z_\star/Z_\odot$ & S/N & N/O & C/N & C/O \\
\hline
\multicolumn{6}{c}{{\it Solar Values}}\\
0.00$\pm$0.05& 1 & 0.17$\pm$0.08 &	0.17$\pm$0.08 &	3.35$\pm$0.08 &	0.57$\pm$0.04\\
\hline
\multicolumn{6}{c}{{\it Sample divided into two metallicity bins}}\\
$\le 0.12$	 & $\le$1.3 & $\sim$0.23 & $\sim$0.16 & 	$\sim$2.84 & 	$\sim$0.45\\
$>$0.12	 & 	>1.3 & $\sim$0.25 & 	$\sim$0.19 & 	$\sim$3.09 & 	$\sim$0.50\\
\hline
\end{tabular}
\end{center}
\end{table}

\subsection{[Mg/Si] vs [Fe/H] and Mg/Si vs C/O}
\label{sec:mgsi_co_fe}

\begin{figure}
\begin{center}
\includegraphics[width=7.5cm,angle=90]{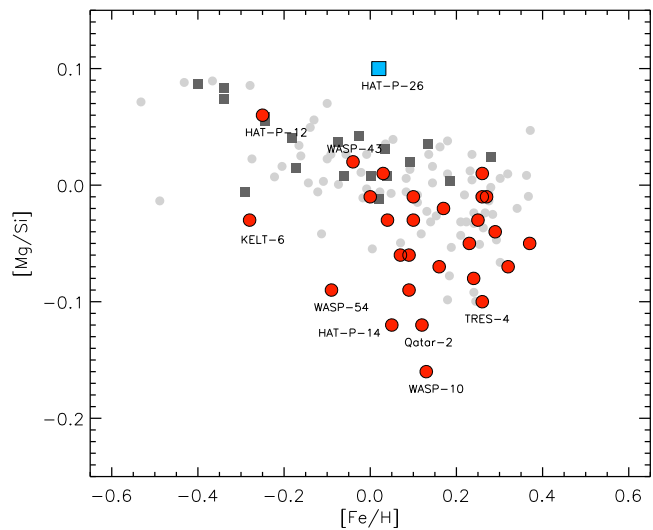}
\caption{[Mg/Si] versus [Fe/H]. Blue square marks the position of the star hosting the planet with the lowest mass in the sample (namely, HAT-P-26). The other stars labeled are those with [Fe/H]<0.0 or with [Mg/Si]<$-$0.1. Mean errors in [Mg/H], [Si/H], and [Fe/H] are similar to each other ($\sim$0.07\,dex). In grey the results by \cite{Adibekyanetal2015} for stars hosting low-mass planets (squares) and high-mass planets (circles) are overplotted.}
\label{fig:mgsi_fe} 
\end{center}
\end{figure}

From Fig.\,\ref{fig:mgsi_fe} we see how the [Mg/Si] ratio depends on stellar iron abundance. We also overplot the results by \cite{Adibekyanetal2015} for planet-hosting stars, keeping in mind that they declare not to find differences between stars with and without high-mass planets. Our targets show values consistent with those by \cite{Adibekyanetal2015}, with a few outliers that are anyway placed at similar positions of undetected planet-hosting stars by the same authors (see their Fig.\,1). Moreover, we note that the lower-mass planet-hosting star in our sample, namely HAT-P-26, shows the higher [Mg/Si] ratio, in agreement with \cite{Adibekyanetal2015}, who declare that low-mass planets are more prevalent around stars with high [Mg/Si]. This could be due to the fact that [Mg/Si] probably plays a very important role in the formation of low-mass planets, e.g. high Mg abundances could mitigate the lower iron abundances or metallicities and make core accretion comparatively more efficient. Moreover, there could be also a dependence of the planetary structure on the Galactic chemical evolution, as for this target we recognized from chemo-dynamical diagnostics a possible thin/thick disk transition origin within the Galactic plane (see Sect.\,\ref{sec:spacemotion}).

\begin{figure}
\begin{center}
\includegraphics[width=7.5cm,angle=90]{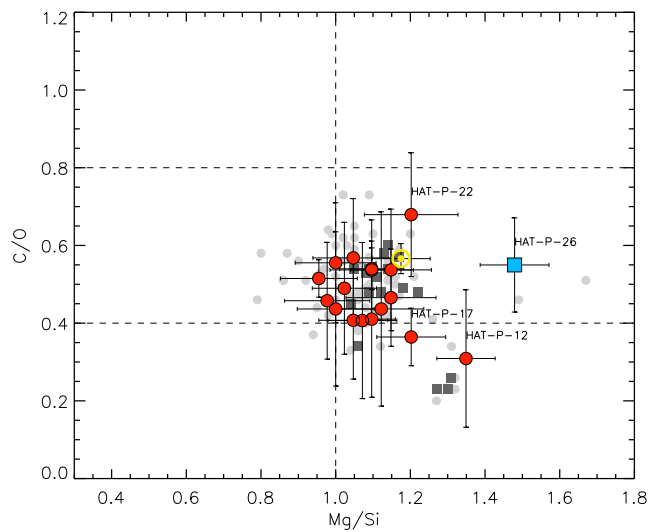}
\caption{C/O versus Mg/Si. Blue square highlights the only star in our sample hosting a planet with mass less than $30\,M_\oplus$. Targets with the highest values of Mg/Si are highlighted. In grey the results by \cite{Suarezetal2018} for stars hosting low-mass planets ($<30\,M_\oplus$; squares) and high-mass planets ($>30\,M_\oplus$; circles) are overplotted. Vertical line represents Mg/Si=1.0, while horizontal lines are plotted 
for C/O=0.4, 0.8, as defined by \cite{Suarezetal2018}. Our solar values of (C/O)$_\odot$=0.57 and (Mg/Si)$_\odot$=1.17 are also represented with a solar symbol (in yellow).}
\label{fig:mgsi_co} 
\end{center}
\end{figure}

A clearer picture can become evident if we consider also volatile elements. While refractory elements (as traced by the abundance ratios Mg/Si\footnote{From now on, the X$_1$/X$_2$ ratio refers to the elemental number ratio: X$_1$/X$_2$=$10^{\log \epsilon(X_1)}/10^{\log \epsilon(X_2)}$, with $\log \epsilon(X_1)$ and $\log \epsilon(X_2)$ absolute abundances.} and Fe/Si ratios) condense close to the star and their abundance ratios remain constant throughout most of the disk, the same is not true for volatile elements like C/O (see \citealt{Thiabaudetal2015a}). Such elemental ratios are important because they govern the distribution and formation of chemical species in the protoplanetary disk, and hence the mineralogy of planets. In Fig.\,\ref{fig:mgsi_co} we show how our stars are distributed in a C/O against Mg/Si plot with respect to the sample of planet-hosting stars in \cite{Suarezetal2018}. Our stars for which all C, O, Mg, Si abundances were measured are mainly concentrated at mean C/O values of $0.48\pm0.09$, with a steep drop-off at super-solar values, and mean Mg/Si values of $1.11\pm0.13$ (similar value is found considering the whole sample for which Mg and Si were derived). These values, together with the wider distribution of Mg/Si when compared with C/O, are indeed consistent with C/O and Mg/Si ratios found by \cite{Breweretal2016} for FGK stars in the solar neighborhood. We also note a tendency for stars cooler than $T_{\rm eff}<5000$\,K to have mean C/O ratios smaller than those of warmer stars, with a difference of $\sim 0.08$. No clear difference is evident for the Mg/Si ratio. 

In region of high C/O, planets form primarily from carbonates, and in regions of low C/O, the Mg/Si determines the types of silicates that dominate the compositions (e.g., \citealt{Breweretal2016}, and references therein). This means that the ratio C/O controls the distribution of Si among carbide and oxide species: if C/O$>0.8$, Si exists in solid form primarily as SiC, and also graphite and TiC will be formed; for C/O ratio below 0.8, Si is present in rock-forming minerals as SiO$_4^{4-}$ or SiO$_2$, serving as seeds for Mg silicates for which the exact composition will be controlled by the Mg/Si value (\citealt{Bondetal2010, Thiabaudetal2015a}). Moreover, \cite{Thiabaudetal2015b} have shown that the condensation of volatile species as a function of radial distance allows for C/O enrichment in specific parts of the protoplanetary disk of up to four times the solar values, leading to the formation of planets that can be enriched in C/O in their envelope up to three times the solar value. This is for instance the case of HD209458b observed by \cite{Giacobbeetal2021} for which a scenario of planet formation beyond the water snowline and migration towards its host star through disk or disk-free migration was hypothesized (see also \citealt{Breweretal2017}). At the same time, Mg/Si governs the distribution of silicates: for Mg/Si$<1$, Mg forms orthopyroxene (MgSiO$_3$) and the excess Si is present as other silicate species such as feldspars (CaAl$_2$Si$_2$O$_8$, NaAlSiO$_8$) or olivine (Mg$_2$SiO$_4$); for Mg/Si values ranging from 1 to 2, Mg is distributed between olivine and pyroxene; for Mg/Si$>2$, all available Si is consumed to form olivine with excess Mg available to bond with other minerals, mostly oxides such as MgO or MgS (\citealt{Bondetal2010, Thiabaudetal2015a}). The peak of the Mg/Si-C/O distribution for our targets is therefore consistent with Si which will take solid form as SiO$_4^{4-}$ and SiO$_2$ and Mg equally distributed between pyroxene and olivine. 

Considering the whole sample of 28 targets for which Mg/Si was determined, the star hosting the planet with mass $< 30\,M_\oplus$ (namely, HAT-P-26) has the highest Mg/Si value ($\sim 1.5$)\footnote{We also note that HAT-P-26 shows the higher values of Mg/Fe and Mg/Ca which have been also proposed as proxy for low-mass planet composition (see \citealt{HinkelUnterborn2019}).}, while 75\% of the higher-mass companion host sample show Mg/Si values between 1.0 and 2.0, which means that Mg is equally distributed between pyroxene and olivine. For the rest of 25\% of high-mass planet hosts with Mg/Si values below 1.0, Mg and Si will form mainly orthopyroxene, whereas the remaining Si will take other forms, such as feldspars or olivine. No stars with 
Mg/Si>2.0 were found either. We remark here that targets like HAT-P-26 and HAT-P-12, with their high values of Mg/Si, are also among those resulting chemically older in our sample. We indeed find a weak trend between Mg/Si and isochronal ages, which could mean that the determination of this chemical ratio could be favored for older (and chromospherically less active) stars (see Sect.\,\ref{sec:ages} for the measurement of chemical and isochronal ages). This kind of conclusions should be confirmed by statistically more significant samples.

If we consider the sample of 18 targets for which both Mg/Si and C/O ratios were measured, 100\% of our planet-hosting stars have C/O values lower than 0.8, with 11\% of the sample having C/O<0.4. This means that Si will be present in rock-forming minerals as SiO$_4^{4-}$ and SiO$_2$. In these cases, silicate mineralogy will be controlled by Mg/Si ratio. Within these 18 targets, 15 high-mass planet hosts show Mg/Si values between 1.0 and 2.0, two targets have Mg/Si<1.0, and the low-mass planet-hosting star HAT-P-26 shows a value of $\sim 1.5$. This supports the finding by \cite{Suarezetal2018}, who claimed that low-mass planets are likely to be found in the 1.0-1.5 Mg/Si regime, although mixed with stars with high-mass planets. 

We also remark here that most targets in our sample show sub-solar values of Mg/Si and C/O ratios. In particular, if we consider the solar values of Mg/Si$_\odot$=1.17$\pm$0.08, C/O$_\odot$=0.57$\pm$0.04, we find that 18/28 (i.e. 64\%) and 11/18 (i.e. 61\%) of our targets show respectively Mg/Si and C/O ratios lower than the solar values at 1$\sigma$ level. As already mentioned in Sect.\,\ref{sec:CNOS}, this highlights once again how the use of solar values as reference stellar abundances could introduce biases in the interpretation of the planetary compositional models (see \citealt{Turrinietal2021a, Turrinietal2021b}).

\subsection{Stellar abundances versus planetary properties}
\label{sec:abundances_plmass}

\begin{figure}
\begin{center}
\includegraphics[width=9.5cm,angle=90]{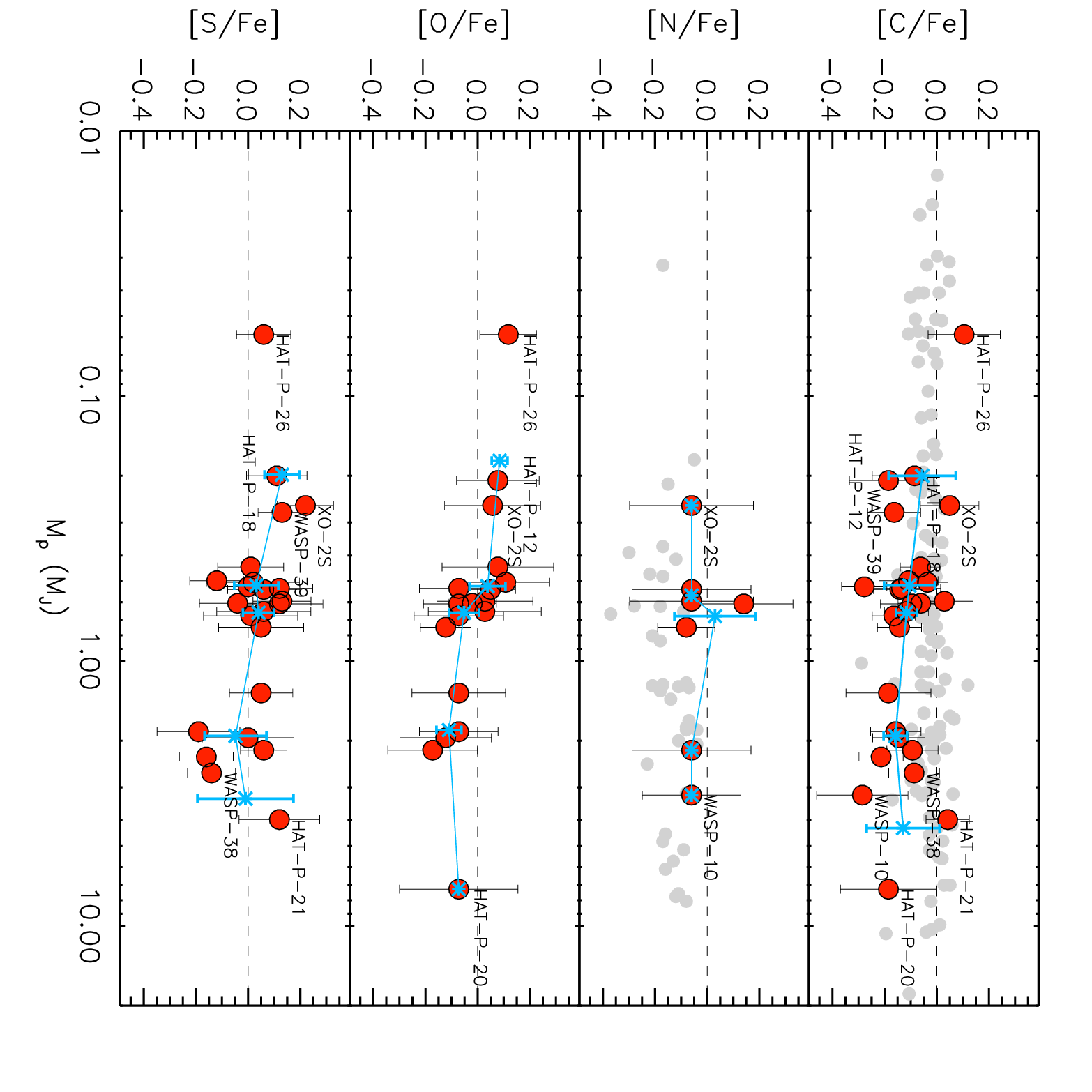}
\caption{[C/Fe], [N/Fe], [O/Fe], and [S/Fe] versus $M_{\rm p}$. Asterisks represent binned values for $M_{\rm p}$ ($<0.4$, 0.4--0.6, 0.6--1.2, 1.2--2.4, $>2.4$\,$M_{\rm Jup}$) and the error bars show the standard deviation for each bin. We labeled the targets with $M_{\rm p}<0.4\,M_{\rm Jup}$ and $M_{\rm p}>2.4\,M_{\rm Jup}$. Grey points for carbon and nitrogen are the values obtained for planet-hosting stars by \cite{suarezandresetal2017} and \cite{suarezandresetal2016}, respectively. Dashed lines represent the solar values.}
\label{fig:mj_cfe_nfe_ofe_sfe} 
\end{center}
\end{figure}

We show in Fig.\,\ref{fig:mj_cfe_nfe_ofe_sfe} [X/Fe] vs planetary mass\footnote{In the cases of multiple planets, such as Kelt-6, XO-2S, and HAT-P-17, we considered the planet 'b', which is the closest planet and that causing the transit (in the cases of Kelt-6 and HAT-P-17). We verified that this choice, motivated by homogeneity reasons, does not change our results and conclusions.} for some volatile elements (i.e. C, N, O, S), with the aim to look for possible relations between stellar abundances and planetary properties. The masses of the planets range between 0.058 and 7.27\,$M_{\rm Jup}$. In the figure, to better visualize possible trends, we represent with asterisks bins at increasing mass steps (i.e. 0.4, 0.6, 1.2, 2.4\,$M_{\rm Jup}$) in order to have similar number of targets per bin. Due to the few targets for which we could measure N abundance, we are not able to draw any conclusion for this element. For the other elements, we see a probable decreasing trend of [C/Fe], [O/Fe], and [S/Fe] with increasing $M_{\rm p}$. For these trends we calculated the Spearman associated statistical significance ($\rho$), finding $\rho \sim 9 \times 10^{-6}$ for the [O/Fe]-$M_{\rm p}$ relation and $\rho \sim 0.07$ for sulphur and carbon. We remark that for the planet XO2-Sb we plot $M_{\rm p} \sin i_{\rm p}$ (instead of $M_{\rm p}$), but we find similar results also when excluding this target, in particular for oxygen. Again, since at least part of these trends could be due to the star hosting the lowest-mass planet, we computed the same statistical significance after excluding HAT-P-26 and we find $\rho$ around 0.18, $6 \times 10^{-5}$, and 0.07 for C, O, S, respectively. This means that a more significant correlation is present for the [O/Fe] ratio vs $M_{\rm p}$ with respect to S and C vs $M_{\rm p}$, with an evident decreasing step towards low [O/Fe] values for $M_{\rm p} > 0.5\,M_{\rm Jup}$. A flat tendency was found for C by \cite{suarezandresetal2017} and a probable increasing trend was found by \cite{suarezandresetal2016} for N. We note that most of the targets hosting low-mass planets and showing higher [O/Fe] are also those resulting chemically old and possibly belonging to the thin-thick disk transition. Moreover, we remark that the number of stars in those papers and in this work is not statistically significant, but if our findings will be confirmed through bigger samples they could imply that the formation of low-mass planets is favored at highest values of stellar volatile elements. A possible, yet speculative, interpretation of these trends could be offered by the giant planet formation process in the framework of the pebble accretion scenario. Giant planets of Jovian or super-Jovian mass are expected to form early in the lifetime of circumstellar disks, when the high disk mass accretion rate is capable of supporting the rapid growth of their cores and sustain their gas accretion rates (see \citealt{Johansenetal2019, Tanakaetal2020} for a discussion). As a result, such massive planets can form also in disks comparatively poorer of the abundant volatile elements O and C. Less massive planets are expected to form over longer timescales, in circumstellar disks characterized by lower mass accretion rates (\citealt{Hartmannetal1998, Johansenetal2019}). Such planets may therefore form more easily around volatile-rich stars, so that the higher abundance of ices can partly compensate the lower disk mass accretion rates supporting their growth. In this framework, the stronger trend observed between the planetary mass and the [O/Fe] ratio could be explained by the larger contribution of O to the mass fraction of heavy elements in stars and their disks (in the Sun O provides about 45\% of the mass of heavy elements, while C and S only 16\% and 2\% respectively; \citealt{Lodders2010}), and its lower volatility compared to C (e.g. \citealt{Turrinietal2021a}, and references therein). This means that any increase in the abundance of O would affect wider orbital regions and would have larger impact on the availability of solid material of planet-forming disks than equal increases in the abundance of C and S, thus resulting more effective in promoting the formation of low-mass planets. On the contrary, the possible trends discussed above cannot be explained as easily in terms of effects linked to the planetary multiplicity, e.g. by the more massive giant planets blocking the flux of pebbles and promoting the formation of low-mass planets on outer orbits. All giant planets in the sample, with the possible exclusion of HAT-P-26, are significantly more massive than the pebble isolation masses typical of the inner regions of circumstellar discs (Johansen et al. 2019), so they should all be capable of blocking the pebble flux at some point during their migration. Furthermore, due to the lower volatility of S with respect to O (e.g. Turrini et al. 2021a,b and references therein), such a process should result in smaller values of S/Fe than O/Fe at larger planetary masses, a signature which is not observed in the data. Specifically, the sublimation of the O frozen as ice in the pebbles (e.g. water ice sublimating once the inward-drifting pebbles cross the water snowline) results in a smaller fraction of O being trapped in the pebbles with respect to S and Fe (which remain in solid form until closer to the star), making the trapping process less effective for O than for S and Fe.

\begin{figure}
\begin{center}
\includegraphics[width=9cm,angle=90]{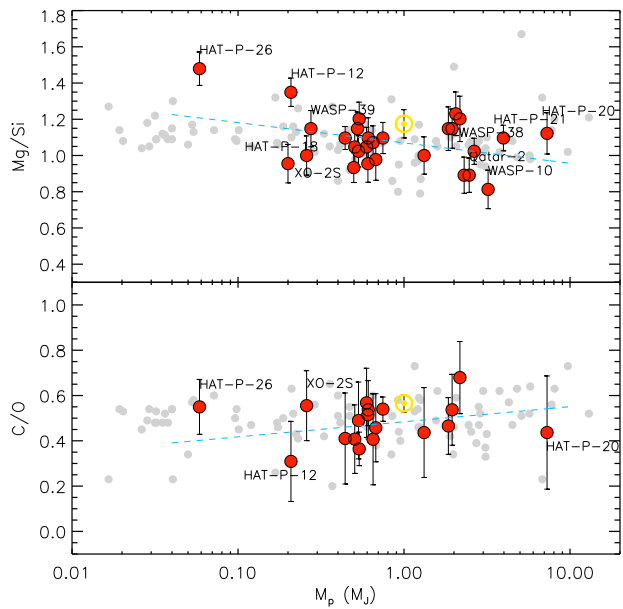}
\caption{Mg/Si and C/O elemental abundance ratios as a function of the planetary mass. Correlation fits are shown with dashed lines. We labeled the targets with $M_{\rm p}<0.4\,M_{\rm Jup}$ and $M_{\rm p}>2.4\,M_{\rm Jup}$. The position of our solar values is marked with the yellow Sun symbol. Grey points are the values of \cite{Suarezetal2018} for planet-hosting stars.}
\label{fig:mj_mgsi_co} 
\end{center}
\end{figure}

We also analyzed Mg/Si and C/O as a function of planetary mass. In Fig.\,\ref{fig:mj_mgsi_co} we see a hint of trend between Mg/Si and $M_{\rm p}$ (with slope of $\sim -0.18$ and a Spearman associated statistical significance $\rho \sim 0.3$, which became $\sim -0.09$ and $0.6$ after excluding HAT-P-26) and a still less evident C/O-$M_{\rm p}$ relation (slope of $\sim 0.16$ and $\rho \sim 0.5$, which became $\sim 0.31$ and $\sim 0.2$ if we do not consider the lowest-mass planet HAT-P-26b). Similar results are obtained after excluding XO-2S for which we know $M_{\rm p} \sin i_{\rm p}$ (and not $M_{\rm p}$). Possible downward trend of Mg/Si-$M_{\rm p}$ and slight increasing trend of C/O-$M_{\rm p}$ relationships were reported by \cite{Misheninaetal2021}, but no conclusion was drawn because of the large scatter of the ratios at certain planetary mass ranges. Very recently, \cite{Tautvaisieneetal2022} find a weak negative C/O slope and a slightly more negative Mg/Si slope toward stars with high-mass planets. Here, we remark that for icy and giant planets, Mg/Si (and also Fe/Si) in stars should be a direct information about the planetary composition, as no differences are expected between star and planet in terms of Mg/Si, as proposed by \cite{Thiabaudetal2015b}. This is because giant planets are mainly formed outside the ice line, a region where all refractory material has condensed. Our findings could imply higher values of Mg/Si for lower mass planets, like HAT-P-26b and HAT-P-12b, which indeed were found to have respectively super-solar and solar or super-solar metallicity (see \citealt{KawashimaMin2021}). We remark here that the solar-metallicity star HAT-P-26 and the metal-poor HAT-P-12 are also targets for which we find old chemical ages and possibility to belong to the Galactic thin-thick disk transition (see Sects.\,\ref{sec:ages}, \ref{sec:spacemotion}), therefore their relatively higher Mg/Si ratios could be related to their position in the Galactic disk.
Regarding C/O, an indirect relation between the planet and the star should be present. This is mainly because final planetary C/O depends on the location and timescale of formation, how much of the atmosphere is accreted from gas versus solids, and how isolated the atmosphere is from mixing with core materials (see, e.g., \citealt{Teskeetal2014} and \citealt{Giacobbeetal2021}, and references therein). In Sect.\,\ref{sec:traceformation} we try to give some view of possible pathways for exoplanets in common with our hosting stars for which C/O and N/O elemental ratios were computed in the literature.

\begin{figure}
\begin{center}
\includegraphics[width=8cm]{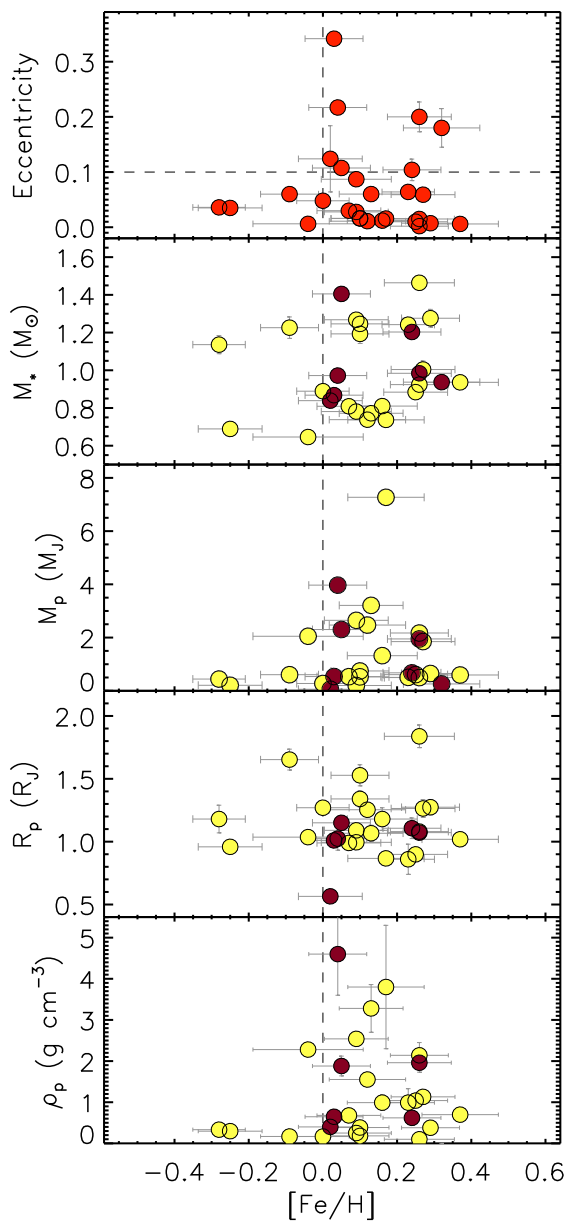}
\caption{{\it From top to bottom:} Orbital eccentricity, stellar mass, planetary mass and radius, and planet density versus stellar [Fe/H]. Horizontal line in the first panel is plotted for $e=0.1$, while vertical dashed lines mark [Fe/H]=0.0. Darker dots in the lower four panels represent stars with $e>0.1$. Note that planet XO-2Sb is shown in the first and third upper panels, while it is not plotted in the last two lower panels because it is not a transiting planet.}
\label{fig:fe_mass_eccentricity_radius} 
\end{center}
\end{figure}

In Fig.\,\ref{fig:fe_mass_eccentricity_radius} we show the distribution of the iron abundance of our stellar sample in terms of the planetary orbit eccentricity/mass/radius/density and stellar mass. We find a tendency for high-eccentricity planets to be around more metal-rich stars, as also found by \cite{DawsonMurray2013} and \cite{Millsetal2019}. Even if we are aware about our not statistically significant sample, we note that the mean [Fe/H], and also mean values of [$\alpha$/H] and Mg/Si, are greater by $\sim 0.03-0.05$\,dex for planets with $e>0.1$, when compared with lower-eccentricity planets. This tendency appears to be not influenced by the planetary radii, because the planet radii of our sample are within a range ($6-21\,R_{\rm E}$) for which no stellar metallicity-planet radius correlation was found (see \citealt{Buchhaveetal2014}). In order to assign a confidence level of our result, we used a one-side $2\times2$ Fisher's exact test\footnote{We used the following web calculator 
(\citealt{Langsrudetal2007}): http://www.langsrud.com/fisher.htm.} (\citealt{Agresti1992}). Choosing the divisions at $e=0.1$ for high-/low- eccentricity orbits and [Fe/H]=0.00 dex for metal-poor/metal-rich stars, we find a $p-value$ of 0.29 as chance that random data would yield this trend, indicating a probability of correlation of 71\%. With this in mind, we also find some evidence for our metal-rich stars to have wider ranges of planetary masses and denser planets (mean $\rho_{\rm P} = 1.68$\,g/cm$^3$) around stars with greater [Fe/H] (and therefore in more eccentric orbits) when compared with planets around more metal-poor stars with $e<0.1$ (mean $\rho_{\rm P} = 1.11$\,g/cm$^3$). We remark here that we are cautious about these possible trends both because the number of our targets is relatively small and also because each correlation depends on the interplay of many planetary/stellar properties/parameters.

\subsection{Can we trace the planet formation scenario?}
\label{sec:traceformation}

\cite{Turrinietal2021a} demonstrated how the joint use of planetary C/N, N/O, and C/O ratios provides useful diagnostics to trace the formation and migration history of giant planets. This result is the direct consequence of the relative volatility of C, O and N in protoplanetary disks: the bulk of O is trapped into solids already in the inner disk regions, the bulk of N remains in gas form until the outermost regions, while C shows an intermediate behavior. The disk gas therefore becomes enriched in N with respect to O and C the farther we move from the star, while solids become increasingly enriched in C and O with respect to N. As a result, the farther from the host star the giant planets start their migration, the more their C/N and N/O ratios will diverge from the stellar values due to the different accretion efficiencies of gas and solids. 

The use of planetary elemental ratios normalized to the respective stellar ratios makes it easier to extract the information on the nature of giant planets and constrain whether their metallicity is dominated by the accretion of gas or of solids (see \citealt{Turrinietal2021a,Turrinietal2021b} for additional discussion). Four of the stars analyzed in the present work (namely, HAT-P-26, WASP-10, HAT-P-12, and WASP-39) host planets for which \cite{KawashimaMin2021} derived metallicity, N/O, and C/O through the spectral disequilibrium retrieval models. Dividing these planetary ratios by those we derived for the hosting stars (which we will label as X/Y$^\ast$\footnote{The X/Y$^\ast$ refers to planetary elemental ratio over stellar elemental ratio: X/Y$^\ast=\frac{X/Y_{\rm P}}{X/Y_{\star}}$}), we can therefore gain insights about the formation pathways of these planets. 

We list in Table\,\ref{tab:plst_elementalratio} the elemental ratios as defined by \cite{Turrinietal2021a}. Values of C/N$^\ast$>C/O$^\ast$>N/O$^\ast$ imply that the budget of heavy elements is dominated by the accretion of solids, while N/O$^\ast$>C/O$^\ast$>C/N$^\ast$ implies its accretion mostly from the disk gas. The authors claim also that in both cases, the separation between the values of the three normalized ratios will increase with the extent of disk-driven migration experienced by the giant planet. Specifically, the farther from the star a giant planet will start its formation, the more significant the difference between its C/N$^\ast$, C/O$^\ast$, and N/O$^\ast$. For our sample of four targets, we note that large uncertainties associated with the measurements of the X/Y$^\ast$ ratio cannot allow us to draw definitive conclusion, therefore our discussion is mainly qualitative and indicative of possible planetary formation scenarios. 

The pattern of the abundance ratios observed for HAT-P-26 and WASP-39 points towards C/N$^\ast$ smaller than N/O$^\ast$, with possible constant and increasing trends between C/N$^\ast$ and C/O$^\ast$, respectively. Given their C/N$^\ast$ and N/O$^\ast$ abundance patterns, these planets likely underwent migration to get to their present orbits (both at $\sim 0.048$\,au) and accreted mostly gas along their path \citep{Turrinietal2021a}. The comparison of the C/N$^\ast$, C/O$^\ast$, and N/O$^\ast$ values is consistent with a scenario where both planets started forming outside the CO$_2$ snowline and accreted most of their gas inward of it \citep{Turrinietal2021a} but where the gas was enriched in O by the evaporation of O-rich ices from the inward drifting dust \citep{BoothIlee2019}. The two host stars have solar iron abundances, while their planets show super-solar and solar metallicity, respectively for HAT-P-26b and WASP-39b (see \citealt{MacDonaldMadhusudhan2019}; \citealt{KawashimaMin2021}). Since the accretion of non-enriched gas would result in sub-stellar metallicity values \citep{Turrinietal2021a,Turrinietal2021b}, the metallicity of these planets appear consistent with the accretion of gas enriched in heavy elements. The estimated abundance patterns tentatively favor the accretion of gas in a disk that underwent chemical reset (Pacetti et al., in prep).

For the planets orbiting HAT-P-12 and WASP-10 we find C/N$^\ast$ greater than N/O$^\ast$, with possible constant and decreasing trends between C/N$^\ast$ and C/O$^\ast$, respectively. The C/N$^\ast$, C/O$^\ast$ and N/O$^\ast$ abundance patterns of the planet orbiting WASP-10 are consistent with extensive migration toward their present orbits (at $\sim 0.038$\,au) and the accretion of heavy elements being dominated by the accretion of solids \citep{Turrinietal2021a}. The high value of the C/O$^\ast$ ratio is suggestive of an additional contribution to the planetary C budget by the accretion of C-enriched gas, which would favor the accretion of gas between the CO$_2$ and CH$_4$ snowlines \citep{BoothIlee2019}. Also the C/N$^\ast$, C/O$^\ast$ and N/O$^\ast$ abundance patterns of the planet orbiting HAT-P-12 are consistent with extensive migration and the accretion of heavy elements being dominated by the accretion of solids. In this case, however, the higher values of the C/O$^\ast$ and N/O$^\ast$ ratios suggest that the planetary budgets of both C and N were affected by the accretion of gas enriched by the evaporation of ices from the inward drifting dust. This would point toward the planet accreting a significant fraction of its gas between the N$_2$ and CO$_2$ snowlines \citep{BoothIlee2019}. In such a scenario, the planet orbiting HAT-P-12 would have started its formation farther out than its counterpart orbiting WASP-10.

\setlength{\tabcolsep}{2pt}
\begin{table*}
\tiny
\caption{Stellar elemental ratios as derived in this work (columns $2\div4$), planetary elemental ratios as computed by \cite{KawashimaMin2021} (columns: $5\div$7), and planetary-over-stellar elemental ratios as defined by \cite{Turrinietal2021a} (columns: $8\div$10).}
\label{tab:plst_elementalratio}
\begin{center}
\begin{tabular}{l rcr | lcl | lcl}
\hline\hline
   Name    & C/N$_\star$  & C/O$_\star$ & N/O$_\star$ & C/N$_{\rm p}$  & C/O$_{\rm p}$ & N/O$_{\rm p}$ &  C/N$^\ast$  & C/O$^\ast$ & N/O$^\ast$   \\
\hline
HAT-P-12 & 2.09$\pm$0.16 & 0.31$\pm$0.18 &  0.15$\pm$0.16 &  5.33$\pm$0.18 & 0.80$\pm$0.15 &  0.15$\pm$0.10 & 2.55$\pm$0.24 & 2.58$\pm$0.23 & 1.00$\pm$0.19  \\ 
WASP-10 & 2.00$\pm$0.21 &  0.37$\pm$0.18 &  0.19$\pm$0.18 &  6.31$\pm$0.32 &  0.82$\pm$0.30 &  0.13$\pm$0.10 & 3.16$\pm$0.38 & 2.22$\pm$0.35  & 0.68$\pm$0.21  \\ 
HAT-P-26 &  3.80$\pm$0.15 &  0.55$\pm$0.12 & 0.14$\pm$0.12 &  1.79$\pm$0.16 &  0.25$\pm$0.13 & 0.14$\pm$0.10 &  0.47$\pm$0.22 &  0.45$\pm$0.18 & 1.00$\pm$0.16  \\ 
WASP-39 & 3.73$\pm$0.12 &  0.51$\pm$0.12 & 0.14$\pm$0.14 &  1.73$\pm$0.16 &  0.26$\pm$0.12 & 0.15$\pm$0.10 & 0.46$\pm$0.20 &  0.51$\pm$0.17 & 1.07$\pm$0.17 \\ 
\hline
\end{tabular}
\end{center}
Notes: $i.$ Planetary C/N were derived by dividing C/O by N/O; $ii.$ When we were not able to measure one of the C, N, O stellar abundances, we considered as [X/H] a value consistent with the abundance of elements with the closer $T_{\rm cond}$ (within the volatile elements C, N, and O; see Table\,\ref{tab:solar_abundances} and Sect.\,\ref{sec:condensation}).
\end{table*}

\section{Conclusions}
\label{sec:conclusions}
In this paper, we present for the first time a wide and comprehensive characterization of a sample of 27 transiting planet host stars within the GAPS programme obtained through a homogeneous and accurate spectroscopic procedure, which is mainly based on high-resolution HARPS-N at TNG spectra (and a few FEROS at ESO spectra). We analyze for the first time this sample with the aim to derive different stellar properties, abundances of many elements, and kinematic information. Our main results can be summarized as follows:
\begin{itemize}
\item We obtain the atmospheric parameters and the abundances of 26 elements from lithium up to europium, together with their kinematic properties. The chemo-kinematic analysis allowed us to recognize that most of the targets appear to belong to the Galactic thin disk, and a few stars likely belong to a thin-thick disk transition. The lithium line is present in seven stars.
\item From analysis of some elemental ratios, most of them including $s$-process and $\alpha$- elements, we derived, for the first time in exoplanet hosts, stellar ages, often consistent with those obtained through theoretical isochrones. 
\item From the analysis of the Mg/Si ratios, we find that most of the targets show values consistent with a distribution of Mg between olivine and pyroxene, and few show Mg forming orthopyroxene. C/O ratios for all targets is lower than 0.8, i.e. compatible with Si present in rock-forming minerals, with a slight tendency of higher values for stars hosting lower-mass planets, even if we remark that at least part of this trend could be related to the position of these targets in the Galactic disk. 
\item Considering volatile elements, we find a tendency for some of them, in particular for the [O/Fe] ratio, to be lower for higher-mass planets. Also in this case, most of the targets hosting low-mass planets and showing high values of [O/Fe] are also those resulting chemo-dynamically old and possibly belonging to the thin-thick disk transition.
\item We find some evidence for high-eccentricity planets to be around more metal-rich stars and also denser planets around stars with higher [Fe/H]. 
\item From our chemo-kinematic analysis, some of the five most interesting targets which can motivate immediate high-precision studies result to be: 
\begin{itemize}
\item[$i.$] the target hosting the lowest mass planet in the sample (i.e. HAT-P-26), with solar [Fe/H] and C/O, high $[\alpha/Fe]$, and the highest values of [Mg/Si] and Mg/Si; it shows $UV$, $TD/D$, $e_{\rm G}$, and $Z_{\rm max}$ compatible with thin-thick disk transition and seem to have migrating from the Galactic inner disk; 
\item[$ii.$] the relatively metal-poor ([Fe/H]$=-0.25$) HAT-P-12 and the metal-rich ([Fe/H]$\sim 0.3-0.4$) XO-2 binary system show enhancements in $[\alpha/Fe]$; they have $UV$ and $TD/D$ compatible with thick disk stars, high galactic eccentricity and large $R_{\rm mean}$ compatible as originating from the inner Galactic disk; 
\item[$iii.$] the solar HAT-P-21 shows $TD/D$ and $Z_{\rm max}$ compatible with thin-thick disk transition and seems to originate from the outer Galactic disk.
\end{itemize}

\item Finally, since detailed knowledge of the formation of a planet requires accurate knowledge of chemical abundances of its host star, we tried to discuss formation and migration mechanisms of those targets for which abundances of planets hosted by stars analyzed in the present work were obtained for the same elemental ratios. We suggest that the planets orbiting around HAT-P-26 and WASP-39 started forming outside the CO$_2$ snowline, while those around HAT-P-12 and WASP-10 probably formed, respectively, between the CO$_2$ and CH$_4$ showlines, and between the N$_2$ and CO$_2$ showlines.
\end{itemize}

We think that analyses like those performed in this work will be necessary for future studies on planetary composition that take into account host star composition, in particular for transiting planet host stars, for which more information about the system formation, migration, and evolution can be retrieved. Metallicity, Mg/Si, C/O, C/N, S/N, N/O ratios are important indicator of planet formation, therefore future high-precision observations are essential to further explore the trend between stellar and planetary properties toward understanding the formation mechanisms of planets. For instance, some of the planets will be observed by {\it JWST}, therefore such a kind of studies are motivated by the prospects of chemical characterization of exoplanets, and how the chemical compositions of planet host stars relate to those of their planets. Forthcoming {\it JWST} observations and other upcoming infrared spectroscopic missions (like {\it Ariel}) will allow us to draw more robust conclusions, in particular for what concerns the level of precision for planetary abundances, useful to provide definitive conclusions on, e.g., planetary formation and evolution. Similarly, precise spectroscopic studies crossed with high-quality photometric data like those acquired with {\it TESS} or in the next future with {\it PLATO} will be useful to clarify the evolutionary stage and to better characterize the planetary system in a global way.

\begin{acknowledgements}
The authors are very grateful to the referee for carefully reading the paper and for his/her useful remarks that allowed us to improve the previous version of the manuscript. We acknowledge the support by INAF/Frontiera through the 
``Progetti Premiali'' funding scheme of the Italian Ministry of Education, University, and Research, by INAF Main Stream funding scheme through the project ``Ariel and the astrochemical link between circumstellar disks and planets'' (C54I19000700005), 
and by the Ariel ASI-INAF agreement n. 2021-5-HH.0. We thank the TNG staff for help with the observations. K.B. also thanks V. Adibekyan, G. Casali, J. Gagn\'e, L. Girardi, and L. Spina for fruitful discussions about some topics of the paper. 
This work is also based on observations collected at the European Southern Observatory under ESO programmes 088.C-0498, 089.C-0444, 090.C-0146. This work has made use of the VALD and NIST atomic line databases, and also of the SIMBAD 
database, operated at CDS (Strasbourg, France). This work has made use of data from the European Space Agency (ESA) mission {\it Gaia} (\url{https://www.cosmos.esa.int/gaia}), processed by the {\it Gaia} Data Processing and Analysis Consortium 
(DPAC, \url{https://www.cosmos.esa.int/web/gaia/dpac/consortium}). Funding for the DPAC has been provided by national institutions, in particular the institutions participating in the {\it Gaia} Multilateral Agreement. This research has made use of the 
NASA Exoplanet Archive, which is operated by the California Institute of Technology, under contract with the National Aeronautics and Space Administration under the Exoplanet Exploration Program.
\end{acknowledgements}

\bibliographystyle{aa} 


\begin{appendix}

\section{Additional tables}

\begin{landscape}
\tiny
\begin{longtable}{lrccrrrrrrrrlrc}
\caption[]{\label{tab:basic_info} Some basic information and parameters of our targets taken from the literature and used for the purpose of this work. Right ascensions, declinations, Gaia\,EDR3 magnitudes and colors, proper motions and parallaxes were taken from \cite{gaiacollaborationetal2021}, spectral types were retrieved from the NASA Exoplanet Archive and SIMBAD, Galactic latitudes and longitudes from SIMBAD, planetary masses and eccentricities from \cite{bonomoetal2017} and the NASA Exoplanet Archive, and the Mittag chromospheric indicators $\log R'_{\rm HK}$ from Claudi et al. (in prep).}\\
\hline\hline
~\\
Name  & $\alpha$ & $\delta$ & Sp.T. & $G$ & $G_{\rm BP}$ & $R_{\rm BP}$ & $\mu_{\alpha}$ & $\mu_{\delta}$ & $\pi$ & $l_{\rm gal}$ & $b_{\rm gal}$ & $M_{\rm p}$  & $e$ & $logR'_{\rm HK}$ \\
     & 	(deg)  & (deg) &  & (mag) &  (mag) &  (mag) & (mas/yr) & (mas/yr) & (mas)	   &	(deg)		 & (deg)	 & ($M_{\rm Jup}$) &  &  \\
\hline
~\\
HAT-P-3  & 206.094 & $+$48.028 & K1 & 11.279 & 11.740 & 10.662 & $-$19.62$\pm$0.01 &  $-$23.973$\pm$0.02 &  7.42$\pm$0.01 & 100.167 &$+$66.693 & 0.609 & 0.010 & $-$4.652 \\ 
HAT-P-4  & 229.991 & $+$36.229 & G1 & 11.063 & 11.370 & 10.596 & $-$21.51$\pm$0.01 &  $-$24.255$\pm$0.02 &  3.11$\pm$0.02 &  58.480 &$+$57.336 & 0.651 & 0.007 & $-$4.868 \\
HAT-P-14 & 260.116 & $+$38.242 & F5 &  9.878 & 10.106 &  9.492 &  $+$2.35$\pm$0.01 &   $-$6.679$\pm$0.01 &  4.45$\pm$0.01 &  62.589 &$+$33.538 & 2.303 & 0.107 & $-$4.802 \\
HAT-P-12 & 209.389 & $+$43.493 & K4 & 12.408 & 12.992 & 11.683 &$+$134.79$\pm$0.01 &  $-$44.229$\pm$0.01 &  7.04$\pm$0.01 &  87.993 &$+$68.876 & 0.208 & 0.035 & $-$4.481 \\ 
HAT-P-15 & 66.248 & $+$39.460 & G5 & 11.716 & 12.252 & 11.024 & $+$14.23$\pm$0.02 &   $-$9.407$\pm$0.02 &  5.19$\pm$0.02 & 161.559 & $-$6.892 & 1.949 & 0.200 & $-$4.793 \\ 
HAT-P-17 & 324.536 & $+$30.488 & K0 & 10.279 & 10.695 &  9.694 & $-$80.28$\pm$0.02 & $-$127.037$\pm$0.02 & 10.82$\pm$0.02 &  80.851 &$-$16.177 & 0.537 & 0.342 & $-$4.870 \\ 
HAT-P-18 & 256.346 & $+$33.012 & K2 & 12.358 & 12.906 & 11.666 & $-$14.00$\pm$0.01 &  $-$36.751$\pm$0.01 &  6.19$\pm$0.01 &  55.652 &$+$35.562 & 0.200 & 0.087 & $-$4.477 \\ 
HAT-P-20 & 111.916 & $+$24.336 & K3 & 10.991 & 11.641 & 10.223 &  $-$5.10$\pm$0.02 &  $-$96.090$\pm$0.02 & 14.01$\pm$0.02 & 194.415 &$+$18.394 & 7.270 & 0.016 & $-$4.022 \\ 
HAT-P-21 & 171.275 & $+$41.028 & G3 & 11.557 & 11.902 & 11.050 &  $-$1.14$\pm$0.02 &  $+$13.523$\pm$0.02 &  3.52$\pm$0.02 & 169.325 &$+$67.464 & 3.970 & 0.217 & $-$4.299 \\ 
HAT-P-22 & 155.681 & $+$50.128 & G5 &  9.525 &  9.932 &  8.952 & $-$26.11$\pm$0.01 &  $+$83.806$\pm$0.02 & 12.27$\pm$0.02 & 163.647 &$+$53.566 & 2.172 & 0.002 & $-$4.878 \\ 
HAT-P-26 & 213.157 & $+$04.059 & K1 & 11.462 & 11.919 & 10.840 & $+$37.74$\pm$0.03 & $-$142.816$\pm$0.02 &  7.00$\pm$0.02 & 346.513 &$+$59.873 & 0.059 & 0.124 & $-$4.847 \\ 
HAT-P-29 & 33.131 & $+$51.778 & F8 & 11.728 & 12.073 & 11.215 &  $-$9.97$\pm$0.02 &   $+$1.790$\pm$0.02 &  3.14$\pm$0.02 & 135.471 & $-$9.108 & 0.676 & 0.104 & $-$4.841 \\
HAT-P-30 & 123.950 & $+$05.836 & G0 & 10.304 & 10.561 &  9.89 & $-$17.23$\pm$0.01 &  $+$23.875$\pm$0.01 &  4.80$\pm$0.02 & 217.412 &$+$21.405 & 0.746 & 0.016 & $-$5.043 \\ 
HAT-P-36 & 188.266 & $+$44.915 & G5 & 12.085 & 12.440 & 11.569 & $-$11.62$\pm$0.01 & $+$8.138$\pm$0.01 &  3.41$\pm$0.01 & 133.414 &$+$71.837 & 1.850 & 0.059 & $-$4.431 \\ 
KELT-6   & 195.982 & $+$30.640 & F8 & 10.198 & 10.442 &  9.791 &  $-$4.96$\pm$0.01 &  $+$15.722$\pm$0.01 &  4.11$\pm$0.02 &  85.775 &$+$85.550 & 0.442 & 0.036 & $-$4.907 \\
Qatar-1  & 303.382 & $+$65.162 & K2 & 12.465 & 13.004 & 11.780 & $+$12.78$\pm$0.01 &  $+$58.064$\pm$0.01 &  5.34$\pm$0.01 &  98.719 &$+$16.424 & 1.321 & 0.012 & $-$4.191 \\ 
Qatar-2  & 207.655 & $-$06.804 & K5 & 13.015 & 13.662 & 12.245 & $-$88.17$\pm$0.02 &  $-$15.187$\pm$0.02 &  5.48$\pm$0.02 & 327.956 &$+$53.168 & 2.470 & 0.011 & $-$4.225 \\ 
TRES-4   & 268.304 & $+$37.211 & F8 & 11.470 & 11.725 & 11.050 &  $-$6.38$\pm$0.02 &  $-$20.891$\pm$0.02 &  1.97$\pm$0.01 &  63.039 &$+$26.994 & 0.498 & 0.015 & $-$5.106 \\
WASP-10  & 348.993 & $+$31.462 & K5 & 12.164 & 12.736 & 11.452 & $+$25.05$\pm$0.01 &  $-$25.366$\pm$0.01 &  7.07$\pm$0.01 & 100.112 &$-$27.144 & 3.210 & 0.060 & $-$4.943 \\ 
WASP-11  & 47.369 & $+$30.673 & K3 & 11.559 & 12.072 & 10.873 &  $+$3.33$\pm$0.07 &  $-$44.433$\pm$0.05 &  7.70$\pm$0.06 & 155.018 &$-$23.463 & 0.532 & 0.030 & $-$4.859 \\ 
WASP-13  & 140.103 & $+$33.882 & G1 & 10.383 & 10.675 &  9.924 &  $-$2.28$\pm$0.02 &  $-$20.072$\pm$0.02 &  4.33$\pm$0.02 & 190.929 &$+$44.536 & 0.525 & 0.016 & $-$4.937 \\
WASP-38  & 243.960 & $+$10.032 & F8 &  9.218 &  9.485 &  8.78 & $-$31.14$\pm$0.02 &  $-$39.167$\pm$0.01 &  7.34$\pm$0.02 &  23.528 &$+$39.033 & 2.648 & 0.028 & $-$4.890 \\ 
WASP-39 & 217.327 & $-$03.444 & G8 & 11.881 & 12.280 & 11.316 & $-$19.04$\pm$0.02 &  $+$0.437$\pm$0.01 & 4.64$\pm$0.01 & 344.403 & $+$51.371 & 0.275 & 0.048 & $-$4.801 \\ 
WASP-43  & 154.908 & $-$09.806 & K7 & 11.894 & 12.660 & 11.040 & $-$41.99$\pm$0.02 &  $-$38.004$\pm$0.02 & 11.47$\pm$0.02 & 252.787 &$+$37.870 & 2.050 & 0.006 & $-$4.030 \\ 
WASP-54  & 205.454 & $-$00.128 & F9 & 10.248 & 10.518 &  9.815 & $-$10.81$\pm$0.02 &  $-$24.744$\pm$0.02 &  3.99$\pm$0.02 & 328.936 &$+$60.175 & 0.606 & 0.060 & $-$4.949 \\
WASP-60  & 356.667 & $+$31.155 & G1 & 12.010 & 12.358 & 11.498 & $+$30.25$\pm$0.02 &   $-$6.005$\pm$0.01 &  2.32$\pm$0.02 & 106.985 &$-$29.702 & 0.505 & 0.064 & $-$4.881 \\
XO-2N    & 117.027 & $+$50.225 & G9 & 10.971 & 11.390 & 10.392 & $-$29.55$\pm$0.02 & $-$154.227$\pm$0.01 &  6.66$\pm$0.02 & 168.290 &$+$29.328 & 0.595 & 0.006 & $-$4.709 \\
XO-2S    & 117.031 & $+$50.216 & G9 & 10.927 & 11.331 & 10.362 & $-$29.31$\pm$0.02 & $-$154.233$\pm$0.01 &  6.67$\pm$0.02 & 168.300 &$+$29.330 & 0.259$^\ast$ & 0.180 & ... \\ 
~\\
\hline
\end{longtable}
\footnotesize{Note:
$^\ast$ This is the value of  $M_{\rm p} \sin i_{\rm p}$.
}
\end{landscape}

\newpage
\pagestyle{empty}
\topmargin 3 cm
\setlength{\tabcolsep}{1.3pt}
\begin{landscape}
\scriptsize
\begin{longtable}{lrrrrrrrrrrrrr}
\caption[]{\label{tab:final_stellar_abund} Final elemental abundances derived in this work. We show in brackets the number of lines and the cases for which we applied the spectral synthesis (s).}\\
\hline\hline
Name & [C/H] & [N/H] & [O/H] & [Na/H] & [Mg/H] & [Al/H] & [Si/H] & [S/H] & [Ca/H] & [Sc/H]& [\ion{Ti}{i}/H] & [\ion{Ti}{ii}/H] & [V/H] \\
 & (dex) & (dex) & (dex) & (dex) & (dex) & (dex) & (dex) & (dex) & (dex)  & (dex) & (dex) & (dex) & (dex) \\
\hline
\multicolumn{14}{c}{{\it Elements with $Z \le 23$}}\\
HAT-P-3  &    0.15$\pm$0.09(3,s) &   0.39$\pm$0.17(s) &    0.18$\pm$0.10(s)   &    0.50$\pm$0.01(2) &    0.27$\pm$0.12(3) &    0.40$\pm$0.06(1) &    0.30$\pm$0.07(12) &    0.37$\pm$0.14(3) &    0.33$\pm$0.12(9)  &	  0.33$\pm$0.07(3) &	0.40$\pm$0.10(17) &    0.40$\pm$0.09(5) &   0.48$\pm$0.09(12) \\
HAT-P-4  &    0.17$\pm$0.03(3,s) &    ...	      &    0.32$\pm$0.20(s)   &    0.16$\pm$0.06(4) &	 0.23$\pm$0.09(4) &    0.23$\pm$0.07(3) &    0.27$\pm$0.08(12) &    0.35$\pm$0.16(3) &    0.27$\pm$0.08(12) &	  0.34$\pm$0.13(3) &	0.27$\pm$0.06(19) &    0.29$\pm$0.09(6) &   0.24$\pm$0.06(11) \\
HAT-P-12 & $-$0.44$\pm$0.12(s)   &    ...	      & $-$0.17$\pm$0.13(s)   & $-$0.25$\pm$0.05(2) & $-$0.18$\pm$0.08(3) & $-$0.01$\pm$0.07(2) & $-$0.24$\pm$0.08(10) &     ...	     & $-$0.18$\pm$0.13(7)  &  $-$0.13$\pm$0.10(3) & $-$0.08$\pm$0.11(16) & $-$0.15$\pm$0.05(2) &   0.01$\pm$0.07(11) \\
HAT-P-14 & $-$0.16$\pm$0.03(3,s) &    ...	      &    ...  	      & $-$0.08$\pm$0.05(4) & $-$0.10$\pm$0.10(4) & $-$0.03$\pm$0.06(2) &    0.02$\pm$0.08(11) & $-$0.11$\pm$0.07(3) &    0.16$\pm$0.09(13) &  $-$0.02$\pm$0.07(3) &	0.04$\pm$0.08(16) &    0.06$\pm$0.11(7) &   0.15$\pm$0.06(3)  \\
HAT-P-15 &    0.11$\pm$0.05(3,s) &    ...	      &    0.14$\pm$0.15(s)   &    0.34$\pm$0.04(2) &	 0.27$\pm$0.11(4) &    0.28$\pm$0.07(3) &    0.28$\pm$0.08(12) &    0.26$\pm$0.15(3) &    0.24$\pm$0.09(11) &	  0.31$\pm$0.13(2) &	0.29$\pm$0.07(20) &    0.40$\pm$0.05(4) &   0.23$\pm$0.08(12) \\
HAT-P-17 & $-$0.11$\pm$0.06(3,s) &$-$0.03$\pm$0.21(s) &    0.08$\pm$0.05(3,s) &    0.03$\pm$0.01(2) &    0.07$\pm$0.08(4) &    0.07$\pm$0.07(2) &    0.06$\pm$0.09(12) &    0.09$\pm$0.11(3) &    0.03$\pm$0.11(10) &	  0.09$\pm$0.13(2) &	0.08$\pm$0.09(21) &    0.10$\pm$0.08(6) &   0.09$\pm$0.06(12) \\
HAT-P-18 &    0.00$\pm$0.13(s)   &    ...	      &    ...  	      &    0.17$\pm$0.05(2) &    0.03$\pm$0.09(3) &    0.17$\pm$0.06(1) &    0.12$\pm$0.09(12) &    0.20$\pm$0.07(1) &    0.15$\pm$0.11(8)  &	  0.11$\pm$0.06(3) &	0.20$\pm$0.08(12) &    0.16$\pm$0.11(3) &   0.23$\pm$0.10(12) \\
HAT-P-20 & $-$0.02$\pm$0.15(s)   &    ...	      &    0.10$\pm$0.20(s)   &     ... 	    &    0.09$\pm$0.09(3) &    0.27$\pm$0.09(2) &    0.11$\pm$0.10(9)  &     ...	     &    0.26$\pm$0.15(3)  &	  0.31$\pm$0.17(3) &	0.29$\pm$0.12(13) &    0.17$\pm$0.11(3) &   0.49$\pm$0.12(11) \\
HAT-P-21 &    0.08$\pm$0.03(3,s) &    ...	      &    ...  	      & $-$0.01$\pm$0.08(4) &    0.07$\pm$0.08(4) &    0.08$\pm$0.09(3) &    0.10$\pm$0.07(11) &    0.16$\pm$0.13(3) &    0.06$\pm$0.08(13) &	  0.12$\pm$0.11(3) &	0.11$\pm$0.06(20) &    0.13$\pm$0.09(5) &   0.04$\pm$0.06(12) \\
HAT-P-22 &    0.17$\pm$0.06(3,s) &   0.20$\pm$0.21(s) &    0.09$\pm$0.15(s)   &    0.33$\pm$0.01(2) &    0.40$\pm$0.13(3) &    0.41$\pm$0.08(3) &    0.39$\pm$0.08(12) &    0.32$\pm$0.04(3) &    0.33$\pm$0.11(9)  &	  0.41$\pm$0.17(3) &	0.42$\pm$0.10(21) &    0.47$\pm$0.08(5) &   0.37$\pm$0.06(12) \\
HAT-P-26 &    0.12$\pm$0.11(2,s) &    ...	      &    0.14$\pm$0.07(3)   &    0.04$\pm$0.04(2) &    0.18$\pm$0.09(3) &    0.25$\pm$0.08(2) &    0.08$\pm$0.08(12) &    0.08$\pm$0.06(3) &    0.07$\pm$0.12(9)  &	  0.18$\pm$0.11(3) &	0.17$\pm$0.11(21) &    0.20$\pm$0.07(4) &   0.20$\pm$0.07(12) \\
HAT-P-29 &    0.08$\pm$0.03(2,s) &    ...	      &    0.17$\pm$0.15(s)   &    0.13$\pm$0.03(4) &    0.13$\pm$0.11(4) &    0.11$\pm$0.10(3) &    0.21$\pm$0.09(11) &    0.25$\pm$0.16(3) &    0.24$\pm$0.08(13) &	  0.18$\pm$0.10(3) &	0.17$\pm$0.06(20) &    0.20$\pm$0.08(5) &   0.16$\pm$0.09(9)  \\
HAT-P-30 & $-$0.04$\pm$0.03(2,s) &   0.02$\pm$0.08(2) & $-$0.02$\pm$0.06(3)   & $-$0.01$\pm$0.06(4) &    0.01$\pm$0.09(4) & $-$0.10$\pm$0.08(2) &    0.04$\pm$0.07(11) &    0.15$\pm$0.14(3) &    0.09$\pm$0.11(14) &	  0.03$\pm$0.08(3) &	0.04$\pm$0.06(19) &    0.05$\pm$0.11(6) &   0.04$\pm$0.08(7)  \\
HAT-P-36 &    0.11$\pm$0.04(3,s) &    ...	      &    0.20$\pm$0.12(s)   &    0.29$\pm$0.03(2) &	 0.25$\pm$0.11(4) &    0.31$\pm$0.07(3) &    0.26$\pm$0.09(11) &    0.08$\pm$0.13(2) &    0.27$\pm$0.10(12) &	  0.22$\pm$0.17(3) &	0.26$\pm$0.06(19) &    0.39$\pm$0.07(5) &   0.17$\pm$0.08(12) \\
KELT-6   & $-$0.34$\pm$0.03(2,s) &    ...	      & $-$0.20$\pm$0.20(s)   & $-$0.31$\pm$0.02(4) & $-$0.25$\pm$0.06(2) & $-$0.42$\pm$0.08(3) & $-$0.22$\pm$0.08(12) & $-$0.27$\pm$0.10(3) & $-$0.17$\pm$0.07(13) &  $-$0.29$\pm$0.08(3) & $-$0.23$\pm$0.06(20) & $-$0.23$\pm$0.06(6) &$-$0.21$\pm$0.06(9)  \\
Qatar-1  & $-$0.03$\pm$0.13(s)   &    ...	      &    0.09$\pm$0.15(s)   &    0.33$\pm$0.03(2) &    0.15$\pm$0.08(3) &    0.32$\pm$0.06(2) &    0.22$\pm$0.10(12) &    0.21$\pm$0.08(1) &    0.18$\pm$0.11(7)  &	  0.34$\pm$0.13(3) &	0.26$\pm$0.10(16) &    0.27$\pm$0.07(2) &   0.35$\pm$0.09(12) \\
Qatar-2  &      ...		 &    ...	      &    ...  	      &    0.32$\pm$0.05(2) &    0.05$\pm$0.11(3) &    0.26$\pm$0.10(3) &    0.17$\pm$0.06(5)  &     ...	     &    0.23$\pm$0.14(3)  &	  0.31$\pm$0.09(2) &	0.26$\pm$0.14(13) &    0.15$\pm$0.09(3) &   0.46$\pm$0.12(11) \\
TRES-4   &    0.15$\pm$0.06(3,s) &    ...	      &    ...  	      &    0.21$\pm$0.08(4) &    0.13$\pm$0.06(3) &    0.26$\pm$0.12(2) &    0.23$\pm$0.09(11) &    0.14$\pm$0.04(2) &    0.26$\pm$0.08(11) &	  0.29$\pm$0.13(2) &	0.19$\pm$0.08(15) &    0.26$\pm$0.08(7) &   0.19$\pm$0.07(6)  \\
WASP-10  & $-$0.16$\pm$0.15(s)   &  0.07$\pm$0.17(s)  &    ...  	      &    0.08$\pm$0.09(2) & $-$0.04$\pm$0.09(3) &    0.14$\pm$0.07(2) &    0.12$\pm$0.09(9)  &     ...	     &    0.10$\pm$0.14(6)  &	  0.15$\pm$0.14(3) &	0.14$\pm$0.13(16) &    0.09$\pm$0.13(2) &   0.19$\pm$0.09(12) \\
WASP-11  & $-$0.07$\pm$0.12(s)   &    ...	      & $-$0.00$\pm$0.12(s)   &    0.22$\pm$0.03(2) &    0.11$\pm$0.09(3) &    0.16$\pm$0.08(2) &    0.17$\pm$0.07(10) &    0.19$\pm$0.09(3) &    0.07$\pm$0.11(8)  &	  0.17$\pm$0.16(3) &	0.08$\pm$0.10(17) &    0.10$\pm$0.11(4) &   0.16$\pm$0.08(12) \\
WASP-13  & $-$0.18$\pm$0.04(3,s) &    ...	      &    ...  	      &    0.04$\pm$0.02(4) &    0.08$\pm$0.08(4) & $-$0.04$\pm$0.13(2) &    0.09$\pm$0.09(12) &    0.10$\pm$0.08(3) &    0.14$\pm$0.08(12) &	  0.06$\pm$0.06(3) &	0.08$\pm$0.06(19) &    0.08$\pm$0.14(7) &   0.07$\pm$0.06(11) \\
WASP-38  &    0.00$\pm$0.05(3,s) &    ...	      &    ...  	      &    0.00$\pm$0.03(4) & $-$0.02$\pm$0.07(2) & $-$0.06$\pm$0.08(2) &    0.04$\pm$0.08(11) & $-$0.05$\pm$0.03(2) &    0.10$\pm$0.08(11) &	  0.00$\pm$0.10(3) &	0.04$\pm$0.07(19) &    0.07$\pm$0.07(6) &   0.11$\pm$0.07(9)  \\
WASP-39  & $-$0.16$\pm$0.07(3,s) &    ...	      &    ...  	      & $-$0.06$\pm$0.08(3) &    0.00$\pm$0.08(4) &    0.01$\pm$0.08(3) &    0.01$\pm$0.09(12) &    0.13$\pm$0.06(3) &    0.01$\pm$0.10(11) &	  0.05$\pm$0.15(2) & $-$0.01$\pm$0.06(21) &    0.02$\pm$0.09(7) &$-$0.04$\pm$0.06(12) \\
WASP-43  &      ...		 &    ...	      &    ...  	      &     ... 	    & $-$0.08$\pm$0.10(2) &    0.19$\pm$0.12(3) & $-$0.10$\pm$0.10(9)  &     ...	     &    0.13$\pm$0.22(3)  &	  0.09$\pm$0.16(3) &	0.08$\pm$0.15(12) & $-$0.06$\pm$0.18(3) &   0.31$\pm$0.09(10) \\
WASP-54  & $-$0.15$\pm$0.04(3,s) &    ...	      & $-$0.11$\pm$0.05(3,s) & $-$0.22$\pm$0.05(4) & $-$0.19$\pm$0.08(4) & $-$0.06$\pm$0.06(3) & $-$0.10$\pm$0.10(11) & $-$0.13$\pm$0.12(2) & $-$0.01$\pm$0.08(14) &  $-$0.07$\pm$0.09(3) & $-$0.10$\pm$0.06(19) & $-$0.05$\pm$0.12(7) &$-$0.09$\pm$0.09(8)  \\
WASP-60  &    0.19$\pm$0.03(3,s) &    ...	      &    0.34$\pm$0.15(s)   &    0.20$\pm$0.07(4) &    0.19$\pm$0.09(4) &    0.14$\pm$0.09(3) &    0.24$\pm$0.08(11) &    0.25$\pm$0.16(3) &    0.24$\pm$0.08(13) &	  0.19$\pm$0.13(3) &	0.18$\pm$0.06(20) &    0.18$\pm$0.10(6) &   0.17$\pm$0.08(10) \\
XO-2N    &    0.40$\pm$0.04(3,s) &   0.31$\pm$0.21(s) &    0.40$\pm$0.15(s)   &    0.48$\pm$0.03(2) &    0.40$\pm$0.11(4) &    0.46$\pm$0.06(2) &    0.45$\pm$0.08(10) &    0.50$\pm$0.04(3) &    0.37$\pm$0.12(9)  &	  0.49$\pm$0.17(3) &	0.46$\pm$0.09(20) &    0.49$\pm$0.09(6) &   0.46$\pm$0.07(12) \\
XO-2S    &    0.37$\pm$0.05(3,s) &   0.26$\pm$0.21(s) &    0.38$\pm$0.15(s)   &    0.45$\pm$0.02(2) &    0.37$\pm$0.11(4) &    0.38$\pm$0.07(2) &    0.44$\pm$0.08(10) &    0.54$\pm$0.03(3) &    0.33$\pm$0.12(9)  &	  0.42$\pm$0.17(3) &	0.36$\pm$0.08(20) &    0.41$\pm$0.10(6) &   0.33$\pm$0.07(12) \\
\hline
Name & [\ion{Cr}{i}/H] & [\ion{Cr}{ii}/H] & [Mn/H] & [Co/H] & [Ni/H] & [Cu/H] & [Zn/H] & [Y/H] & [Zr/H] & [Ba/H]& [La/H] & [Nd/H] & [Eu/H] \\
 & (dex) & (dex) & (dex) & (dex) & (dex) & (dex) & (dex) & (dex) & (dex)  & (dex) & (dex) & (dex) & (dex) \\
\hline
\multicolumn{14}{c}{{\it Elements with $Z > 23$}}\\
HAT-P-3  &   0.40$\pm$0.10(13) &   0.40$\pm$0.12(6) &	 ...		  &   0.41$\pm$0.10(5) &    0.37$\pm$0.08(37) &    0.29$\pm$0.10(1) &	 0.25$\pm$0.10(3) &    0.24$\pm$0.08(5) &   0.33$\pm$0.05(3) &   0.18$\pm$0.07(2) &    0.42$\pm$0.08(2) &    0.20$\pm$0.09(3) & 0.41$\pm$0.08(1)\\ 
HAT-P-4  &   0.26$\pm$0.07(15) &   0.28$\pm$0.06(6) &	 0.18$\pm$0.06(3) &   0.30$\pm$0.07(5) &    0.28$\pm$0.09(45) &    0.21$\pm$0.12(4) &	 0.22$\pm$0.09(3) &    0.30$\pm$0.09(6) &   0.20$\pm$0.07(3) &   0.38$\pm$0.08(2) &    0.37$\pm$0.08(2) &    0.28$\pm$0.02(2) & 0.19$\pm$0.08(1)\\ 
HAT-P-12 &$-$0.20$\pm$0.11(10) &   0.14$\pm$0.09(1) & $-$0.30$\pm$0.04(2) &$-$0.19$\pm$0.07(4) & $-$0.29$\pm$0.10(40) & $-$0.38$\pm$0.15(2) & $-$0.28$\pm$0.05(3) & $-$0.42$\pm$0.10(3) &$-$0.27$\pm$0.07(1) &$-$0.38$\pm$0.09(2) &    0.08$\pm$0.11(1) & $-$0.15$\pm$0.18(2) & 0.02$\pm$0.03(1)\\
HAT-P-14 &   0.01$\pm$0.07(12) &   0.02$\pm$0.09(6) & $-$0.13$\pm$0.06(2) &   0.08$\pm$0.12(2) & $-$0.02$\pm$0.08(33) & $-$0.10$\pm$0.13(3) & $-$0.15$\pm$0.08(3) &    0.11$\pm$0.09(5) &   0.16$\pm$0.06(3) &   0.36$\pm$0.08(2) &    0.04$\pm$0.06(2) &    0.16$\pm$0.06(2) &   ...	\\ 
HAT-P-15 &   0.31$\pm$0.06(14) &   0.32$\pm$0.09(6) &	 0.38$\pm$0.04(3) &   0.26$\pm$0.09(5) &    0.33$\pm$0.08(43) &    0.34$\pm$0.10(3) &	 0.35$\pm$0.11(3) &    0.34$\pm$0.11(6) &   0.26$\pm$0.06(3) &   0.31$\pm$0.09(2) &    0.26$\pm$0.06(2) &    0.18$\pm$0.06(3) & 0.25$\pm$0.08(1)\\ 
HAT-P-17 &   0.08$\pm$0.08(15) &   0.08$\pm$0.08(6) &	 0.12$\pm$0.08(3) &   0.04$\pm$0.09(5) &    0.05$\pm$0.08(43) &    0.08$\pm$0.11(2) &	 0.06$\pm$0.06(3) & $-$0.03$\pm$0.09(5) &   0.07$\pm$0.07(3) &$-$0.02$\pm$0.07(2) &    0.06$\pm$0.08(2) & $-$0.02$\pm$0.06(3) & 0.14$\pm$0.09(1)\\
HAT-P-18 &   0.19$\pm$0.10(11) &   0.30$\pm$0.07(3) &	 ...		  &   0.06$\pm$0.10(5) &    0.12$\pm$0.09(37) &    0.03$\pm$0.13(2) &	 0.06$\pm$0.05(3) &    0.05$\pm$0.11(4) &   0.03$\pm$0.18(2) &   0.06$\pm$0.07(2) &    0.30$\pm$0.14(2) &    0.14$\pm$0.13(3) &    ...  \\
HAT-P-20 &   0.34$\pm$0.11(9)  &   0.56$\pm$0.07(3) &	 ...		  &   0.35$\pm$0.10(4) &    0.27$\pm$0.09(31) &    0.33$\pm$0.11(1) &	 0.27$\pm$0.05(2) &    0.24$\pm$0.13(2) &   0.15$\pm$0.07(1) &   0.14$\pm$0.08(2) &	  ...	        &    0.19$\pm$0.02(1) & 0.32$\pm$0.05(1)\\
HAT-P-21 &   0.07$\pm$0.06(15) &   0.05$\pm$0.07(6) & $-$0.04$\pm$0.06(3) &   0.07$\pm$0.08(5) &    0.03$\pm$0.08(46) &    0.07$\pm$0.13(4) &	 0.12$\pm$0.09(3) & $-$0.06$\pm$0.07(6) &$-$0.06$\pm$0.09(3) &   0.12$\pm$0.06(2) &    0.15$\pm$0.04(2) &    0.10$\pm$0.06(2) & 0.19$\pm$0.06(1)\\ 
HAT-P-22 &   0.34$\pm$0.07(13) &   0.38$\pm$0.12(6) &	 0.42$\pm$0.08(3) &   0.34$\pm$0.10(5) &    0.35$\pm$0.08(40) &    0.42$\pm$0.09(2) &	 0.37$\pm$0.10(3) &    0.28$\pm$0.11(5) &   0.27$\pm$0.07(3) &   0.29$\pm$0.09(2) &    0.33$\pm$0.06(2) &    0.26$\pm$0.05(3) & 0.38$\pm$0.04(1)\\ 
HAT-P-26 &   0.10$\pm$0.10(13) &   0.13$\pm$0.07(5) &	 0.07$\pm$0.07(2) &   0.10$\pm$0.10(5) &    0.07$\pm$0.08(42) &    0.09$\pm$0.13(2) &	 0.06$\pm$0.06(3) & $-$0.08$\pm$0.08(5) &   0.18$\pm$0.12(3) &$-$0.04$\pm$0.10(2) &    0.14$\pm$0.12(2) &    0.07$\pm$0.07(3) & 0.19$\pm$0.08(1)\\ 
HAT-P-29 &   0.21$\pm$0.06(15) &   0.21$\pm$0.07(6) &	 0.15$\pm$0.06(3) &   0.16$\pm$0.06(4) &    0.19$\pm$0.07(46) &    0.16$\pm$0.11(4) &	 0.19$\pm$0.09(3) &    0.27$\pm$0.08(6) &   0.25$\pm$0.07(3) &   0.33$\pm$0.08(2) &    0.37$\pm$0.09(2) &    0.25$\pm$0.03(3) & 0.09$\pm$0.06(1)\\ 
HAT-P-30 &   0.08$\pm$0.08(15) &   0.09$\pm$0.09(6) & $-$0.01$\pm$0.07(3) &   0.10$\pm$0.10(2) &    0.02$\pm$0.07(45) & $-$0.03$\pm$0.12(4) &	 0.09$\pm$0.09(3) &    0.13$\pm$0.08(6) &   0.13$\pm$0.05(3) &   0.31$\pm$0.05(2) &    0.13$\pm$0.06(2) &    0.11$\pm$0.05(3) & 0.16$\pm$0.09(1)\\ 
HAT-P-36 &   0.31$\pm$0.06(15) &   0.33$\pm$0.09(5) &	 0.31$\pm$0.10(3) &   0.22$\pm$0.10(4) &    0.29$\pm$0.08(41) &    0.24$\pm$0.13(4) &	 0.33$\pm$0.22(3) &    0.36$\pm$0.11(6) &   0.26$\pm$0.07(3) &   0.36$\pm$0.10(2) &    0.27$\pm$0.06(2) &    0.20$\pm$0.09(2) & 0.41$\pm$0.07(1)\\ 
KELT-6   &$-$0.32$\pm$0.07(14) &$-$0.34$\pm$0.06(6) & $-$0.44$\pm$0.05(3) &$-$0.19$\pm$0.09(4) & $-$0.32$\pm$0.06(45) & $-$0.34$\pm$0.12(4) & $-$0.36$\pm$0.05(3) & $-$0.37$\pm$0.07(6) &$-$0.26$\pm$0.04(3) &$-$0.12$\pm$0.08(2) & $-$0.18$\pm$0.11(2) & $-$0.26$\pm$0.06(3) & $-$0.14$\pm$0.05(1)\\ 
Qatar-1  &   0.29$\pm$0.09(11) &   0.45$\pm$0.10(3) &	 ...		  &   0.26$\pm$0.10(4) &    0.23$\pm$0.09(38) &    0.23$\pm$0.15(2) &	 0.17$\pm$0.06(3) &    0.04$\pm$0.11(3) &   0.22$\pm$0.07(2) &   0.09$\pm$0.09(2) &    0.36$\pm$0.13(1) &    0.22$\pm$0.14(3) &    ...  \\
Qatar-2  &   0.25$\pm$0.12(10) &   0.30$\pm$0.10(1) &	 ...		  &   0.27$\pm$0.10(4) &    0.16$\pm$0.09(37) &    0.25$\pm$0.11(2) &	      ...	  & $-$0.02$\pm$0.14(2) &	...	     &   0.07$\pm$0.09(2) &	  ...	        &    0.10$\pm$0.01(2) &    ...  \\
TRES-4   &   0.21$\pm$0.07(12) &   0.28$\pm$0.06(6) &	 0.19$\pm$0.08(3) &   0.22$\pm$0.07(1) &    0.22$\pm$0.07(40) &    0.21$\pm$0.10(4) &	 0.19$\pm$0.08(3) &    0.24$\pm$0.07(5) &   0.25$\pm$0.09(3) &   0.30$\pm$0.09(2) &    0.30$\pm$0.08(2) &    0.22$\pm$0.04(3) &   ...	 \\ 
WASP-10  &   0.23$\pm$0.13(12) &   0.49$\pm$0.10(3) &	 ...		  &   0.09$\pm$0.09(4) &    0.18$\pm$0.08(38) &    0.06$\pm$0.11(2) &	 0.06$\pm$0.07(2) &    0.19$\pm$0.10(3) &   0.01$\pm$0.03(1) &   0.18$\pm$0.13(2) &    0.21$\pm$0.06(1) &    0.04$\pm$0.02(1) & 0.12$\pm$0.06(1)\\
WASP-11  &   0.15$\pm$0.09(12) &   0.23$\pm$0.08(4) &	 ...		  &   0.11$\pm$0.10(5) &    0.15$\pm$0.09(43) &    0.17$\pm$0.11(2) &	 0.15$\pm$0.06(3) &    0.04$\pm$0.12(4) &   0.16$\pm$0.20(2) &$-$0.06$\pm$0.10(2) &    0.13$\pm$0.06(1) &    0.01$\pm$0.12(3) & 0.12$\pm$0.04(1)\\
WASP-13  &   0.08$\pm$0.06(15) &   0.05$\pm$0.06(6) &	 0.00$\pm$0.09(3) &   0.08$\pm$0.08(5) &    0.06$\pm$0.07(44) &    0.08$\pm$0.11(4) &	 0.09$\pm$0.10(3) &    0.05$\pm$0.08(6) &   0.09$\pm$0.04(3) &   0.21$\pm$0.05(2) &    0.29$\pm$0.04(2) &    0.08$\pm$0.02(3) & 0.12$\pm$0.06(1)\\ 
WASP-38  &   0.03$\pm$0.06(14) &   0.03$\pm$0.07(6) & $-$0.04$\pm$0.04(3) &   0.08$\pm$0.07(4) &    0.04$\pm$0.07(43) &    0.03$\pm$0.11(4) &	 0.03$\pm$0.09(3) &    0.07$\pm$0.08(5) &   0.08$\pm$0.04(3) &   0.19$\pm$0.05(2) &    0.18$\pm$0.10(2) &    0.10$\pm$0.09(3) & 0.00$\pm$0.09(1)\\ 
WASP-39  &   0.03$\pm$0.07(15) &   0.02$\pm$0.07(6) & $-$0.04$\pm$0.06(3) &$-$0.03$\pm$0.09(5) &    0.00$\pm$0.09(46) & $-$0.05$\pm$0.13(4) &	 0.01$\pm$0.08(3) &    0.03$\pm$0.10(6) &   0.00$\pm$0.07(3) &   0.02$\pm$0.09(2) &    0.10$\pm$0.04(2) &    0.00$\pm$0.08(3) & 0.19$\pm$0.05(1)\\ 
WASP-43  &   0.08$\pm$0.12(9)  &   0.04$\pm$0.08(1) &	 ...		  &   0.13$\pm$0.06(4) &    0.05$\pm$0.10(28) &    0.01$\pm$0.13(1) & $-$0.26$\pm$0.09(2) & $-$0.04$\pm$0.08(1) &$-$0.19$\pm$0.04(1) &$-$0.20$\pm$0.07(2) &	  ...	        &    0.14$\pm$0.01(1) & 0.25$\pm$0.08(1)\\
WASP-54  &$-$0.14$\pm$0.07(15) &$-$0.14$\pm$0.07(6) & $-$0.25$\pm$0.06(3) &$-$0.04$\pm$0.09(4) & $-$0.14$\pm$0.07(42) & $-$0.20$\pm$0.11(4) & $-$0.14$\pm$0.05(3) & $-$0.05$\pm$0.08(6) &$-$0.03$\pm$0.06(3) &   0.16$\pm$0.05(2) &    0.07$\pm$0.06(2) & $-$0.04$\pm$0.03(3) & $-$0.09$\pm$0.09(1)\\ 
WASP-60  &   0.23$\pm$0.06(15) &   0.24$\pm$0.07(6) &	 0.19$\pm$0.09(3) &   0.25$\pm$0.09(5) &    0.22$\pm$0.08(46) &    0.24$\pm$0.12(4) &	 0.28$\pm$0.10(3) &    0.26$\pm$0.07(6) &   0.23$\pm$0.04(3) &   0.26$\pm$0.05(2) &    0.29$\pm$0.09(2) &    0.19$\pm$0.06(3) & 0.29$\pm$0.09(1)\\ 
XO-2N    &   0.46$\pm$0.09(13) &   0.50$\pm$0.12(6) &	 0.59$\pm$0.09(3) &   0.49$\pm$0.09(5) &    0.46$\pm$0.09(40) &    0.44$\pm$0.09(1) &	 0.44$\pm$0.09(3) &    0.41$\pm$0.11(5) &   0.43$\pm$0.09(3) &   0.38$\pm$0.10(2) &    0.55$\pm$0.06(2) &    0.35$\pm$0.09(3) & 0.40$\pm$0.04(1)\\ 
XO-2S    &   0.39$\pm$0.08(13) &   0.46$\pm$0.12(6) &	 0.54$\pm$0.08(3) &   0.40$\pm$0.08(5) &    0.40$\pm$0.09(40) &    0.41$\pm$0.09(1) &	 0.44$\pm$0.10(3) &    0.36$\pm$0.11(5) &   0.33$\pm$0.08(3) &   0.30$\pm$0.08(2) &    0.45$\pm$0.06(2) &    0.23$\pm$0.04(3) & 0.35$\pm$0.04(1)\\ 
\hline
\end{longtable}
\end{landscape}

\setlength{\tabcolsep}{4pt}
\begin{table*}
\caption{Stellar kinematic properties as derived in the present work. Columns list the $UVW$ velocities, the thick disk-to-thin disk probability, the Galactic eccentricity, the maximum vertical distance above the Galactic plane, and 
the difference between the mean and the current Galactocentric distances.}
\label{tab:space_motion}
\begin{center}
\begin{tabular}{lrrrrccr}
\hline\hline
    Name & $U$ & $V$ & $W$ & $TD/D$ & $e_{\rm G}$ & $Z_{\rm max}$  & $\Delta(R_{\rm mean}-R_{\rm GC}$)\\
     & 	(km/s)  & (km/s) & (km/s) & & &  (kpc)  &  (kpc)   \\
\hline
HAT-P-3   &  $-$6.50$\pm$0.03 & $-$13.95$\pm$0.09 &  $-$7.22$\pm$0.20 & 0.01 & 0.086 & 0.1783 & $-$0.54\\
HAT-P-4   &    3.918$\pm$0.18 & $-$30.88$\pm$0.18 &    24.89$\pm$0.20 & 0.04 & 0.173 & 0.5305 & $-$1.01 \\
HAT-P-12  & $-$55.42$\pm$0.10 & $-$78.87$\pm$0.12 &  $-$0.63$\pm$0.19 & 3.73 & 0.404 & 0.1598 & $-$2.17 \\
HAT-P-14  &  $-$9.52$\pm$0.09 &  $-$2.88$\pm$0.18 &  $-$7.84$\pm$0.13 & 0.01 & 0.039 & 0.1822 & $-$0.15 \\
HAT-P-15  & $-$43.61$\pm$0.19 &	  9.03$\pm$0.07 &     6.05$\pm$0.05 & 0.02 & 0.093 & 0.1055 & 0.42 \\
HAT-P-17  &    57.04$\pm$0.08 &	 18.10$\pm$0.20 & $-$14.38$\pm$0.10 & 0.05 & 0.274 & 0.2545 & 1.32 \\
HAT-P-18  &     9.43$\pm$0.10 & $-$13.17$\pm$0.14 &     3.91$\pm$0.12 & 0.01 & 0.121 & 0.1378 & $-$0.47 \\
HAT-P-20  &    11.32$\pm$0.38 & $-$11.97$\pm$0.11 & $-$12.41$\pm$0.13 & 0.01 & 0.121 & 0.1811 & $-$0.41 \\
HAT-P-21  &    4.394$\pm$0.11 &	 25.98$\pm$0.12 & $-$46.58$\pm$0.26 & 0.52 & 0.169 & 1.2478 & 1.48 \\
HAT-P-22  & $-$26.33$\pm$0.14 &	 44.72$\pm$0.06 &     3.10$\pm$0.19 & 0.08 & 0.221 & 0.1447 & 2.27 \\
HAT-P-26  &    66.50$\pm$0.18 & $-$45.10$\pm$0.22 & $-$27.74$\pm$0.21 & 0.50 & 0.351 & 0.5654 & $-$0.96 \\
HAT-P-29  &    17.61$\pm$0.16 &	  8.99$\pm$0.16 &     7.80$\pm$0.05 & 0.01 & 0.134 & 0.1400 & 0.50 \\
HAT-P-30  & $-$60.33$\pm$0.17 &	  9.90$\pm$0.14 &    18.61$\pm$0.10 & 0.05 & 0.147 & 0.3742 & 0.58 \\
HAT-P-36  & $-$24.21$\pm$0.07 &	 10.82$\pm$0.07 & $-$13.54$\pm$0.21 & 0.02 & 0.057 & 0.3996 & 0.47 \\
KELT-6   & $-$23.49$\pm$0.05 &	 25.18$\pm$0.07 &     6.83$\pm$0.21 & 0.02 & 0.125 & 0.3533 & 1.14 \\
Qatar-1  & $-$52.17$\pm$0.09 & $-$34.71$\pm$0.24 &    10.76$\pm$0.09 & 0.06 & 0.207 & 0.1914 & $-$1.14 \\
Qatar-2  & $-$72.15$\pm$0.24 & $-$35.94$\pm$0.19 &  $-$2.65$\pm$0.20 & 0.11 & 0.251 & 0.1810 & $-$1.05 \\
TRES-4   &    31.47$\pm$0.33 & $-$24.06$\pm$0.21 &     0.96$\pm$0.16 & 0.02 & 0.209 & 0.2734 & $-$0.70 \\
WASP-10  & $-$14.40$\pm$0.05 &  $-$1.86$\pm$0.19 & $-$11.48$\pm$0.10 & 0.01 & 0.022 & 0.1537 & $-$0.11 \\
WASP-11  & $-$11.82$\pm$0.18 &  $-$3.57$\pm$0.16 & $-$15.41$\pm$0.18 & 0.01 & 0.033 & 0.2131 & $-$0.16 \\
WASP-13  & $-$14.67$\pm$0.15 &  $-$9.79$\pm$0.10 &     9.92$\pm$0.15 & 0.01 & 0.054 & 0.2485 & $-$0.39 \\
WASP-38  &  $-$6.52$\pm$0.18 & $-$20.32$\pm$0.09 &     5.40$\pm$0.16 & 0.01 & 0.117 & 0.1391 & $-$0.76 \\
WASP-39  & $-$56.70$\pm$0.13 &	 10.81$\pm$0.05 & $-$31.77$\pm$0.16 & 0.12 & 0.141 & 0.6214 & 0.60 \\
WASP-43  & $-$13.43$\pm$0.06 &	  2.92$\pm$0.18 & $-$14.20$\pm$0.15 & 0.01 & 0.023 & 0.2109 & 0.09 \\
WASP-54  &  $-$5.45$\pm$0.13 & $-$15.70$\pm$0.13 &  $-$7.19$\pm$0.20 & 0.01 & 0.096 & 0.2652 & $-$0.59 \\
WASP-60  & $-$50.74$\pm$0.46 & $-$38.84$\pm$0.28 &  $-$5.89$\pm$0.19 & 0.06 & 0.222 & 0.2096 & $-$1.28 \\
XO2-N    & $-$87.53$\pm$0.19 & $-$78.23$\pm$0.25 &  $-$2.35$\pm$0.12 &13.48 & 0.436 & 0.1036 & $-$2.01\\
XO2-S    & $-$87.06$\pm$0.19 & $-$78.15$\pm$0.24 &  $-$2.31$\pm$0.11 &13.01 & 0.435 & 0.1035 & $-$2.01 \\
\hline
\end{tabular}
\end{center}
Note: A positive/negative difference in $\Delta(R_{\rm mean}-R_{\rm GC})$ greater/smaller than $+1/-1$\,kpc indicates possible migration from the outer/inner Galactic disk.
\end{table*}



\section{Comparison with previous works}

\begin{figure*}[h]
\centering
\includegraphics[width=6cm,angle=90]{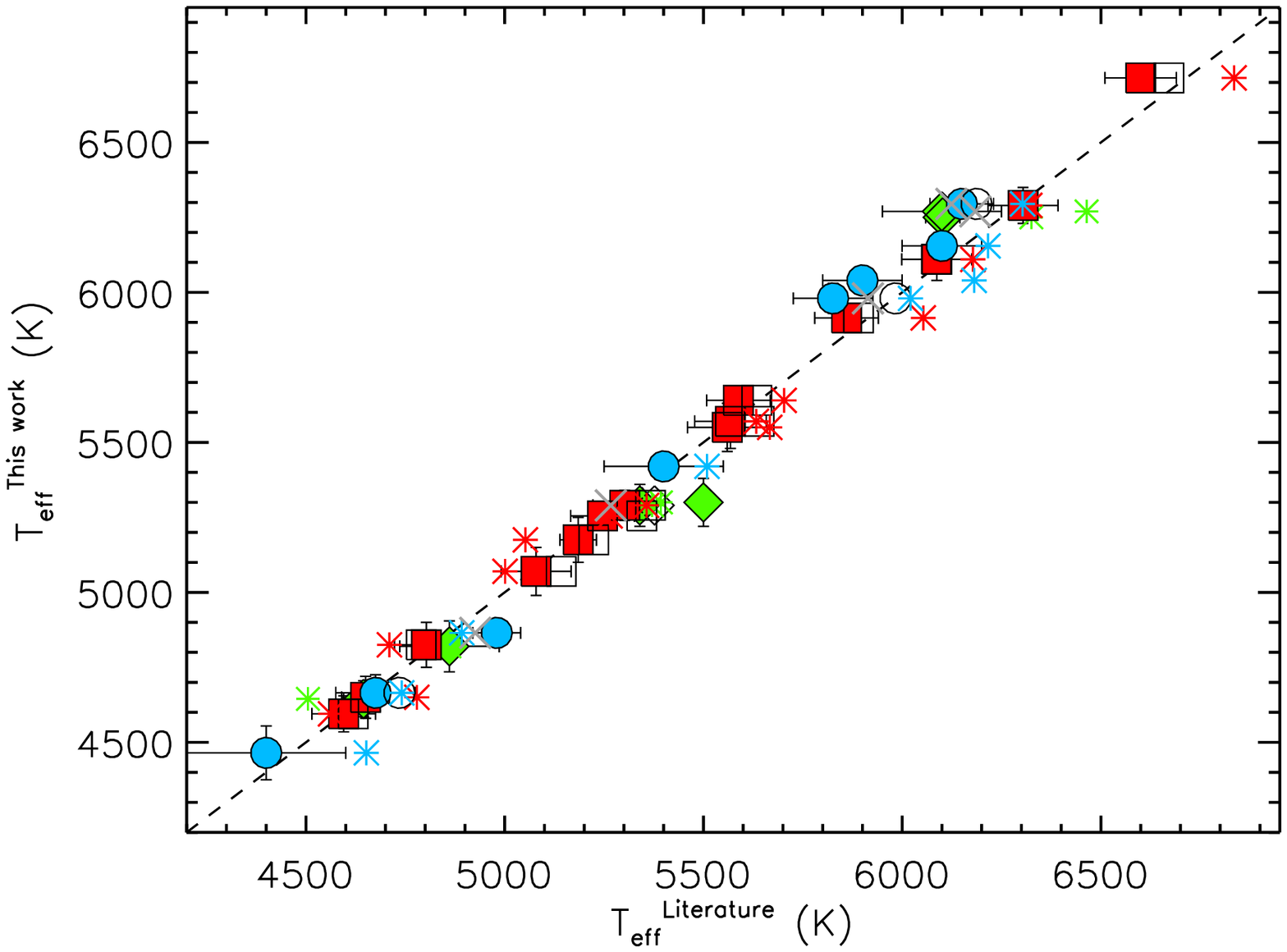}
\includegraphics[width=6cm,angle=90]{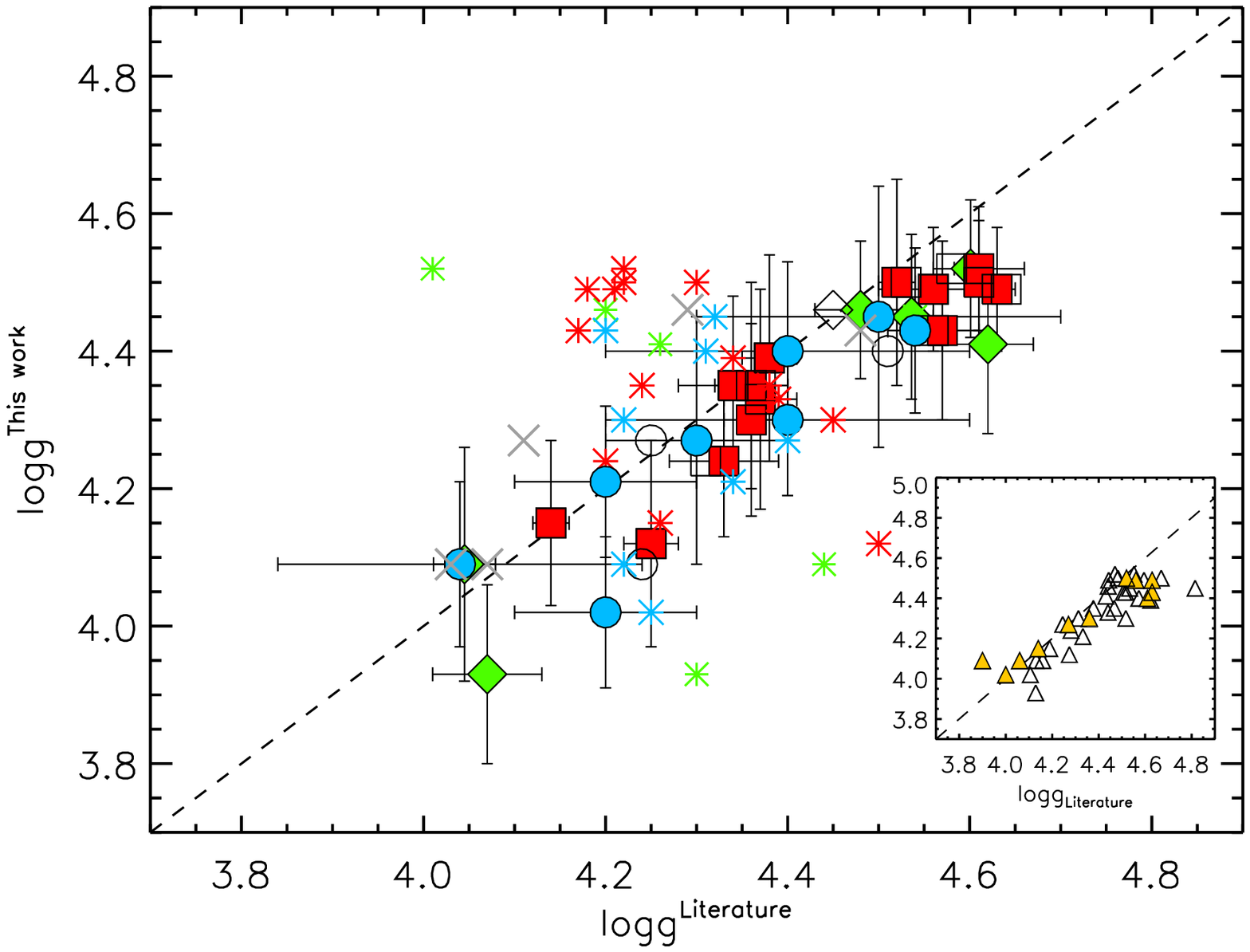}
\includegraphics[width=6cm,angle=90]{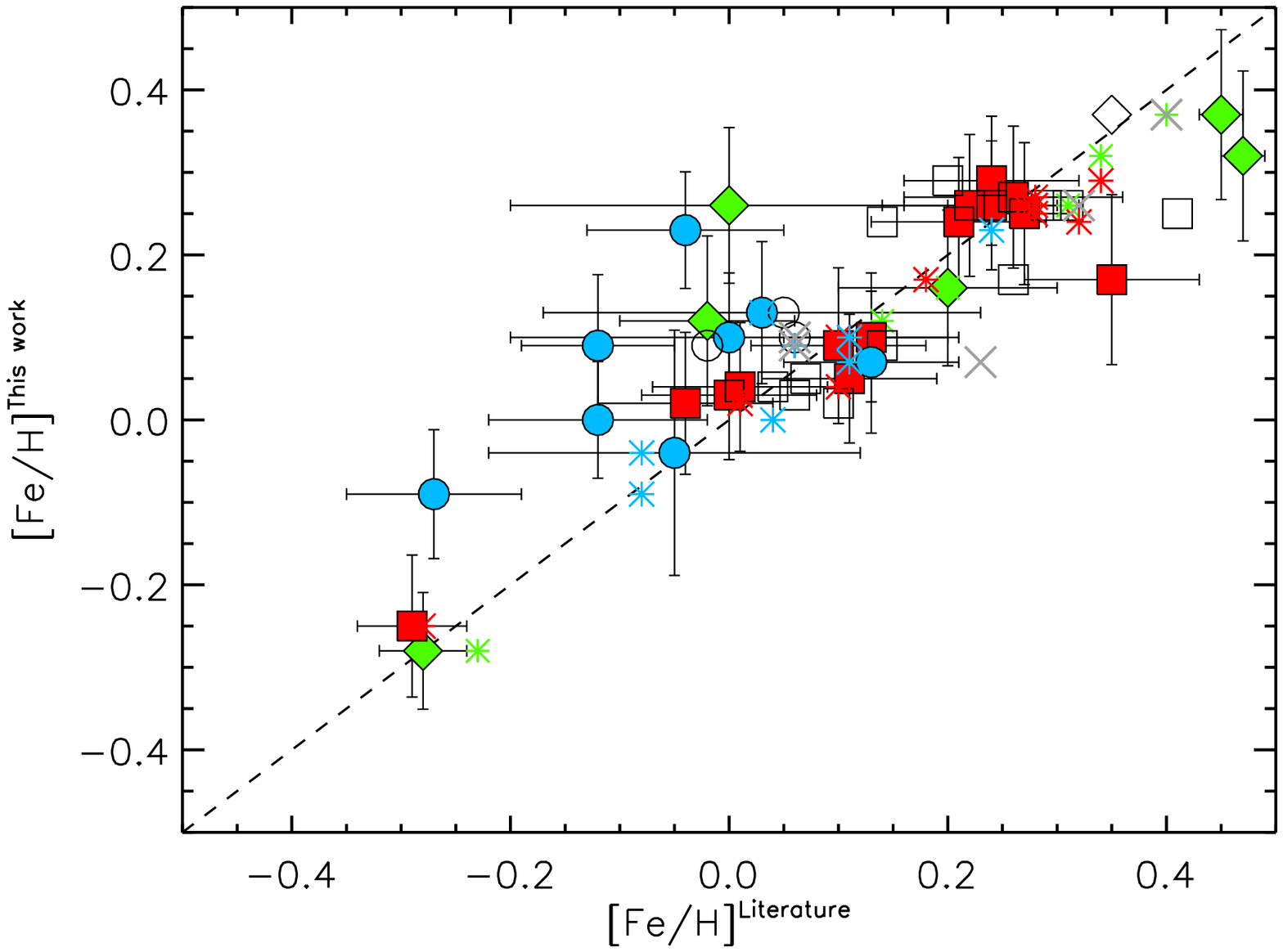}
\includegraphics[width=6cm,angle=90]{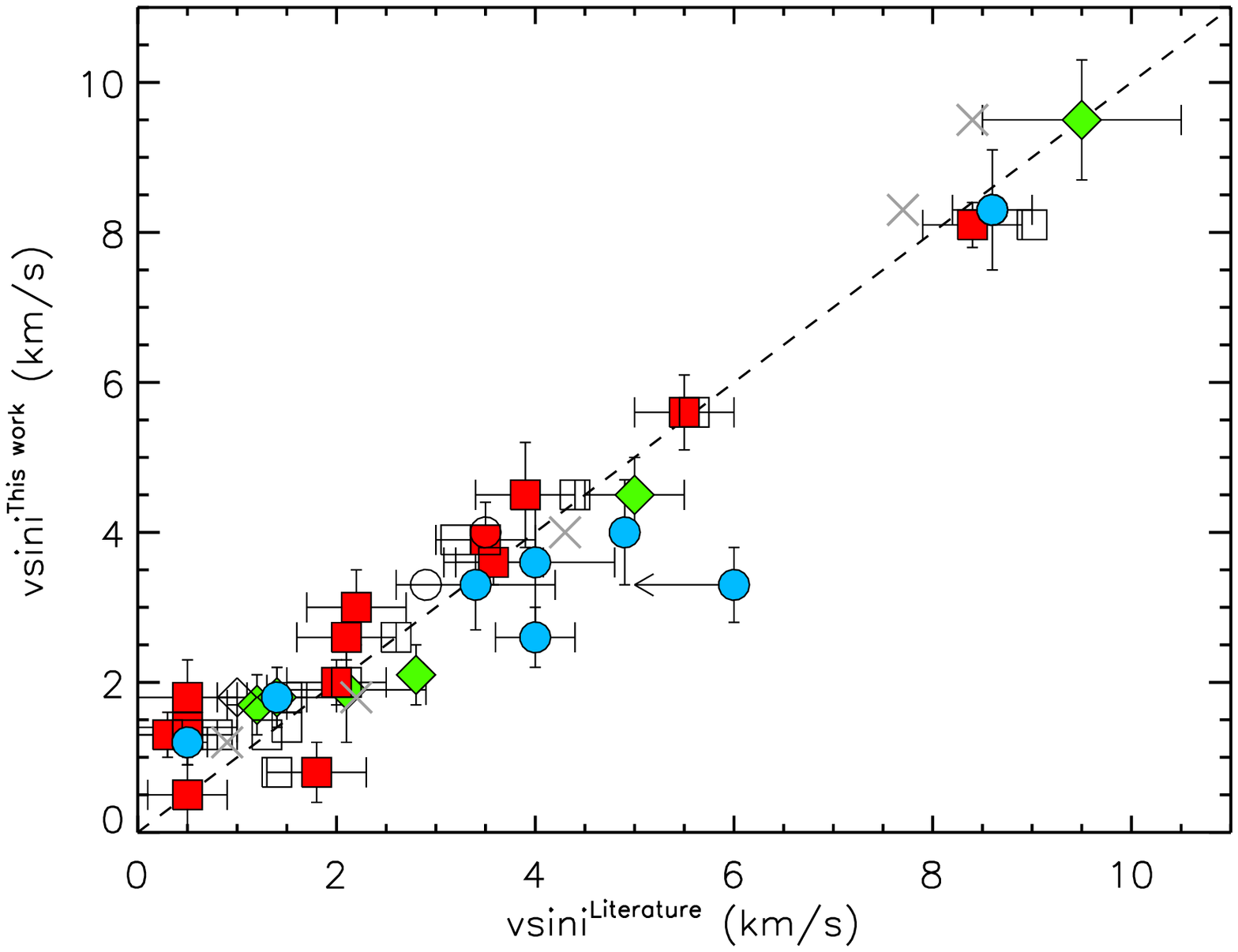}
\caption{Comparison between our stellar parameters (\teff, \logg, \feh, \vsini) and those from the literature. Filled and open symbols represent the comparisons with the discovery papers and \cite{Torresetal2012}, respectively (squares mark the HAT-P sample, circles the WASP sample, and diamonds the remaining sample). Asterisks represent the comparison with the SWEET-Cat (in the \logg\,plot the inset shows the comparison with the measurements obtained with the light curves by \citealt{Mortieretal2013} - filled yellow triangles - or through Gaia parallaxes by \citealt{Sousaetal2021} - open triangles). Crosses mark the values obtained by \cite{Breweretal2016}.}
\label{fig:param_comp} 
\end{figure*}

\section{[X/Fe] versus [Fe/H]}

\begin{figure*}[h]
\begin{center}
\includegraphics[width=14cm,angle=90]{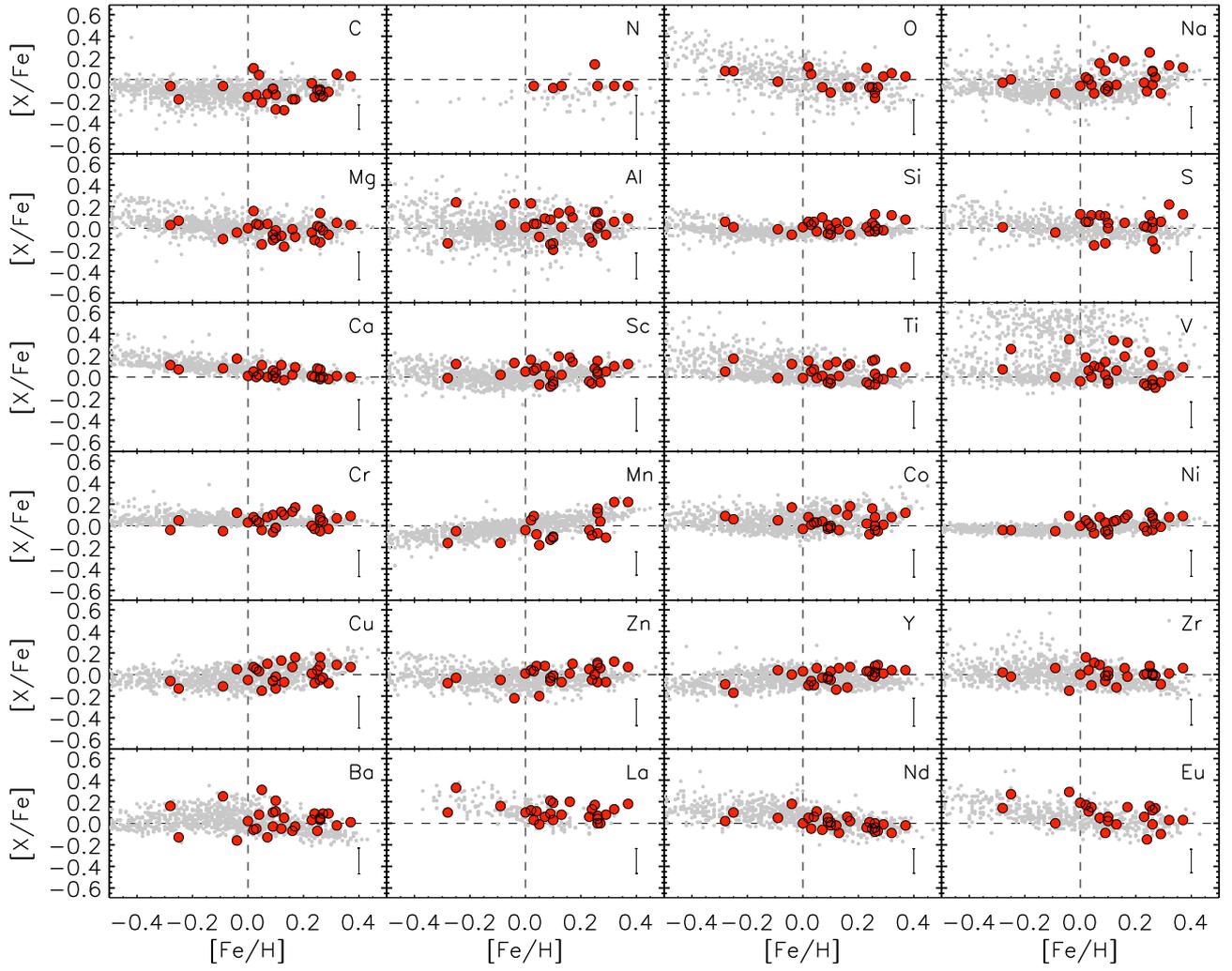}
\caption{[X/Fe] versus [Fe/H] for our sample. See Fig.\,\ref{fig:abundances_FeH} for the references of the overplotted grey points. Mean error bars are shown on the bottom-right of each panel. Solar values are marked with dashed lines.}
\label{fig:abundancesFe_FeH} 
\end{center}
\end{figure*}

\end{appendix}

\end{document}